\documentclass[aps,prl,twocolumn,superscriptaddress, 10pt]{revtex4-2} 

\usepackage{tabularx}
\makeatletter
\def\hlinewd#1{%
\noalign{\ifnum0=`}\fi\hrule \@height #1 %
\futurelet\reserved@a\@xhline}
\makeatother
\usepackage{scrextend}
\usepackage{amsmath}
\usepackage{epsfig}
\usepackage{blindtext}
\usepackage[normalem]{ulem}
\usepackage{microtype}
\usepackage{subfigure}
\usepackage{siunitx}
\usepackage{multirow}

\usepackage[T1]{fontenc}
\usepackage[utf8]{inputenc}
\usepackage{csquotes}
\usepackage{amsmath,amssymb,amsthm}
\usepackage{graphicx}
\usepackage{xcolor}
\usepackage{orcidlink} 

\numberwithin{equation}{section}
\renewcommand{\theequation}{\arabic{section}.\arabic{equation}}

\usepackage[dvipsnames]{xcolor}

\usepackage[labelfont=bf]{caption}

\usepackage[export]{adjustbox}

\usepackage{hyperref}
\hypersetup{
    colorlinks=true,
    linkcolor=MidnightBlue,
    citecolor=ForestGreen,
    urlcolor=RoyalBlue
}

\renewcommand{\theequation}{\arabic{equation}}
\makeatletter
\@removefromreset{equation}{section}
\makeatother

\newcommand{\dd}[0]{\mathrm{d}}

\newcommand{\github}[1]{\href{https://github.com/mathieukaltschmidt/#1}{\includegraphics[width=9pt, valign=c]{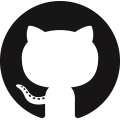}\ #1}}

\newcommand{\githubj}[1]{\href{https://github.com/veintemillas/#1}{\includegraphics[width=9pt, valign=c]{figures/github.png}\ #1}}

\newcommand{\n}{\mathfrak{n}}

\newcommand{\ctheta}{{\psi}}
\newcommand{\hpsi}{{\widehat{\psi}}}
\newcommand{\cmass}{{m_\ctheta}}

\renewcommand\({\left(}
\renewcommand\){\right)}
\renewcommand\[{\left[}

\newcommand{\be}{\begin{equation}}
\newcommand{\ee}{\end{equation}}
\newcommand{\bea}{\begin{eqnarray}}
\newcommand{\eea}{\end{eqnarray}}

\begin{document}

\preprint{KANAZAWA-26-05}

\newcommand{\maintitle}{Amnesia in the Axion Misalignment Landscape}

\title{\maintitle}

\date{\today}

\author{Kierthika Chathirathas~\orcidlink{0009-0005-5041-9867}}
\affiliation{Institute for Astroparticle Physics (IAP), Karlsruhe Institute of Technology (KIT), 76131 Karlsruhe, Germany}

\author{Cem Eröncel~\orcidlink{0000-0002-9308-1449}}
\affiliation{\.Istinye University, Faculty of Engineering and Natural Sciences, 34396,
\.Istanbul, Türkiye}

\author{Mathieu Kaltschmidt~\orcidlink{0000-0002-6470-5371}}
\affiliation{CAPA \& Departamento de F\'isica Te\'orica,
			Universidad de Zaragoza,
			C. Pedro Cerbuna 12,
			50009 Zaragoza,
			Spain}
			
\author{Javier Redondo~\orcidlink{0000-0002-1044-8197}}
\affiliation{CAPA \& Departamento de F\'isica Te\'orica,
			Universidad de Zaragoza,
			C. Pedro Cerbuna 12,
			50009 Zaragoza,
			Spain}
			
\author{Ken'ichi Saikawa~\orcidlink{0000-0002-9205-9813}}
\affiliation{Institute for Theoretical Physics, 
			Kanazawa University, 
            Kakuma-machi, Kanazawa, 
			Ishikawa 920-1192, 
			Japan}

\begin{abstract}
Pre-inflationary axion cosmologies admit a broad range of initial velocities and angles that
generate a landscape of misalignment histories, whose nonlinear fate and
small-scale structure remain uncertain. We use lattice simulations to follow
the QCD axion in the kinetic misalignment scenario through the QCD crossover
into the non-relativistic regime, towards matter-radiation equality. In
the meV mass range, this landscape contains a continuum of
\enquote{summit-landing} solutions sensitive to both the initial angle and
velocity.
Nonlinear fragmentation reduces the comoving axion number and modifies the
relation between the misalignment history and present relic abundance. For
sufficiently large axion masses, strong self-interactions drive the
small-scale spectrum towards the attractor found for post-inflationary
axions, so that distinct early histories can lead to similar late-time
structures.
\end{abstract}

\maketitle

\section{Introduction}\label{sec:intro}

The hypothetical quantum chromodynamics (QCD) axion~\cite{Weinberg:1977ma, Wilczek:1977pj}, pseudo-Nambu-Goldstone boson of the spontaneous breaking of the Peccei-Quinn (PQ) symmetry~\cite{Peccei:1977hh, Peccei:1977ur}, would simultaneously solve the strong CP problem while standing out as a compelling cold dark matter (CDM) candidate~\cite{Preskill:1982cy, Abbott:1982af, Dine:1982ah} with rich phenomenology~\cite{Arza:2026rsl,OHare:2024nmr,Marsh:2015xka, Sikivie:2006ni}. Despite a broad experimental program, including laboratory searches, helioscopes and haloscopes~\cite{Graham:2015ouw, Irastorza:2018dyq, Berlin:2024pzi, Baryakhtar:2025jwh}, a sizeable part of the parameter space remains unexplored~\cite{AxionLimits}. Particularly relevant for this work, the meV mass range has repeatedly been identified as a \enquote{focal point} of axion physics, see Refs.~\cite{Raffelt:2011ft, Cicoli:2026fqp}. 

The kinetic misalignment mechanism~\cite{Co:2019jts,Chang:2019tvx} in the
pre-inflationary scenario, where the PQ symmetry is broken before inflation
and never restored, naturally populates the meV range. A large initial field
velocity allows the axion to cross many periods of its potential before
being trapped by QCD. The evolution resembles a \emph{cosmic roulette}: the
initial angle $\Theta_1$ and velocity $v$ determine how many barriers are
crossed and whether the field stops close to a summit, so nearby initial
conditions can lead to sharply different trapping and fragmentation
histories. Kinetic misalignment should therefore be viewed as a
two-dimensional axion misalignment landscape in $(\Theta_1,v)$, rather than
as a mechanism controlled by the initial velocity alone.

The pre-inflationary axion inherits the small primordial adiabatic perturbations generated by
the inflaton, which nonlinearities of the QCD potential can amplify to
$\mathcal O(1)$ values and fragment the homogeneous condensate
~\cite{Eroncel:2022vjg}. The effect of kinetic fragmentation on the relic
abundance has been studied previously~\cite{Eroncel:2022vjg}, and its
potential to seed miniclusters in the pre-inflationary scenario was explored
in Ref.~\cite{Eroncel:2022efc}. However, the subsequent nonlinear evolution
of this small-scale structure was not followed. Since post-inflationary
axions instead inherit order-one inhomogeneities from causally disconnected
regions and topological defects, previous studies suggested that the
pre- and post-inflationary scenarios could produce distinguishable
minicluster populations~\cite{Arvanitaki:2019rax,Eroncel:2022efc,
Gorghetto:2024vnp}. Whether self-interactions preserve this memory of the
primordial history is the central question of this work. Addressing it
requires high-performance-computing simulations that resolve the nonlinear
fragmentation around trapping and follow the subsequent spectral evolution
across the QCD crossover towards matter--radiation equality (MRE).

In this Letter, we first map the misalignment landscape and identify two summit-landing
branches, showing explicitly that the relic abundance can be highly
sensitive to $\Theta_1$ as well as to $v$. Three-dimensional simulations
around the QCD epoch then quantify how fragmentation changes the axion number
for sufficiently small $f_A$, corresponding to large axion masses and
velocities. Most importantly, by following the post-fragmentation spectrum
towards MRE, we find that sufficiently strong
self-interactions drive it towards the attractor previously found for
post-inflationary axions~\cite{Gorghetto:2024vnp}. Thus, a rich early-time
misalignment landscape can lead either to universal late-time small-scale
structure or, when self-interactions are inefficient, to surviving memory of
the initial conditions. This has direct implications for axion miniclusters
and other small-scale probes of axion dark matter~\cite{HOGAN1988228,
Kolb_1993,Kolb:1993hw,Vaquero:2018tib,Pierobon:2023ozb}.
\section{Misalignment and Summit Landings}\label{sec:homogeneous}

\begin{figure}[tbp]
\centering
\includegraphics[width=\columnwidth]{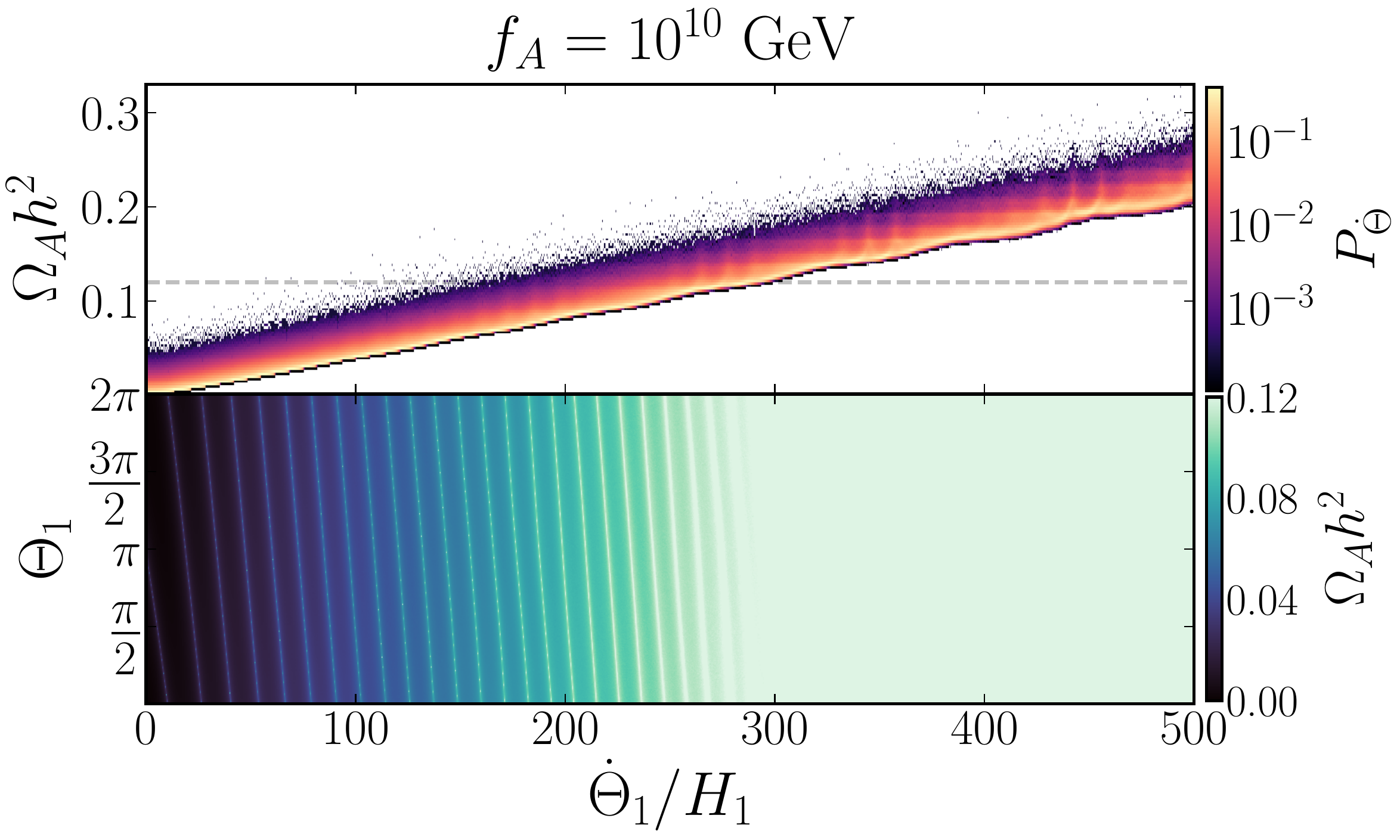}
\caption{\small\emph{Lower panel:} Axion CDM density $\Omega_Ah^2$ as a function of the zero mode initial conditions $(\dot{\Theta}_1/H_1,\Theta_1)$ for $f_A = 10^{10}$~GeV, neglecting fluctuations. \emph{Upper panel:} The corresponding density distribution $P_{\dot{\Theta}}\equiv P(\Omega_Ah^2|\dot{\Theta}_1)$, obtained by marginalising the lower panel over $\Theta_1$ with a uniform prior and normalising at each $\dot{\Theta}_1$.}
\label{fig:grid}
\end{figure}

The QCD axion $A$ is often expressed via the misalignment angle $\theta \equiv A/f_{A}$, with $f_{A}$ the axion decay constant. In conformal time, $d\eta = dt/R$, where $R$ is 
the Friedmann-Lemaître-Robertson-Walker (FLRW) scale factor, its early Universe dynamics are governed by
\begin{equation}
\theta'' + 2{\cal H}\theta'-\triangle \theta + m_A^2R^2\sin\theta = G,
\end{equation}
where ${\cal H}=R'/R$,   $m_A^2(T)=\chi(T)/f_A^2$ is the temperature-dependent axion mass with $\chi(T)$ the QCD topological susceptibility, and $G$ is the source term from gravity. We focus on the evolution around Universe temperatures of order GeV.  We assume radiation domination and use the lattice QCD results from Ref.~\cite{Borsanyi:2016ksw} for $\chi(T)$ and thermal degrees of freedom. Primes (dots) denote derivatives with respect to conformal (physical) time.

We first consider the zero mode $\Theta = \langle \theta \rangle$, neglecting initially small perturbations. The dynamical time scale is $\eta_1$ defined by $m_A=H(\eta_1)$, after which the QCD potential effectively pushes $\Theta\to 0$ and eventually drives damped harmonic oscillations, which behave as cold dark matter~\cite{Abbott:1982af, Preskill:1982cy, Dine:1982ah}. The abundance today can be expressed parametrically as 
\begin{equation}
\begin{aligned}
\Omega_Ah^2 &\sim 0.12 \left(\frac{f_A}{10^{12}\rm GeV}\right)^\frac{\n+6}{\n+4}(1-\cos\Theta_s)\\&\times
\left(1+v^\frac{\n+4}{\n+6}+\frac{\n+4}{6}\log\frac{\pi}{\pi-|\Theta_s|}\right)^\frac{\n+6}{\n+4}, 
\end{aligned}
\end{equation}
in terms of the initial velocity $v\equiv\dot\Theta_1/H_1$ and the stopping angle $\Theta_s$, defined for moving solutions by the first turning point, $\Theta'(\eta_s)=0$. For $v=0$, we formally identify $\Theta_s=\Theta(\eta\to0)$.
The main term, involving $f_A$ explicitly, arises because the axion energy is the QCD vacuum energy $\propto \chi$ driven to damped oscillations, and the larger $f_A$ is, the later they start and the less damped they become until today. The fact that $\chi\propto T^{-\n}$ with $\n\sim 8$ down to $T_c\sim 150$ MeV~\cite{Borsanyi:2016ksw} enhances this effect. A large initial velocity $v\gg 1$ delays the onset of the oscillations until the field can be ``trapped'' by the QCD potential, thus enhancing the yield proportionally to $v$ and implying a large stopping angle $\Theta_s>\pi/2$. Finally, if the stopping angle is very close to $\pi$, there is a logarithmic delay of the oscillation onset that also enhances the yield.  
The above formula covers the kinetic misalignment~\cite{Co:2019jts, Chang:2019tvx} ($v\gg1$) and the standard misalignment mechanisms~\cite{Abbott:1982af, Preskill:1982cy, Dine:1982ah} ($v=0$), including the large-misalignment case~\cite{Arvanitaki:2019rax}, where $\Theta$ starts close to the hilltop. The dependence of $\Omega_A$ on the initial conditions $\Theta_1=\Theta(\eta_1)$ and velocity $v$ computed numerically shows these trends, see Fig. \ref{fig:grid}.

\begin{figure}[tbp]
\centering
\includegraphics[width=\columnwidth]{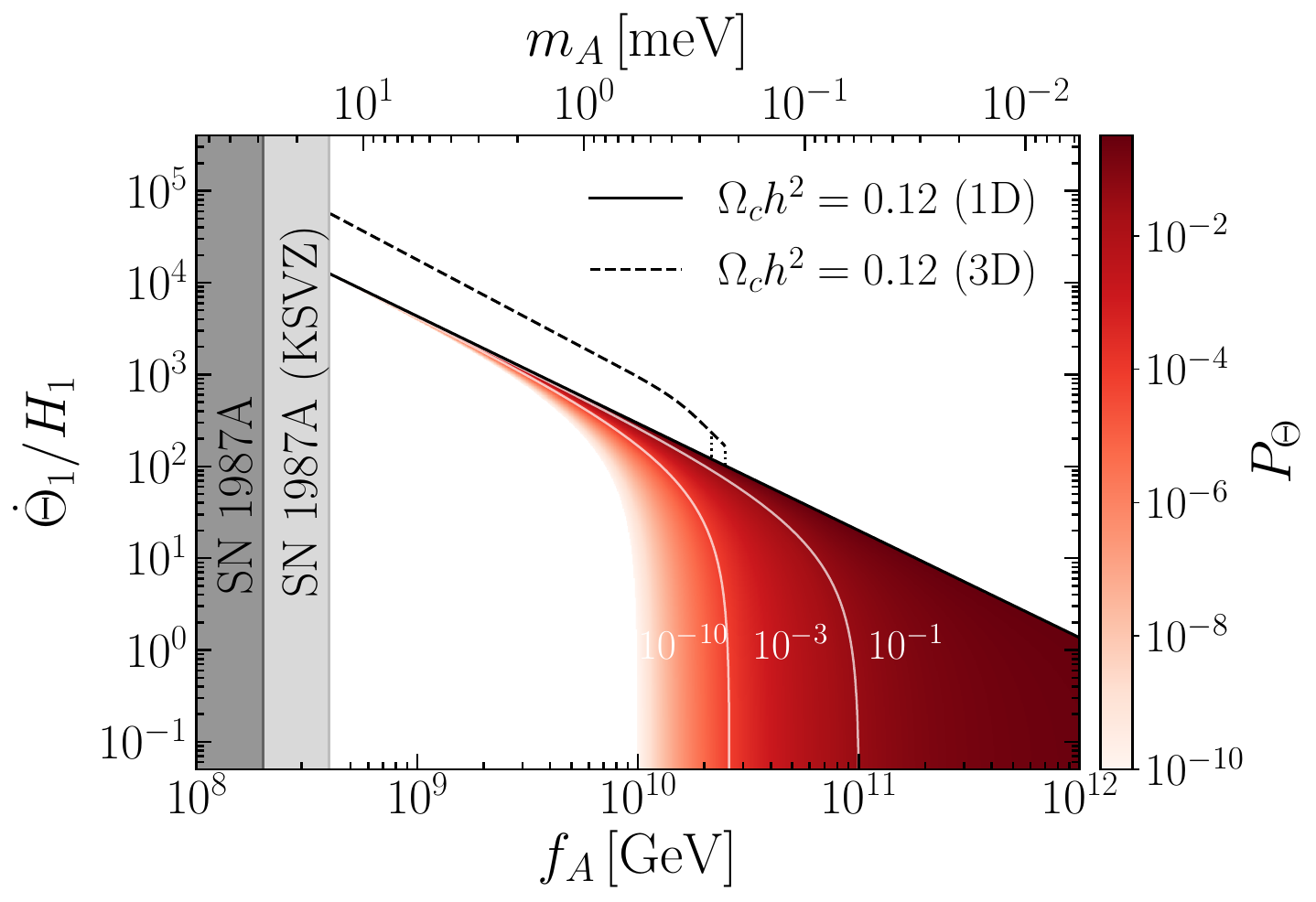}
\caption{\small Axion zero mode velocity reproducing the observed CDM relic density from zero mode (1D, solid) and 3D simulations including adiabatic fluctuations (dashed).  The red shaded region shows the fine-tuning probabilities $P_{\Theta}$ together with some reference lines in white. 
We also show excluded regions for KSVZ axions~\cite{Raffelt:2006cw, Caputo:2024oqc} and general models~\cite{Springmann:2024ret}.
}
\label{fig:finetuning}
\end{figure}

\begin{figure*}[tbh]
    \centering
    \includegraphics[width=\linewidth]{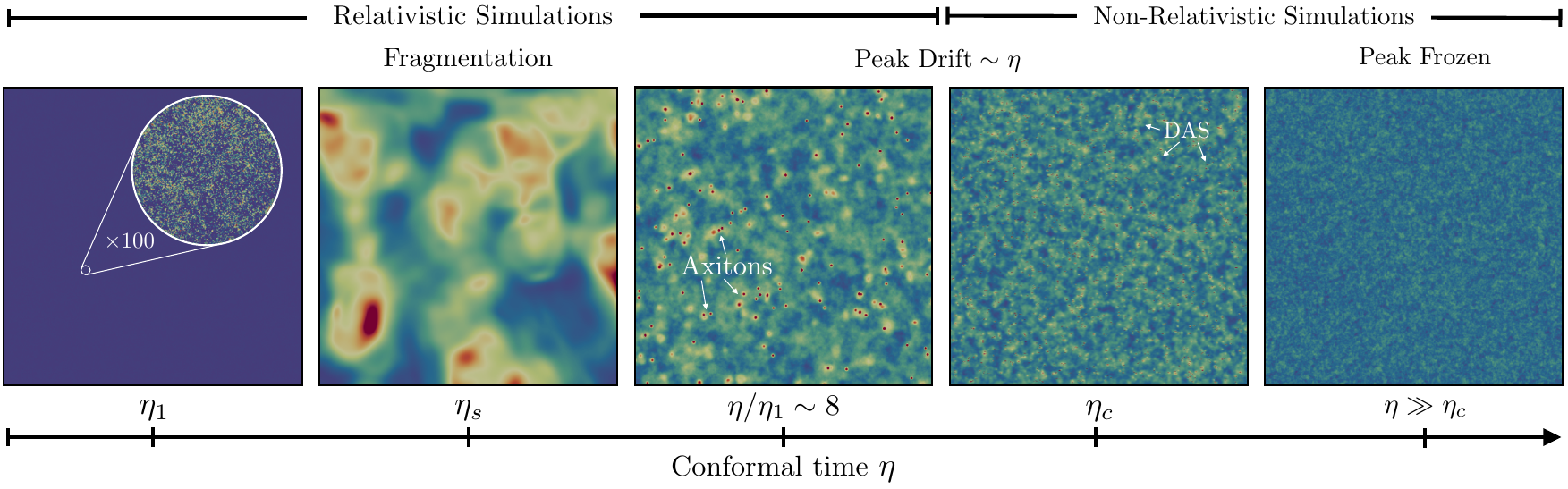}
    \caption{\small Evolution of the energy density contrast, $\delta=\rho_A/\langle\rho_A\rangle -1$, projected along the line of sight for a typical kinetic misalignment simulation. The colormap saturates at $\delta\sim \mathcal{O}(1)$. The initially small fluctuations at $\eta_1$ (here $\delta\sim10^{-2}$) are amplified, and the field fragments at $\eta_s$. A growing population of axitons  forms, and we must stop our relativistic simulations once these shrink to sub-grid sizes. Afterwards, simulations of the non-relativistic field beyond $\eta_c$ are possible once wave-collapse into dense axion stars (DAS) is halted at the lattice resolution, preventing sub-grid modes from contaminating the evolution. After $\eta_c$ the dynamics effectively freeze until MRE.}
    \label{fig:history}
\end{figure*}

Previously, it was known that for $f_A\lesssim 10^{12}$ GeV, the measured relic density $\Omega_ch^2=0.12$~\cite{Planck:2018vyg} can be obtained either through the \emph{large} misalignment mechanism, i.e.  $\Theta_1\sim \pi$~\cite{Arvanitaki:2019rax}, or the kinetic misalignment mechanism with $v=v_{\rm KM}\gg 1$. However, we can identify a continuum of intermediate solutions, where a smaller initial velocity drives the field to stopping close to $\pi$ by  lucky coincidence, and thus to a combined enhancement of the relic density. We call these solutions \emph{summit landings} and they interpolate between the misalignment and kinetic misalignment regimes.   
Since for each $v$, there is a value of $\Theta_1$ for which $\Theta_s=\pi$, the field resting there for an unconstrained amount of time can result in an arbitrarily large $\Omega_A$. 
There are two values of $\Theta_1$, generally very close to each other, for which the CDM yield corresponds to observations. The upper panel of Fig.~\ref{fig:grid} shows 
the probability distribution for $\Omega_A$ marginalised over $\Theta_1$, reflecting the linear $v$-dependence and that summit landings at smaller $v$ exist with reasonable probabilities. The level of fine-tuning, estimated analytically as a function of $f_A$, is shown in Fig.~\ref{fig:finetuning}.

This observation highlights a higher level of complexity than previously expected. For each $f_A$, we can have multiple values of $v$ yielding the correct relic density. The different zero-mode histories will amplify fluctuations differently and thus, in general, we cannot conclude that axion CDM substructure is defined only by $f_A$. This is particularly relevant for $f_A>10^{10}$ GeV, for which summit landings are not unlikely. 
At small $f_A$, however, they are extremely fine-tuned, and when neglecting them we should be able to unambiguously predict  the small scale properties of axion CDM. Most remarkably, in the small $f_A$ regime we need large $v$, which implies that axion amplitudes will be large until late, and as it turns out the substructure is strongly determined by nonlinear dynamics. 

The pioneering studies of Refs.~\cite{Eroncel:2022efc,Eroncel:2024rpe} considered only the linear evolution of fluctuations, rendering them inaccurate at small $f_A$. By also neglecting summit landings, they remain incomplete at large $f_A$ as well.

\section{Fragmentation and Relic Abundance}\label{sec:fragmentatation}

In the pre-inflationary scenario, quantum fluctuations of the inflaton~\footnote{Isocurvature fluctuations originating from the axion field ~\cite{Kobayashi:2013nva, Planck:2018jri} are model dependent and strongly constrained~\cite{Planck:2018jri} so we neglect them here.} are imprinted on the axion through gravity as adiabatic curvature and temperature perturbations~\cite{Weinberg:2003sw, Allali:2025pja}.
\nocite{Kobayashi:2013nva,Planck:2018jri}
As a minimal assumption, we seed the axion field with the near scale-invariant spectrum, measured at large length scales, $k^{-1}\sim (1-10^4)\,\rm Mpc$, by CMB and large-scale structure probes \cite{Planck:2018vyg, eBOSS:2020yzd, DES:2021wwk}, extrapolated to the small scales $k_1\equiv\mathcal{H}_1\sim {\rm mpc}^{-1}$ relevant for our simulations. 

These fluctuations are initially tiny, so we can consider the linearized evolution of comoving Fourier modes,
\begin{equation}
\ctheta_k'' + w_k^2\,\ctheta_k = G_k,
\label{eqn:eom_modes}
\end{equation}
with comoving field $\ctheta_k=\theta_k R$, conformal frequency $w_k^2 = k^2 + m_A^2R^2\cos\Theta - R''/R$ set by the
average field $\Theta$, and $G_k\propto \Phi_k(0)$ the gravitational field perturbation.  
The adiabatic initial conditions are
$\theta_k(\eta\to0)=-(v/2)(R_1/R)\Phi_k(0)$ and
$\theta_k'(\eta\to0)=-\mathcal H\theta_k$
\cite{Eroncel:2022vjg}.

Fluctuations grow by extracting energy from the zero mode until it is
considerably depleted or the modes decouple. The gravitational source can
provide an additional seed for this amplification.
The former case is referred to as \emph{fragmentation} and marks the end of the linear regime. The energy transfer proceeds via \emph{tachyonic} and \emph{parametric resonances}~\cite{Fonseca:2019ypl, Morgante:2021bks, Eroncel:2022vjg} and the \emph{gravitational source term} \cite{Sikivie:2021trt}, for details see the SupM~\cite{supM}.

To satisfy $\Omega_Ah^2=0.12$ for $f_A < 2\times 10^{10}\,\mathrm{GeV}$, we need $v\gtrsim 100$ for which the field fragments quickly (around trapping) before the gravitational effects from $G_k$ can affect the evolution so we neglect it~\footnote{This can be estimated from the linear evolution and also checked a posteriori}, see however~\cite{Hardy:2026tgz,Angulo:2026auo}.  
To follow the resulting nonlinear evolution we resort to 3D simulations with the \githubj{jaxions} code~\cite{Jaxions:2026} to discretise the axion field on a 3D Cartesian grid with up to $8192^3$ lattice sites, prepare the adiabatic initial conditions at $\eta_1$ and evolve them with a 4th (6th) order precision scheme in time (space). 

Fragmentation transfers energy from the zero mode into modes with nonzero
momentum $k/R\sim m_A$ and, through nonlinear number-changing processes,
reduces the comoving axion number and relic density.
Since much later, when amplitudes are small, nonlinearities are dominated by the leading quartic self-interaction, which conserves axion number, the number density should approach a constant that determines $\Omega_Ah^2$.    
We find that fragmentation lowers the prediction from the homogeneous (1D) field evolution by a dilution factor $D\equiv\Omega_A^{\rm 3D}/\Omega_A^{\rm 1D}\sim0.2 - 0.3$
at high velocity, up to a factor of $\sim 4$, consistent with Ref.~\cite{Fasiello:2025ptb}. 
A naive estimate from the lowest order number-violating process $4a\to2a$ gives $D\sim1/\sqrt2$, nearly a factor
of three off. Thus, fragmentation is truly nonlinear and non-perturbative. 

Calibrating $D(v)$ using the 3D runs yields the velocity required for any $f_A$, or equivalently mass $m_A$, to account for the observed CDM, as shown in Fig.~\ref{fig:finetuning}. Because $D(v)$ drops off steeply
for $v\gtrsim100$, the condition $\Omega_A h^2=0.12$ is degenerate in a narrow
window $2.15\times10^{10}\lesssim f_A/\mathrm{GeV}\lesssim2.48\times10^{10}$, where two velocities reproduce the observed abundance. 

At lower velocities, tuned initial conditions instead lead to summit landings, where the tachyonic instability at $\Theta=\pi$ amplifies fluctuations strongly, which can further suppress the abundance. We present one example of a summit landing case in the SupM~\cite{supM}, although we stress that a complete numerical treatment of this regime is very complicated and resource-intensive. 

\section{Post-fragmentation Evolution}\label{sec:post-fragmentation}
Our $v\gtrsim 100$ simulations show that tachyonic and parametric resonance fragment the zero mode close to $\eta_s$ into axions with $k\lesssim (m_A R)_s$. The resulting field amplitudes are large, so nonlinear self-interactions further affect the field. 
We expect wave-collapse into long-lived oscillons, which in this context are called \emph{axitons}~\cite{Kolb:1993hw,Kolb:1994xc,Vaquero:2018tib}, and a slow number-conserving evolution due to $2a\to 2a$ scattering.  

The timeline of a typical simulation is shown in Fig.~\ref{fig:history}. 
We see the amplification of fluctuations at stopping, axitons which grow in number and decrease in size as $\sim (m_A R)^{-1}$ for as long as $\chi$ grows ($\eta<\eta_c$, where $\eta_c$ is the conformal time at which $m_A$ reaches its zero-temperature value), while the characteristic scale associated with the main spectral peak decreases---at a much slower rate---and localized objects eventually disperse.   

We record the comoving number spectrum $N(k) \propto R^3\frac{\partial n_A}{\partial\log k}$ and the dimensionless power spectrum $\Delta^2(k)$ to study the evolution.  
The main peak $k_p\sim (m_AR)_s$ evolves towards the ultraviolet (UV), first following the comoving mass $k_p\simeq m_AR$ and then more slowly as the peak axions become non-relativistic (NR). Even then, a secondary feature due to axitons is visible at $k\sim m_AR$.  Beyond the peaks, the spectrum  falls off abruptly to the level of the adiabatic fluctuations. The fast growth of the axion mass $m_AR\propto \eta^{\n/2+1\sim 5}$ pushes these axitons to subgrid sizes relatively fast, and the additional feature is pushed beyond the Nyquist wavenumber $k_{\mathrm{Nyq}}$. Failing to resolve axitons produces UV white noise which contaminates results in the IR, so we terminate our relativistic simulations when the comoving axiton size reaches the grid resolution.  

We continue our study by simulating the non-relativistic field $\Upsilon$,  where $ \psi =  (\Upsilon e^{-i \int^\eta m_A R \dd\eta'} + \text{h.c.})/\sqrt{2 m_A R}$, which obeys a modified Gross-Pitaevskii equation.   We modify the nonlinear evolution by using an unphysically small value of $\chi$ in the resummed potential term, preventing wave collapse from reaching unresolved scales and producing regulated dense-axion-star (DAS)-like configurations~\cite{Visinelli:2017ooc}. This allows to extend the evolution beyond the QCD cross-over, where $\n\to 0$ and the field is expected to freeze, through many orders of magnitude of free-streaming until matter-radiation equality ($\eta_{\rm eq}\sim 10^{10}\eta_1$), where gravity would become relevant.  

\begin{figure}[t]
    \centering
    \includegraphics[width=\columnwidth]{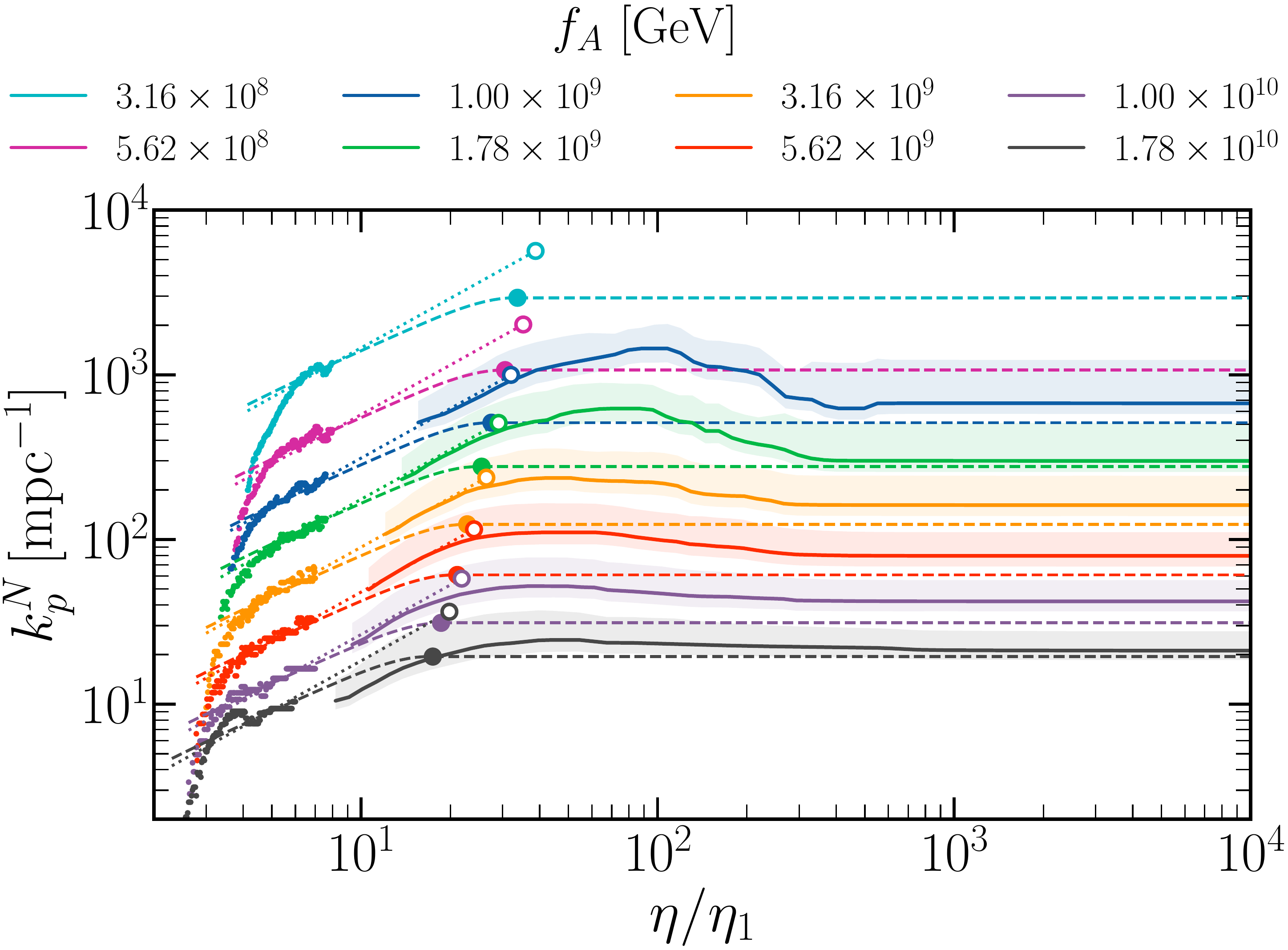}
    \caption{\small Evolution of the axion spectrum peak $k_p^{N}$ from relativistic ($\eta<8\eta_1$) and non-relativistic ($\eta>8\eta_1$) simulations over the theory model $\sim\sqrt{m_A}/R$ (dashed) and pure linear growth Eq.~\ref{eq:kp_pred} (dotted). Shaded areas show $25\%$ of the actual full-width-at-half-max (FWHM) of the fitted peak. The extrapolated $\eta_c$ is shown as filled/white circle.
    }
    \label{fig:kpeak_fit}
\end{figure}

\paragraph{Peak Drift from Self-Interactions.}
Non-relativistic $2a\to 2a$ scattering with coupling constant $\lambda=\chi(T)/f_A^4$ redistributes axions in momentum at a rate
$\Gamma \simeq \lambda^2 n_A^2 R^2/64m_A^3k_p^2$~\cite{Gorghetto:2024vnp}.
The rate is faster than the Universe expansion due to the strong temperature-dependence of the coupling, $\Gamma/H\propto m_A/k_p^2R^2\propto \eta^{\n/2-2}/k_p^2$.  
Scattering shifts the spectral peak $k_p$ towards the UV, which feeds back
into the scattering rate by lowering $\Gamma/H$.
Since the peak can only grow as long as $\Gamma/H\sim 1$, we expect an attractor behaviour where $k_p$ tracks $\Gamma/H\sim 1$ as a function of time,  
\begin{equation}
   k_p\sim 2.93\, \mathrm{mpc}^{-1}
   \left(\frac{10^{10}\, \rm GeV}{f_A}\right)^{1.3426}
    \left(\frac{m_A}{H_1}\right)^{1/2}\frac{R_1}{R}
   \label{eq:kp_pred}.
\end{equation}

Since at high $T$, $\sqrt{m_A}/R\propto \eta^{\n/4-1}\sim \eta$, we predict that $k_p$ grows linearly~\footnote{A self-similar phase space argument (SupM~\cite{supM}) yields the same trend.} until $\n\to 0$ at $T_c$, where $\chi$ stops growing and scattering decouples, see SupM~\cite{supM}.  

In Fig.~\ref{fig:kpeak_fit}, we show the evolution of the axion spectrum peak, $k^N_p$, in our simulations and the theoretical prediction from Eq.~\eqref{eq:kp_pred}.  
Since Eq.~\eqref{eq:kp_pred} can only be correct up to ${\cal O}(1)$ factors from the spectral shape and the peak drift efficiency, we multiply the prediction by a free coefficient $k^N_p=C_{\rm fit}k_p^{\rm pred}$ and fit it to the late relativistic evolution, finding $C_{\rm fit}\sim 0.5-1$, as expected.
The $\Upsilon$ simulations start when all modes are safely non-relativistic, with a spectrum peaked at the $k_p$ extrapolated linearly from the relativistic peak drift model. 
The agreement between model and simulation is very good, given the rough model, the wide peak (we show 25\% of the FWHM), the moderate statistics and the low resolution of the NR simulations. The peak grows, decreases slightly after the cross-over (dots and squares) and settles to $k_p^N(\eta_c\to\eta_{\rm eq})\simeq 30\,{\rm mpc}^{-1} (10^{10\,}{\rm GeV}/f_A)^{1.29} $ for $f_A \in (3\times 10^8,10^{10})\,{\rm GeV}$. 
Free-streaming should disperse localised structures (wave coherence) below the length scale of $k_{\rm fs}=2.5 {\rm mpc}^{-1}(10^{10}{\rm GeV}/f_A)^{1/2}$. The main peak stays constant through the free-streaming period until MRE, unlike DAS related features, which being coherent and small, diffuse away, as the physical axitons would. The spectrum of density fluctuations $\Delta^2$ peaks at a factor of $k_p^\Delta= 1.96(6)k_p^N$ as it would for uncorrelated noise, suggesting that the peak is largely due to incoherent waves at MRE. 
 
\section{Discussion}\label{sec:discussion}

The late-time peak evolution of post-inflationary axions produced by the decay
of topological defects was derived in Ref.~\cite{Gorghetto:2024vnp}; see also
Refs.~\cite{Gorghetto:2018myk,Gorghetto:2020qws,Buschmann:2019icd,
Buschmann:2021sdq,Saikawa:2024bta,Kim:2024wku,Benabou:2024msj,
Correia:2025nns}. The argument is sufficiently general to apply also to the
post-fragmentation evolution studied here. Indeed, for
$f_A>5\times10^9$ GeV, where Ref.~\cite{Gorghetto:2024vnp} reports results,
our extrapolated peak positions agree despite the completely different
initial conditions. At smaller $f_A$ they remain compatible, within sizable
uncertainties, with a naive extrapolation of those results. Thus, after
fragmentation, the distribution---and most importantly its peak
position---is driven towards a value determined by $f_A$ and $\Omega_A$ but
largely independent of the previous history. Kinetic misalignment therefore
evolves towards the same attractor previously found for post-inflationary
axions, and the outcome seems equivalent to that of the post-inflationary
scenario studied in Ref.~\cite{Gorghetto:2024vnp}.

Whether the initial conditions remain observable is controlled by how
efficiently self-interactions act before decoupling at $\eta_c$. For
$f_A<2\times10^{10}$ GeV, the high scattering rate drives the spectrum to the
attractor and erases most of its history, as accurately seen in our
simulations. This range can extend to $f_A\sim5.7\times10^{10}$ GeV even when
scattering is efficient only during an interval close to $\eta_c$. At larger
$f_A$, however, self-interactions are weaker and sensitivity to the initial
conditions $(\Theta_1,v)$ can survive, leading to more variable predictions.
Both variables matter: $\Theta_1$ can drive the field to a summit landing and
enhance both $\Omega_A$ and the possibility of fragmentation. Several
plausible, non-fine-tuned initial conditions yield the observed relic density
through either pure kinetic misalignment or summit-landing solutions. For a
fixed $\Omega_A$, however, the amplitude of the zero-mode oscillations after
trapping is stable, so the fluctuation-growth histories cannot differ
enormously. In this weak-scattering regime the effect of $G_k\ne0$ must also
be taken into account.

\begin{figure}[t]
\centering
\includegraphics[width=\columnwidth]{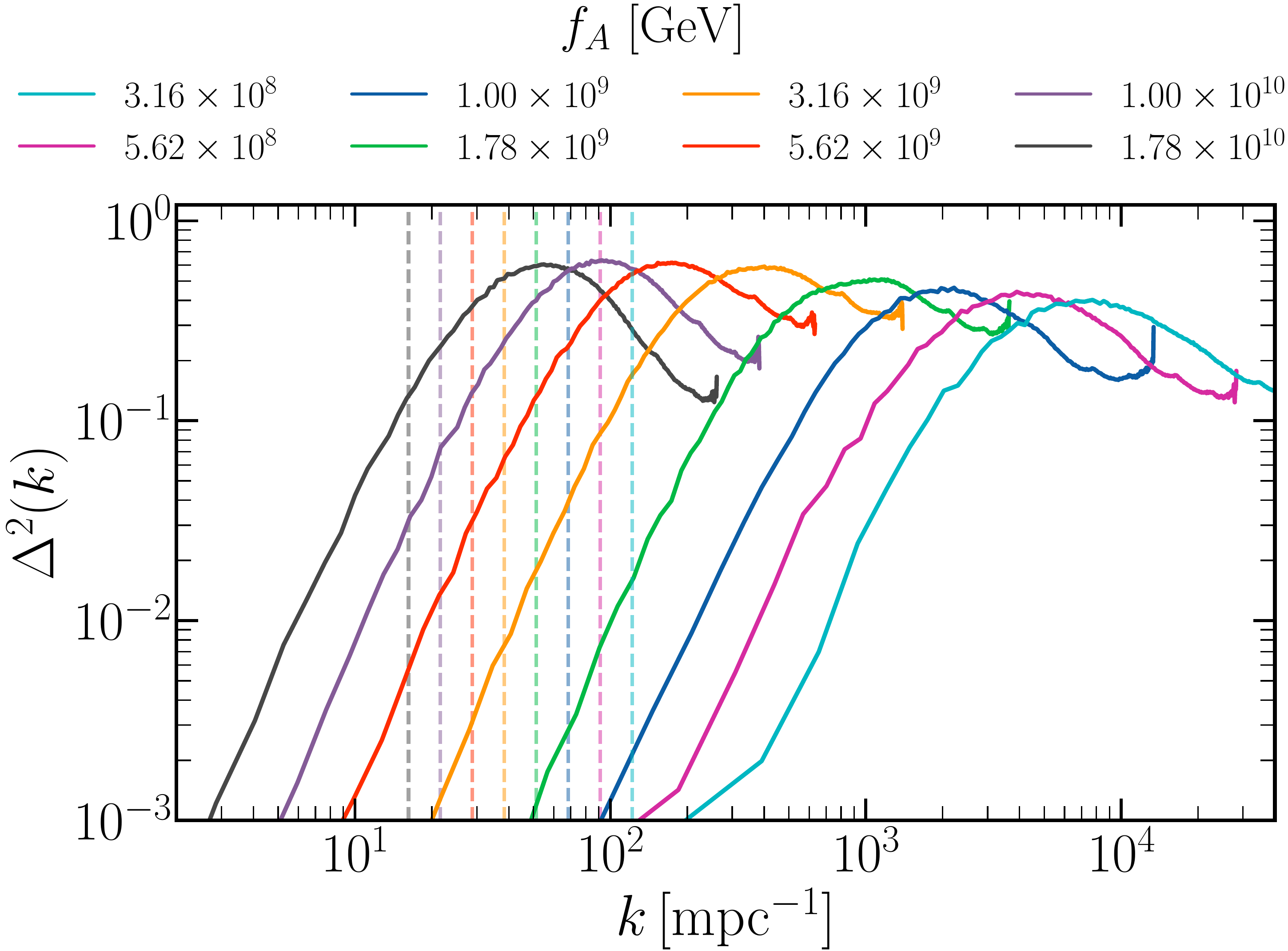}
\caption{\small Dimensionless power spectrum $\Delta^2(k)$ at times around
MRE before gravitational collapse. The peak lies at a wavenumber larger than
the Jeans scale $k_p^{\Delta}>k_J$ (dashed lines), the condition under
which quantum pressure prevents collapse.}
\label{fig:ps_fA}
\end{figure}

The consequences for small-scale structure follow from the spectra of
density fluctuations at matter--radiation equality shown in
Fig.~\ref{fig:ps_fA}. They reach values of order $0.5$, but their peaks lie at
wavenumbers larger than the quantum Jeans scale
$k_J\equiv(16\pi G\rho_{\rm CDM}m_A^2)^{1/4}\sim
21\,\mathrm{mpc}^{-1}(10^{10}\,\mathrm{GeV}/f_A)^{0.5}$. Large density fluctuations would
collapse promptly into axion miniclusters, whereas quantum pressure prevents
modes with $k>k_J$ from collapsing~\cite{HOGAN1988228,Kolb_1993}. For
$f_A\lesssim6\times10^{10}$ GeV, the peak of the power spectrum is comparable to or larger than the Jeans scale
and gravitational collapse is therefore strongly affected by quantum
pressure. The ratio $r=k_p/k_J$ controls the delay relative to the standard
matter--radiation-equality collapse redshift,
$z_{\rm coll}\sim z_{\rm eq}/(1+r)$, lowering the central densities. These
are further reduced by the unavoidable presence of a large fraction of
axion stars~\cite{Gorghetto:2024vnp}. Since our spectra at MRE are seemingly equivalent within the uncertainties to those of Ref.~\cite{Gorghetto:2024vnp} - we find  
$k_p^N/k_J|_{\rm MRE}\sim1.78(10^{10}\,\mathrm{GeV}/f_A)^{0.73}$ while they find $\sim(10^{10}\,\mathrm{GeV}/f_A)^{0.5}$ -- the structure-formation analysis
of that reference applies \emph{mutatis mutandis} to our scenario.

This convergence determines whether the cosmological histories can be
distinguished. If the spectral index of the instantaneous axion-emission
spectrum is $q\gg1$, as in
Refs.~\cite{Gorghetto:2018myk,Gorghetto:2020qws,Kim:2024wku}, the
post-inflationary scenario requires $f_A\sim10^{10}$ GeV. Self-interactions
then erase the spectral differences, and the post-inflationary and kinetic
misalignment scenarios would not be distinguishable through the structure of
miniclusters and axion stars. For $q=1$, as in
Refs.~\cite{Buschmann:2021sdq,Saikawa:2024bta,Correia:2025nns}, the
post-inflationary scenario instead requires
$f_A\gtrsim5.7\times10^{10}$ GeV, where self-interactions are less important and
the two histories could remain distinguishable.

\vspace{1em}
\begin{acknowledgments}
\begin{center}
\textbf{Acknowledgments}
\end{center}

This article is based on work from COST Action COSMIC WISPers (CA21106), supported by COST (European Cooperation in Science and Technology).

K.C. would like to thank the Department of Theoretical Physics at the University of Zaragoza for hosting their research stay and acknowledges the support of the European Consortium for Astroparticle Theory in the form of an Exchange Travel Grant.

The work of M.\,K. and J.\,R. is supported by the grants PGC2022-\-126078NB-\-C21 and PID2024-\-160228NB-I00 funded by MCIN/\-AEI/\-10.13039\-501100011033 and ERDF - A way of making Europe, and DGA-\-FSE 2023-\-E21-\-23R by the Government of Arag\'{o}n, Spain, and the European Union NextGenerationEU Recovery and Resilience Program on Astrof\'{i}sica y F\'{i}sica de Altas Energ\'{i}as CEFCA-\-CAPA-\-ITAINNOVA. 

Furthermore, M.\,K. thanks the Department of Physics and the Wisconsin IceCube Particle Astrophysics Center (WIPAC) at the University of Wisconsin–Madison, USA, for their hospitality during an extended research stay. This stay was supported by the Government of Aragón, Spain, through a travel grant from the program \enquote{Subvenciones para la movilidad de personal investigador predoctoral en formación - Convocatoria 2025} (Ref. MVE\_\-05\_\-25), as part of the PhD fellowship specified in DGA-ORDEN-\-CUS/702/2022.

The work of K.\,S. is supported by JSPS KAKENHI Grant Numbers JP24K07015 and JP26K00695.

Computations were performed in part on the HPC systems Raven and Viper at the Max Planck Computing and Data Facility.

\end{acknowledgments}

\bibliographystyle{bib/utphys}
\bibliography{bib/kini}

\onecolumngrid
\appendix
\setcounter{equation}{0}
\setcounter{figure}{0}
\setcounter{table}{0}
\renewcommand{\theequation}{S\arabic{equation}}
\renewcommand{\thefigure}{S\arabic{figure}}
\renewcommand{\thetable}{S\arabic{table}}

\makeatother

\begin{center}
\makeatletter
\large{Supplemental Material for the Letter \\[.5em] \textbf{\@title}}\\[1em]
\normalsize
   Kierthika Chathirathas, Cem Eröncel, Mathieu Kaltschmidt, Javier Redondo, and Ken'ichi Saikawa 
\makeatother
\end{center}

This Supplemental Material (SupM) provides additional technical details of the calculations and the lattice simulations leading to the results presented in the main Letter. 

\section{QCD Axion Cosmology}\label{app:qcd-axion-cosmology}

Here we show how we compute the thermal history of the early Universe relevant for the cosmological evolution of the QCD axion. 
The reader can consult and reuse our \texttt{python} module at \github{qcd-axion-cosmology}. 
The evolution of the scale factor $R$ and temperature $T$ with conformal time, defined by $d\eta=dt/R$, is obtained from the Friedmann equation $H^2=8\pi G\rho_r/3$ together with radiation-energy conservation, $\dot\rho_r+3HTs=0$.
We write the energy and entropy densities as $\rho_r=(\pi^2/30)g_*(T)T^4$ and
$s=(2\pi^2/45)g_{*S}(T)T^3$, where $g_*(T)$ and $g_{*S}(T)$ denote the effective numbers of relativistic degrees of freedom contributing to the energy and entropy densities, respectively. The coupled evolution equations for $R$ and $T$ can then be written in the convenient form
\begin{equation}
\frac{d \log R}{d \log \eta} = R \eta H, \qquad 
\frac{d \log T}{d \log \eta} = -\frac{R \eta H}{1+\frac{1}{3}\frac{d\log g_{*S}}{d\log T}}.
\end{equation}
We use initial conditions above the electroweak scale,  $T\gg 246$ GeV,
where we assume $R\eta H=1$ because $g_*$ and $g_{*S}$ are assumed to be constant. 
The scale factor normalisation is 
\begin{equation}
    R(T_i)=R(T_0)\left(\frac{g_{*S}(T_0)T_0^3}{g_{*S}(T_i)T_i^3}\right)^{1/3},
\end{equation}
with $R(T_0)=1$ and $T_0=2.7255$ K~\cite{Fixsen:2009abc}.  The effective numbers of relativistic degrees of freedom are built using the lattice results of Ref.~\cite{Borsanyi:2016ksw} and using \texttt{nudec\_BSM}
\cite{Escudero:2025kej} to track the non-instantaneous decoupling of $\nu_e$ and $\nu_{\mu,\tau}$.
The topological susceptibility $\chi(T)$, which fixes the axion mass via
$m_A^2(T) = \chi(T)/f_A^2$, is also built from the results of Ref.~\cite{Borsanyi:2016ksw}, matched with the chiral perturbation theory result of \cite{GrillidiCortona:2015jxo} at low $T$ and extrapolated as $\chi(T)\propto T^{-8.16}$ beyond the
lattice range. 
The module tabulates the relevant parameters $T(\eta), R(\eta), H(\eta)$ and $\chi(T(\eta))$ in a file \texttt{jaxi-cosmo.txt}.

The QCD cosmology table is read by \githubj{jaxions}~\cite{Jaxions:2026} to calculate
$\eta_1, R_1$, and $H_1$, as a function of $f_A$ by solving $m_A(\eta_1) = H(\eta_1)$ and
evaluating $R_1(\eta_1)$. The results are in good agreement with those presented in Ref.~\cite{Vaquero:2018tib}, which we recall here for convenience. Using $m_A=5.7\,\mu\mathrm{eV}\,(10^{12}\,\mathrm{GeV}/f_A)$, they read
\begin{align}
T_1 &\simeq 2.524\,\mathrm{GeV}
\left(\frac{10^{10}\,\mathrm{GeV}}{f_A}\right)^{0.1638},\\
H_1 &\simeq 7.85\times10^{-3}\,\mu\mathrm{eV}
\left(\frac{10^{10}\,\mathrm{GeV}}{f_A}\right)^{0.338},\label{eq:H1_QCD}\\
R_1
&\simeq3.37\times10^{-14}
\left(\frac{f_A}{10^{10}\,\mathrm{GeV}}\right)^{0.1712},\label{eq:R1_QCD}\\
\eta_1 &\simeq 0.0241\,\mathrm{pc}
\left(\frac{f_A}{10^{10}\,\mathrm{GeV}}\right)^{0.167}.
\end{align}

Additionally, the tools to run the 1D zero-mode simulations and to prepare the initial spectra used as input for the \githubj{jaxions} runs are available in a second repository, \github{kinetic-misalignment}.

\section{Dark Matter Abundance}\label{app:abundance}
In this section, we derive analytic estimates for the evolution of the homogeneous axion field and the resulting relic abundance in the kinetic misalignment mechanism. When the effective relativistic degrees of freedom are constant, radiation domination implies $R''=0$. Their temperature dependence introduces only small corrections that are not relevant for the following analytic discussion, although we retain them in the numerical calculations described in the previous section. In the $R''=0$ approximation, choosing the origin of conformal time at the Big Bang gives the convenient time--scale-factor variable
\begin{equation}
\tau\equiv\frac{\eta}{\eta_1}=\frac{R}{R_1}.
\end{equation}
We assume the QCD axion Lagrangian
\begin{equation}
\mathcal{L}_A=\sqrt{-g}\left[
\frac{f_A^2}{2}g^{\mu\nu}\partial_\mu\theta\,\partial_\nu\theta
-\chi(T)(1-\cos\theta)\right].
\end{equation}
For the homogeneous mode $\Theta(t)$, the kinetic and potential energy densities are
\begin{equation}
\rho_K=\frac12 f_A^2\dot\Theta^2,
\qquad
\rho_V=\chi(T)(1-\cos\Theta),
\end{equation}
and $m_A^2(T)=\chi(T)/f_A^2$. We denote the zero-temperature susceptibility by
$\chi_0\equiv\chi(0)=(75.5\,\mathrm{MeV})^4$~\cite{GrillidiCortona:2015jxo}.
At early times, $\eta\ll\eta_1$, the equation of motion for the zero mode $\Theta$ is given by
\be
\Theta''+2 {\cal H}\Theta'=0.
\ee
The general solution of this equation in radiation-domination
is $\Theta=A/R+B$, where $A$ and $B$ are constants. Here, $A$ controls 
the early kination solution and the amplitude of the adiabatic fluctuations. We normalise $R_0\equiv R(\eta_0)=1$
and define the dimensionless velocity parameter as the field velocity in units of Hubble at the onset of the QCD potential,
\begin{equation}
v \equiv \frac{\dot\Theta_1}{H_1}.
\label{eqn:A_coeff}
\end{equation}
Since $\dot\Theta=-A\,\dot R/R^2$ implies $\dot\Theta/H=-A/R$, the early solution
reads $A=-v\,R_1$, i.e. $\dot\Theta/H=v\,R_1/R$. The second coefficient
$B$ is the misalignment angle in the zero-velocity case or the asymptotic value of $\Theta$ in the kinetic misalignment, neglecting the role of the QCD potential. 

The kinetic energy density therefore evolves as
\begin{equation}
\rho_K=\frac12 f_A^2\dot\Theta^2
=\frac12 f_A^2\,v^2\Big(\frac{R_1}{R}\Big)^{2}H^2\,\propto\,R^{-6},
\end{equation}
redshifting as kination. 

The potential energy due to the QCD potential grows as
$\rho_V\sim \chi(R)$, which at high temperatures $T>T_c=150$ MeV grows as $\chi_0 (T_c/T)^{\n}\propto R^\n$ and at some point dominates the dynamics of the zero mode.

The kination phase will end when the field stops ($\Theta'=0$) at $R=R_s$ due to the force of the QCD potential. We estimate the stopping time by equating the kinetic energy to the characteristic potential scale, $\rho_K(R_s)\simeq\chi(R_s)$. This is an order-one trapping criterion rather than the exact equality $\rho_K=\rho_V$, which would retain a factor $1-\cos\Theta_s$. Using
$H=H_1(R_1/R)^2$, $m_A(R_1)=H_1$, and hence
\begin{equation}
\rho_K(R)=\frac12 f_A^2H_1^2v^2\left(\frac{R_1}{R}\right)^6,
\qquad
\chi(R)=f_A^2H_1^2\left(\frac{R}{R_1}\right)^\n,
\end{equation}
we find
\begin{equation}
\Big(\frac{R_s}{R_1}\Big)^{\frac{\n+6}{2}}\simeq\frac{|v|}{\sqrt2}\,.
\label{eqn:stop}
\end{equation}
If the initial velocity is large enough for the field to undergo at least one summit crossing, $\Theta=(2j+1)\pi$, before trapping, capture occurs only after its kinetic energy has fallen below the potential-barrier scale. In the limit $m_A(R_s)\gg H_s$, Hubble friction can be neglected during the short evolution between the last summit crossing and the turning point. Numerically evolving this trajectory for $\n=8$, with initial conditions $\Theta=-\pi$ and positive summit-crossing velocity, yields the lower bound $|\Theta_s|\gtrsim1.35$ modulo $2\pi$. Thus, the stopping angle is generically large and the harmonic approximation $1-\cos\Theta_s\simeq\Theta_s^2/2$ is not initially justified. These nonlinearities strongly enhance the growth of fluctuations and facilitate fragmentation, as discussed in the following sections. For the present homogeneous abundance estimate, however, we neglect their effect on the subsequent evolution.
At stopping, the energy density is stored only in the potential $\rho_s=\chi(R_s)(1-\cos\Theta_s)$, and the axion number
density is $n_s=\rho_s/m_A(R_s)=m_A(R_s)f_A^2(1-\cos\Theta_s)$.

After stopping, the zero mode falls towards the minimum and undergoes damped oscillations. For histories with at least one summit crossing, the motion is initially nonlinear, but the oscillation amplitude decreases rapidly because of the fast growth of the axion mass and Hubble damping. The evolution therefore enters the linear regime shortly after trapping.

Once the oscillation amplitude is small, the equation of motion becomes 
\begin{equation}
\Theta''+2 {\cal H}\Theta' +m_A^2R^2 \Theta=0,
\end{equation}
which admits the WKB solution
\begin{equation}
\Theta(\eta)=\bar\Theta(\eta)\sin(\Omega_0+\alpha)\quad\text{with}\quad
\bar\Theta(\eta)=\bar\Theta_s\left(\frac{m_A(R_s)\,R_s^3}{m_A(R)\,R^3}\right)^{1/2}\quad\text{and}\quad\Omega_0=\int^\eta d\eta'\,m_A R(\eta').
\label{eqn:wkb}
\end{equation}
The WKB amplitude implies adiabatic conservation of the comoving axion number, $nR^3=\text{const}$, with
$n=\frac12 m_A\bar\Theta^2f_A^2$.
As an approximation, we extrapolate the WKB solution back to the stopping time and match it directly to $n_s$. We denote the stopping time and the onset of harmonic oscillations by $\tau_s$ and $\tau_h$, respectively. This matching neglects the short nonlinear transient between them. In our radiation-dominated power-law approximation, the dimensionless conformal frequency is $\omega(\tau)=m_A R\eta_1\simeq\tau^{\n/2+1}$. The duration of this transient is therefore of order one local oscillation time,
\begin{equation}
\Delta\tau_{\rm nl}\equiv\tau_h-\tau_s
=\mathcal{O}(\omega_s^{-1})
=\mathcal{O}\!\left(\tau_s^{-\frac{\n}{2}-1}\right).
\end{equation}
Since the homogeneous abundance scales as $\Omega_Ah^2\propto\tau_s^{(\n+6)/2}$, the corresponding fractional matching uncertainty is
\begin{equation}
\frac{\Delta(\Omega_Ah^2)}{\Omega_Ah^2}
=\mathcal{O}\!\left(\frac{\n+6}{2}\frac{\Delta\tau_{\rm nl}}{\tau_s}\right)
=\mathcal{O}\!\left(\frac{\n+6}{2}\frac{H_s}{m_A(R_s)}\right)
=\mathcal{O}\!\left[\frac{\n+6}{2}
\left(\frac{|v|}{\sqrt2}\right)^{-\frac{\n+4}{\n+6}}\right].
\label{eq:wkb-matching-error}
\end{equation}
This estimate quantifies only the error from matching the short homogeneous nonlinear transient; it does not include number-changing effects from fragmentation. The rapidly growing axion mass and Hubble damping reduce the amplitude quickly, making the evolution approximately linear soon after $R_s$. The matching then fixes the effective
amplitude at trapping, $\bar\Theta_s^2\equiv2(1-\cos\Theta_s)$, which reduces to
$\bar\Theta_s=\Theta_s$ in the harmonic limit. Before the QCD cross-over, $m_A\propto R^{\n/2}$ and Eq.~\eqref{eqn:wkb} gives $\bar\Theta\propto R^{-(\n+6)/4}$. Once the axion mass saturates, the amplitude redshifts as $\bar\Theta\propto R^{-3/2}$. Using $\chi(R_s)=\chi_0(R_s/R_c)^\n$, comoving-number conservation gives the present-day abundance step by step as
\begin{equation}
\begin{aligned}
n_0
&=n_s\left(\frac{R_s}{R_0}\right)^3
=m_A(R_s)f_A^2(1-\cos\Theta_s)\left(\frac{R_s}{R_0}\right)^3,\\
\rho_{A,0}=m_{A,0}n_0
&=\sqrt{\chi_0\chi(R_s)}\,(1-\cos\Theta_s)
\left(\frac{R_s}{R_0}\right)^3\\
&=\chi_0\left(\frac{R_s}{R_c}\right)^{\frac{\n}{2}}
(1-\cos\Theta_s)\left(\frac{R_s}{R_0}\right)^3,\\
\Omega_A h^2=\frac{\rho_{A,0}}{\rho_c}
&=\frac{\chi_0}{\rho_c}\left(\frac{R_1}{R_c}\right)^{\frac{\n}{2}}
\left(\frac{R_1}{R_0}\right)^{3}(1-\cos\Theta_s)
\left(\frac{R_s}{R_1}\right)^{\frac{\n+6}{2}}.
\end{aligned}
\label{eqn:relic_supm}
\end{equation}
Here 
\begin{equation}
\rho_c\equiv\frac{\rho_{\rm crit}}{h^2}
=\frac{3H_{100}^2}{8\pi G},
\qquad
H_{100}\equiv100\,\mathrm{km}\,\mathrm{s}^{-1}\,\mathrm{Mpc}^{-1}. \label{eq:rho_c_definition}
\end{equation}
We can rewrite $(R_1/R_c)^{\n/2}(R_1/R_0)^3\propto R_1^{(\n+6)/2}\propto
f_A^{(\n+6)/(\n+4)}$, where we used $R_1\propto f_A^{2/(\n+4)}$, obtained from
$m_A(R_1)=H_1$. Further, using Eq.~\eqref{eqn:stop}, the last factor becomes $v/\sqrt{2}$, yielding
\begin{align}
 \Omega_Ah^2 
\label{eqn:omega1}
&\simeq 4.08\times 10^{-4} \left(\frac{f_A}{10^{10}\ \rm GeV}\right)^\frac{\n+6}{\n+4} v
\equiv 0.12\,\frac{v}{v_{\rm KM}},\\
v_{\rm KM}
&\simeq 294\left(\frac{10^{10}\ \rm GeV}{f_A}\right)^\frac{\n+6}{\n+4}, 
\end{align}
where we estimated $\cos\Theta_s\sim 0$ because we have $|\Theta_s|>1.35$ and in this regime we do not consider/need $\Theta_s$ tuned to the hilltop (see next section).  
Equation~\eqref{eqn:omega1} describes the non-tuned kinetic-misalignment solution. The first part of the following section derives the complementary summit-landing solution, in which the field crosses its last potential maximum with a small residual velocity and the resulting delay of trapping enhances the abundance. We then use this result to quantify the fine-tuning probability.

\section{Summit Landings and Fine-tuning Probability}\label{app:fine-tuning}
At fixed normalised initial velocity $v=\dot\Theta_1/H_1$, there is a critical initial angle $\Theta_1^\star$ for which the homogeneous field asymptotically approaches its last potential maximum, $\Theta=(2j+1)\pi$. Initial angles close to $\Theta_1^\star$ give rise to two summit-landing branches. In the first, the field overshoots the summit with a small residual velocity and departs on the other side. In the second, it stops just before reaching the summit and subsequently rolls back. Both trajectories spend a long time near the unstable maximum, delaying the onset of oscillations and enhancing the relic abundance, but their boundary conditions and their mappings to the initial angle $\Theta_1$ differ. We first consider the overshoot branch and return to the pre-summit stopping branch below.

Using the dimensionless conformal time $\tau\equiv\eta/\eta_1$ introduced above, let $\tau_\ell$ denote the time of the last summit crossing. For the overshoot branch, we choose this summit to be $\Theta=-\pi$ and define
\begin{equation}
\delta\equiv\Theta+\pi.
\end{equation}
Close to the maximum, the linearised equation of motion is
\begin{equation}
  \delta'' + \frac{2}{\tau}\,\delta' - \tau^{\n+2}\,\delta = 0 .
\end{equation}
Neglecting slowly varying prefactors, its growing and decaying solutions are
\begin{equation}
\delta\propto e^{\pm f(\tau)},
\qquad
f(\tau)\equiv\frac{2}{\n+4}\tau^{\frac{\n+4}{2}}.
\end{equation}
For the overshoot branch, the boundary conditions at the summit are
\begin{equation}
\delta(\tau_\ell)=0,
\qquad
\delta'(\tau_\ell)=v_\ell>0,
\end{equation}
where $v_\ell$ is the residual conformal velocity. They give
\begin{equation}
\delta(\tau)=\epsilon_+\left(e^{f-f_\ell}-e^{f_\ell-f}\right),
\label{eq:overshoot}
\end{equation}
with
\begin{equation}
\epsilon_+\equiv\frac{v_\ell}{2f'_\ell},
\qquad
f_\ell\equiv f(\tau_\ell).
\end{equation}
We define $\tau_d$ by $\Theta_d=-\pi/2$, or equivalently $\delta_d=\pi/2$, marking departure from the hilltop regime. This is not the stopping time: it is the end of the delayed evolution close to the summit. Keeping only the growing solution gives
\begin{equation}
  \tau_d = \left(\tau_\ell^{\frac{\n+4}{2}}
       + \frac{\n+4}{2}\,\log\left(\frac{\delta_d}{\epsilon_+}\right)\right)^{\!\frac{2}{\n+4}} .
  \label{eq:taus}
\end{equation}
We evaluate the generic abundance formula in Eq.~\eqref{eqn:relic_supm} at the matching time $\tau_h$, when ordinary harmonic oscillations begin. After the field leaves the hilltop regime at $\tau_d$, reaching this matching point takes at most an order-one number of local oscillation times. We therefore approximate $\tau_h=\tau_d+\mathcal{O}(\omega_d^{-1})=\tau_d+\mathcal{O}(\tau_d^{-\n/2-1})$, which induces a relative error of order $(\n+6)H_d/[2m_A(R_d)]$ in the homogeneous abundance, as already estimated in Eq.~\eqref{eq:wkb-matching-error}.
Approximating the time of the last summit crossing by the characteristic trapping time,
$\tau_\ell\simeq\tau_t=(v^2/2)^{1/(\n+6)}$, and inserting Eq.~\eqref{eq:taus}, we obtain for the overshoot branch
\begin{equation}
  \Omega_A h^2 \simeq \frac{0.12}{v_{\rm KM}}
  \left[v^{\frac{\n+4}{\n+6}}
        +\frac{\n+4}{2^{\frac{\n+8}{2(\n+6)}}}
        \log\frac{\delta_d}{\epsilon_+}\right]^{\frac{\n+6}{\n+4}},
  \label{eq:omega_full}
\end{equation}
where we used the normalized non-tuned prediction in Eq.~\eqref{eqn:omega1} and absorbed the factors $\sqrt2$ and $1/2$ into the logarithmic coefficient. Equation~\eqref{eqn:omega1} is recovered when the logarithmic term vanishes.

For the pre-summit stopping branch, we instead choose the summit at $\Theta=\pi$ and define the positive distance
\begin{equation}
\delta\equiv\pi-\Theta.
\end{equation}
The turning point is the stopping point itself, $\tau_-=\tau_s$, with $\Theta_s\simeq\pi$. Defining $\Delta_-\equiv\pi-\Theta_s>0$, the boundary conditions are
\begin{equation}
\delta(\tau_-)=\Delta_-,
\qquad
\delta'(\tau_-)=0.
\end{equation}
They give
\begin{equation}
\delta(\tau)=\epsilon_-\left(e^{f-f_-}+e^{f_--f}\right),
\label{eq:presummit}
\end{equation}
with
\begin{equation}
\epsilon_-\equiv\frac{\Delta_-}{2},
\qquad
f_-\equiv f(\tau_-).
\end{equation}
The departure time is therefore
\begin{equation}
\tau_{d,-}=\left(\tau_-^{\frac{\n+4}{2}}
+\frac{\n+4}{2}\log\frac{\delta_d}{\epsilon_-}\right)^{\!\frac{2}{\n+4}}.
\label{eq:taus-minus}
\end{equation}
Approximating $\tau_-=\tau_s\simeq\tau_t$, the pre-summit and overshoot branches have the same logarithmic enhancement, with $\epsilon_-$ replacing $\epsilon_+$, despite their different boundary conditions.

Away from either fine-tuned summit-landing trajectory, the logarithmic term is subdominant and Eq.~\eqref{eq:omega_full} reduces directly to the non-tuned relation in Eq.~\eqref{eqn:omega1}, which also defines $v_{\rm KM}$.
The same yield can be obtained at smaller $v$ through the logarithmic enhancement. Defining $p\equiv(\n+4)/(\n+6)$ and matching either summit-landing branch to the non-tuned solution at $v=v_{\rm KM}$ gives
\begin{equation}
  \log\left(\frac{\delta_d}{\epsilon_\pm}\right)
  = \frac{2^{1-p/2}}{\n+4}
  \left(v_{\rm KM}^{p}-v^{p}\right).
  \label{eq:epsilon-target}
\end{equation}
We now relate these local hilltop parameters to the initial-angle intervals. Close to the critical initial angle, the mappings are linear,
\begin{equation}
\epsilon_+=C_+\Delta\Theta_1^{(+)},
\qquad
\epsilon_-=C_-\Delta\Theta_1^{(-)},
\end{equation}
where $C_+$ and $C_-$ need not be identical. For a flat prior $dP/d\Theta_1=1/(2\pi)$, the probability that a random initial angle at fixed $v$ and $f_A$ gives $\Omega_Ah^2\geq0.12$ is therefore
\begin{equation}
P_\Theta=\frac{\Delta\Theta_1^{(+)}+\Delta\Theta_1^{(-)}}{2\pi}
=\frac{\epsilon_{0.12}}{2\pi}\left(\frac{1}{C_+}+\frac{1}{C_-}\right),
\end{equation}
where Eq.~\eqref{eq:epsilon-target} fixes the common limiting value $\epsilon_{0.12}$. Approximating the two Jacobians by $C_+\simeq C_-\simeq1$ and using $\delta_d=\pi/2$ gives
\begin{equation}
  P_\Theta \;=\; \frac{1}{2}\,
  \exp\!\left[-\frac{2^{1-p/2}}{\n+4}
  \left(v_{\rm KM}^{p}-v^{p}\right)\right],
  \qquad\text{for}\ v\le v_{\rm KM},
  \label{eq:PTheta}
\end{equation}
Equation~\eqref{eq:PTheta} is reliable only in the fine-tuned regime, where $\epsilon_{0.12}\ll\delta_d$ and hence $P_\Theta\ll1$. Its formal continuation to the non-tuned line $v=v_{\rm KM}$ gives $P_\Theta=1/2$, but this value has no physical significance because both the linear hilltop solution and the local mapping to $\Delta\Theta_1$ then cease to apply. The physical angular probability instead becomes of order unity near this line. The coloured region is truncated at the $P_\Theta=10^{-10}$ contour, excluding cases that require extremely large fine-tuning. For $\n=8$ and $v\ll v_{\rm KM}$, the exponent is dominated by $-2^{-3/7}v_{\rm KM}^{6/7}/6$, while the $v^{6/7}$ term is subdominant. Consequently, contours are nearly vertical (set by $f_A$) and bend up only near the line. Since
$v_{\rm KM}\propto f_A^{-7/6}$ grows at small $f_A$, the suppression
$\sim\exp\left({-2^{-3/7}v_{\rm KM}^{6/7}/6}\right)$ results in a sharp low-$f_A$ edge.

Naturally, we refuse to solve the strong CP fine-tuning problem by introducing an even more severe tuning of $\Theta_1$. We thus restrict ourselves to the region of parameter space where the tuning width of $\Theta_1$ yielding the measured CDM abundance $\Omega_c h^2 = 0.120(1)$~\cite{Planck:2018vyg} has a
non-negligible probability.

Fig.~\ref{fig:grid} illustrates the dependence on the initial conditions for $f_A=10^{10}\,\mathrm{GeV}$. We solve for $x\equiv\Omega_Ah^2$ on a uniform grid in $(\Theta_1,v)$. The lower panel shows the resulting value $x(\Theta_1,v)$ at every grid point. At each fixed $v$, we histogram the values obtained while varying $\Theta_1$; the upper panel displays this histogram in the $(v,x)$ plane, with the height of each bin encoded by colour. Because the grid is uniform in $\Theta_1$, the normalized histogram is the conditional probability density associated with a flat prior on the initial angle,
\begin{equation}
 P(x\mid v)\equiv P_{\dot\Theta}(x\mid v)
 \equiv \int_0^{2\pi}\frac{d\Theta_1}{2\pi}\,
 \delta_{\rm D}\!\left[x-x(\Theta_1,v)\right]
 =\frac{1}{2\pi}\sum_i
 \left|\frac{\partial x}{\partial\Theta_1}\right|^{-1}_{\Theta_{1,i}},
 \label{eq:PThetadot}
\end{equation}
where the sum runs over all roots $x(\Theta_{1,i},v)=x$ and $\int dx\,P(x\mid v)=1$. We denote this density by $P_{\dot\Theta}$ in Fig.~\ref{fig:grid}; it describes the full abundance distribution at fixed velocity and is distinct from the integrated fine-tuning probability. The latter is its upper tail,
\begin{equation}
 P_\Theta(x_c\mid v)=\int_{x_c}^{\infty}dx\,
 P(x\mid v),
\end{equation}
with $x_c=0.12$ for the observed CDM abundance used above.

\section{Linear Fluctuation Growth and Gravitational Source Term \boldmath$G_k$}\label{app:G_k}
In this section we discuss the initial conditions for the fluctuations, their linear growth, amplification through tachyonic and parametric resonances and the effect of the source term. 
We find that for $v\gtrsim 150$, fragmentation should occur at or before trapping, where the effects of the gravitational term are negligible.

In conformal Newtonian gauge and in the absence of anisotropic stress, the perturbed metric is given by
\begin{equation}
    \label{eq:perturbed-metric}
    ds^2=R^2(\eta)\left[-\left(1-2\Phi\right)d\eta^2 + (1+2\Phi)\delta_{ij}dx^i dx^j\right].
\end{equation}
At linear order in the field and metric perturbations, the EOM for the Fourier modes reads
\begin{equation}
    \label{eq:eom-fourier-modes}
    \theta_k''+2\mathcal{H}\theta_k'+\left[k^2 +R^2 m_A^2(T)\cos\Theta \right]\theta_k = -4\Theta' \Phi_k' + R^2 m_A^2(T) \sin\Theta \left[ 2\Phi_k - \frac{d \log\chi}{d \log T}\frac{\delta T}{T}\right].
\end{equation}
The temperature fluctuations $\delta T/T$ can be related to the perturbations in the energy density of the radiation bath $\rho_{\text{rad}}$ through $\delta \rho_{\text{rad}}/ \rho_{\text{rad}} = 4(\delta T/T)$. The latter are in turn related to the curvature perturbations as
\begin{equation}
    k^2 \Phi_k + 3\mathcal{H}(\Phi_k'+\mathcal{H}\Phi_k)=\frac{3}{2}\mathcal{H}^2 \frac{\delta \rho_{\text{rad}}}{\rho_{\text{rad}}}
\label{eqn:poisson}
\end{equation}
Substituting Eq.~\eqref{eqn:poisson} into Eq. \eqref{eq:eom-fourier-modes}, setting $\frac{d\log\chi}{d\log T} = -\n$
and $\cmass=m_AR$, we obtain
\begin{equation}
    \theta_k''+2\mathcal{H}\theta_k'+\left[k^2 +\cmass^2\cos\Theta\right]\theta_k = G_k,
\end{equation}
with the gravitational source
\begin{equation}
    G_k = -4\Theta'\Phi_k' 
    + \cmass^2 \sin\Theta 
    \left[ 2\Phi_k + \frac{\n}{2}\left(1+\tfrac{1}{3}(k\eta)^2\right)\Phi_k 
    + \frac{\n}{2}\, \eta\, \Phi_k'\right].
    \label{eq:Gk-full}
\end{equation}
In radiation domination the modes evolve as
\begin{equation}
\Phi_k = 3\,\Phi_k(0)\,\frac{\sin x- x\cos x}{x^3}\,,\qquad x = \frac{k \eta}{\sqrt{3}}\,.
\label{eqn:phi-mode-evol}
\end{equation}
It remains constant on super-horizon scales and decays as $\Phi_k\propto x^{-2}$ after horizon entry.

The primordial amplitude $\Phi_k(0)$ is measured on large length scales, $k^{-1}\sim(1$--$10^4)\,\mathrm{Mpc}$, by CMB and large-scale-structure probes~\cite{Planck:2018vyg,eBOSS:2020yzd,DES:2021wwk}, and is consistent with the nearly scale-invariant spectrum
\begin{equation}
 |\Phi_k(0)|^2=\left(\frac{2}{3}\right)^2|\mathcal{R}_k|^2
 =\left(\frac{2}{3}\right)^2\frac{2\pi^2}{k^3}A_s
 \left(\frac{k}{k_{\rm piv}}\right)^{n_s-1},
 \label{eqn:spectrum}
\end{equation}
where $\mathcal{R}_k$ is the comoving-curvature perturbation, $A_s=2.1\times10^{-9}$, $k_{\rm piv}=0.05\,\mathrm{Mpc}^{-1}$, and $n_s\simeq0.965$~\cite{Planck:2018vyg,Tristram:2023haj,ACT:2020gnv}. We extrapolate this spectrum to the small scales $k_1\equiv\mathcal{H}_1\sim4.14\times10^7\,\mathrm{Mpc}^{-1}(10^{10}\,\mathrm{GeV}/f_A)^{0.167}$ relevant here~\cite{Vaquero:2018tib,Borsanyi:2016ksw}, finding $(k/k_{\rm piv})^{n_s-1}\sim0.49(1)$ at the scales relevant for our analysis.

For the numerical evolution we normalize the conformal field by its value of
the scale factor at $\eta_1$,
\begin{equation}
 \hpsi_k\equiv \frac{R}{R_1}\theta_k
 =\frac{\psi_k}{R_1},
 \qquad \psi_k\equiv R\theta_k.
\end{equation}
The hat will henceforth distinguish the order-one field used by the
relativistic code from the convention $\psi=R\theta$ used in the main Letter.
Its equation of motion follows from Eq.~\eqref{eq:eom-fourier-modes},
\begin{equation}
 \hpsi_k''+\left[k^2+R^2m_A^2(T)\cos\Theta
 -\frac{R''}{R}\right]\hpsi_k=\frac{R}{R_1}G_k.
 \label{eq:psi-eom}
\end{equation}
Thus the rescaling removes the Hubble-friction term and turns each linear mode
into a frictionless oscillator. In exact radiation domination $R''=0$; more
generally, we retain the small $R''/R$ correction in the numerical evolution.
We therefore evolve $\hpsi_k$ and its conformal derivative $\hpsi_k'$ rather
than $\theta_k$ and $\theta_k'$.

\paragraph{Initial Conditions.}
The source $G_k$ has two contributions: the velocity-dependent term $-4\Theta'\Phi_k'$, which determines the early-time behavior, and the mass-dependent terms, which become relevant only for $\eta\gtrsim\eta_1$. At earlier times the axion mass is negligible and, in radiation domination, $R\propto\eta$ and $\Theta'=C/\eta^2$, with $C\equiv\eta^2\Theta'=v\eta_1$. The leading early-time expansion of Eq.~\eqref{eqn:phi-mode-evol} gives
\begin{equation}
 \Phi_k'=-\frac{k^2\eta}{15}\Phi_k(0)
 +\mathcal{O}(k^4\eta^3).
\end{equation}
Neglecting the mass terms, the mode equation becomes
\begin{equation}
 \theta_k''+\frac{2}{\eta}\theta_k'+k^2\theta_k
 =\frac{4C\Phi_k(0)}{15\eta}k^2,
 \label{eq:early-free-modes}
\end{equation}
whose general solution is the sum of homogeneous and inhomogeneous pieces,
\begin{equation}
 \theta_k^{\rm hom}=\frac{a_k\cos(k\eta)+b_k\sin(k\eta)}{\eta},
 \qquad
 \theta_k^{\rm inh}=\frac{4C\Phi_k(0)}{15\eta}.
 \label{eq:early-inhomogeneous}
\end{equation}
On super-horizon scales the adiabatic perturbation satisfies~\cite{Eroncel:2022vjg}
\begin{equation}
 \theta_k^{\rm ad}\simeq-\Phi_k(0)\frac{\Theta'}{2\mathcal H}
 =-\frac{C\Phi_k(0)}{2\eta}.
 \label{eq:adiabatic-ic}
\end{equation}
Matching the general solution to this behavior fixes
\begin{equation}
 a_k=-\left(\frac12+\frac4{15}\right)C\Phi_k(0)
 =-\frac{23}{30}C\Phi_k(0),
 \qquad b_k=0,
\end{equation}
where $b_k=0$ follows from neglecting an independent primordial axion fluctuation. For the rescaled field, the early-time solution is therefore
\begin{equation}
 \hpsi_k= v\Phi_k(0)
 \left[\frac4{15}-\frac{23}{30}\cos(k\eta)\right],
 \qquad
 \hpsi_k'=\frac{23}{30}kv\Phi_k(0)\sin(k\eta).
 \label{eq:early-psi-modes}
\end{equation}
The small difference between $23/30\simeq0.767$ and $0.78$ arises because the leading expansion of $\Phi_k'$ used above is being extrapolated through horizon entry. Solving with the full radiation transfer function gives $0.78$~\cite{Eroncel:2022vjg}, a correction of only about $1.7\%$. This effect is included in our initial spectrum. Using the super-horizon coefficient $0.5$ instead would reduce the initial fluctuation amplitude, but would not change the physical amplification mechanisms. In the strongly nonlinear fragmentation regime studied here, the subsequent growth rapidly erases this difference in normalization, leaving the final outcome essentially unchanged.
For each realization, we draw the Fourier amplitudes $\Phi_k(0)$ from a Gaussian random field with zero mean and variance given by Eq.~\eqref{eqn:spectrum}, imposing $\Phi_{-k}(0)=\Phi_k^*(0)$ so that the position-space field is real. At the scales relevant for our simulations, where $(k/k_{\rm piv})^{n_s-1}\simeq0.49$, the corresponding dimensionless amplitudes are
\begin{equation}
 \left[\frac{k^3}{2\pi^2}|\Phi_k(0)|^2\right]^{1/2}
 \simeq2.14\times10^{-5},
 \qquad
 \left[\frac{k^3}{2\pi^2}|\hpsi_k|^2\right]^{1/2}
 \simeq1.67\times10^{-5}\,v,
\end{equation}
where the second result includes the coefficient $0.78$. These adiabatic axion modes provide the initial fluctuation field at $R=R_1$.
\paragraph{Initial amplitude and linear regime.}
The integrated variance of the adiabatic seed follows directly from
Eq.~\eqref{eqn:spectrum}. Approximating the mild tilt by its value at the
scales of interest gives
\begin{equation}
 \sqrt{\langle\theta^2\rangle_{\rm ad}}
 \simeq 1.67\times10^{-5}\,v
 \left(\ln\frac{k_{\rm max}}{k_1}\right)^{1/2}.
 \label{eq:theta-rms-seed}
\end{equation}
The maximal duration of kination considered in
Ref.~\cite{Eroncel:2022vjg} defines the physical cutoff
\begin{equation}
 \frac{k_{\rm kin}}{k_1}
 \equiv\frac{10^5k_*}{k_1}
 =10^5\left(\frac{v}{2}\right)^{-2/(\n+6)}
 =10^5\left(\frac{v}{2}\right)^{-1/7}
 \qquad(\n=8),
 \label{eq:seed-cutoff}
\end{equation}
above which the primordial spectrum depends on the pre-kination history. We
therefore keep the physical cutoff $k_{\rm max}\leq k_{\rm kin}$ explicit,
normalizing it to the maximal value $k_{\rm kin}=10^5k_*$. For this maximal
choice $\ln(k_{\rm max}/k_1)$ is approximately 10. The initial fluctuations
therefore become
nonlinear when
\begin{equation}
 v\gtrsim v_{\rm lin}
 \equiv\frac{5.99\times10^4}
 {\sqrt{\ln(k_{\rm max}/k_1)}}
 \simeq1.9\times10^4.
 \label{eq:seed-linear-threshold}
\end{equation}
Thus, sufficiently large velocities already produce order-one spatial
fluctuations before the QCD potential becomes dynamically important. Their
subsequent response to the rapidly growing potential is intrinsically
nonlinear, even without an intermediate linear amplification stage.

The same seed can also carry an appreciable axion number before becoming
nonlinear. For a freely propagating relativistic mode, the field and momentum
quadratures give
\begin{equation}
 \frac{dN_{\rm ad}}{d\ln k}
 \equiv R^3\frac{dn_{\rm ad}}{d\ln k}
 \simeq f_A^2R_1^2\,k\,\Delta_\theta^2(k),
 \qquad
 \Delta_\theta^2(k)\equiv
 \frac{k^3}{2\pi^2}|\theta_k|^2.
 \label{eq:seed-number-spectrum}
\end{equation}
With $\Delta_\theta^2\simeq(1.67\times10^{-5}v)^2$, its integral is
UV dominated,
\begin{equation}
 N_{\rm ad}\simeq
 f_A^2R_1^2 k_{\rm max}
 (1.67\times10^{-5}v)^2.
 \label{eq:seed-number}
\end{equation}
By contrast, the non-tuned homogeneous kinetic solution has
$N_{\rm KM}\simeq f_A^2H_1R_1^3v/\sqrt2$. Using $k_1=H_1R_1$ and
Eq.~\eqref{eq:seed-cutoff}, we find
\begin{equation}
 \frac{N_{\rm ad}}{N_{\rm KM}}
 \simeq\sqrt2\,(1.67\times10^{-5})^2v
 \frac{k_{\rm max}}{k_1}
 \simeq4.35\times10^{-5}v^{6/7}
 \left(\frac{k_{\rm max}}{10^5k_*}\right).
 \label{eq:seed-number-ratio}
\end{equation}
Assuming that this comoving number is subsequently conserved, its present
abundance is therefore
\begin{equation}
 \Omega_A^{\rm ad}h^2
 \simeq1.78\times10^{-8}
 \left(\frac{f_A}{10^{10}\,\mathrm{GeV}}\right)^{7/6}
 v^{13/7}
 \left(\frac{k_{\rm max}}{10^5k_*}\right).
 \label{eq:seed-abundance}
\end{equation}
Along the kinetic-misalignment CDM line, $v=v_{\rm KM}$, this becomes
\begin{equation}
 \Omega_A^{\rm ad}h^2
 \simeq6.8\times10^{-4}
 \left(\frac{10^{10}\,\mathrm{GeV}}{f_A}\right)
 \left(\frac{k_{\rm max}}{10^5k_*}\right).
 \label{eq:seed-abundance-km}
\end{equation}
For the maximal cutoff $k_{\rm max}=10^5k_*$, the unamplified seed is
consequently negligible near
$f_A=10^{10}\,\mathrm{GeV}$, but already contributes
$\Omega_A^{\rm ad}h^2\simeq7\times10^{-3}$ at $f_A=10^9\,\mathrm{GeV}$ and
about $2\times10^{-2}$ at $f_A\simeq3\times10^8\,\mathrm{GeV}$. Below this
scale Eq.~\eqref{eq:seed-linear-threshold} is reached and the linear estimate
cannot be extrapolated quantitatively. Nevertheless, it shows that the
primordial adiabatic fluctuations provide a potentially important
conserved-number floor for the subsequent nonlinear QCD evolution. The
estimate is explicitly cutoff dependent because relativistic number is
weighted toward the highest occupied momenta.

\subsubsection{Amplification Mechanisms}

\paragraph{Tachyonic intervals in a rotating background.}
Once the QCD potential becomes relevant, the frequency of the rescaled modes
is
\begin{equation}
 \omega_{k,\Theta}^2(\eta)=
 k^2+m_A^2R^2\cos\Theta-\frac{R''}{R}.
 \label{eq:tachyonic-frequency}
\end{equation}
Consider an individual interval $(\eta_{j,-},\eta_{j,+})$ during which
$\omega_{k,\Theta}^2<0$. Away from its turning points, the growing WKB
solution gives the within-window amplification
\begin{equation}
 {\cal T}_{k,j}^{\rm tach}\sim e^{X_{k,j}},
 \qquad
 X_{k,j}\equiv
 \int_{\eta_{j,-}}^{\eta_{j,+}}d\eta\,
 \sqrt{-\omega_{k,\Theta}^2(\eta)}.
 \label{eq:single-tachyonic-window}
\end{equation}
Only modes satisfying
\begin{equation}
 k^2<\cmass^2|\cos\Theta|+\frac{R''}{R}
 \label{eq:tachyonic-window-band}
\end{equation}
enter the window. Since $\cmass=m_AR\propto R^{\n/2+1}$ before the QCD
crossover, it increases steeply from one passage to the next. The latest
negative-curvature interval consequently gives the largest $X_{k,j}$ and
dominates the transient tachyonic amplification.

However, $ {\cal T}_{k,j}^{\rm tach}$ is the gain measured only across the
negative-frequency window, not the Floquet multiplier of the complete
rotation. Multiplying the individual factors $e^{X_{k,j}}$ neglects the
subsequent positive-frequency evolution. For a rotating kinetic-misalignment
background, the growth during the negative-curvature part of a rotation is
largely compensated during the remainder of the cycle.

This cancellation can be shown exactly when expansion and the variation of
the mass are neglected over one rotation. Defining the conformal mass
$\cmass\equiv m_AR$, the
background and perturbation equations are
\begin{equation}
 \Theta''+\cmass^2\sin\Theta=0,
 \qquad
 \hpsi_k''+\left(k^2+\cmass^2\cos\Theta\right)\hpsi_k=0.
 \label{eq:rotating-hill-system}
\end{equation}
Differentiating the background equation shows that
\begin{equation}
 \hpsi_0(\eta)=\frac{R}{R_1}\Theta'(\eta)
 \label{eq:rotation-zero-mode}
\end{equation}
is an exact $k=0$ perturbation. For a rotational solution $\Theta'$ is
nonzero and periodic, so its Floquet multiplier is unity. The perturbation
equation has no first-derivative term and hence has a conserved Wronskian;
the product of its two Floquet multipliers is one. The second multiplier is
therefore also unity and
\begin{equation}
 \mu_{k=0}=0.
 \label{eq:rotation-zero-floquet}
\end{equation}
Thus, the tachyonic portions of a complete rotation do not produce a net
exponential instability.

The same result organizes the finite-momentum spectrum. The mode equation
can be written as the periodic eigenvalue problem
\begin{equation}
 \left[-\frac{d^2}{d\eta^2}
 -\cmass^2\cos\Theta(\eta)\right]\hpsi_k
 =k^2\hpsi_k.
 \label{eq:rotation-hill-operator}
\end{equation}
Because the $k=0$ eigenfunction $\Theta'$ has no nodes, it is the lowest
periodic band edge. Small positive $k^2$ therefore enters a stable band.
Exponential growth at finite momentum occurs only in the gaps of this Hill
operator and is properly classified as parametric resonance, rather than as
the accumulation of separate tachyonic bursts.

In the expanding QCD background the cancellation is not exact, but it
remains an good local description when $m_A/H\gg1$: the background
parameters change only by a relative amount
$\mathcal O(H/m_A)$ per rotation. Earlier, when the rotation is faster than
the mass scale, the intervals with $\omega_{k,\Theta}^2<0$ are short, which suppressed the amplification. The residual transient amplification is naturally suppressed and might require huge number of summits to turn a linear adiabatic seed into a fragmented field.
This might explain why we do not observe fragmentation driven by tachyonic
intervals in the rotating kinetic-misalignment solutions, see also \cite{Morgante:2021bks,Fonseca:2019ypl}.

The argument does not apply to a summit landing or hilltop trajectory. These
histories contain an isolated negative-curvature interval which is not
followed by the compensating part of a rotation. If
$\Theta=\pi-\delta$ remains close to the unstable maximum, then
\begin{equation}
 \delta''-\cmass^2\delta\simeq0,
 \qquad
 \Delta\eta_{\rm hill}\simeq
 \cmass^{-1}\ln\frac1{\epsilon},
 \label{eq:hilltop-dwell}
\end{equation}
where $\epsilon$ is the initial distance from the summit and $\cmass$ has
been treated as approximately constant during the estimate. A fluctuation
with $k<\cmass$ then grows by
\begin{equation}
 {\cal T}_k^{\rm hill}\sim
 \exp\left[\sqrt{\cmass^2-k^2}\,
 \Delta\eta_{\rm hill}\right]
 \simeq
 \epsilon^{-\sqrt{1-k^2/\cmass^2}}.
 \label{eq:hilltop-tachyonic-growth}
\end{equation}
The amplification can therefore be parametrically large, approaching
$1/\epsilon$ for the infrared modes. This applies both to a
large-misalignment initial condition placed near the hilltop and to a
summit-landing trajectory that dynamically arrives there with very small
velocity. In the latter case the growing fluctuations eventually backreact
and regulate the formally divergent homogeneous hilltop enhancement. These
long, non-periodic windows must be distinguished from the short
negative-curvature intervals of the rotating branch. 

\paragraph{Parametric resonance.}
The oscillating zero mode can parametrically amplify fluctuations through
the finite-$k$ instability bands identified above. In the small-amplitude
regime we use the WKB solution in Eq.~\eqref{eqn:wkb} and define a common
phase notation for the zero mode and fluctuations,
\begin{equation}
 \Theta(\eta)=\bar\Theta(\eta)\cos\!\left[\Omega_0(\eta)+\alpha\right],
 \qquad
 \Omega_k(\eta)\equiv\int^\eta d\eta'\,\omega_k(\eta'),
 \qquad
 \omega_k^2\equiv k^2+m_A^2R^2-\frac{R''}{R}.
 \label{eq:parametric-background}
\end{equation}
In particular, $\omega_0\simeq\cmass$ and hence
$\Omega_0\simeq\int^\eta d\eta'\,\cmass$, in agreement with
Eq.~\eqref{eqn:wkb}. Absorbing the constant phase $\alpha$ into the origin of
$\Omega_0$, over a single oscillation $\bar\Theta$ and $\cmass$ change
adiabatically and
\begin{equation}
 \cos\Theta\simeq
 1-\frac{\bar\Theta^2}{4}
 \left(1+\cos2\Omega_0\right).
\end{equation}
Neglecting the small $R''/R$ correction, the mode equation becomes
\begin{equation}
 \frac{d^2\hpsi_k}{d\Omega_0^2}
 +\left[A_k-2q\cos2\Omega_0\right]\hpsi_k=0,
 \qquad
 A_k=1+\frac{k^2}{\cmass^2}-\frac{\bar\Theta^2}{4},
 \qquad
 q=\frac{\bar\Theta^2}{8}.
 \label{eq:parametric-mathieu}
\end{equation}
The first Mathieu band and its maximum conformal-time Floquet rate are
therefore
\begin{equation}
 \frac{\bar\Theta^2}{8}
 <
 \frac{k^2}{\cmass^2}
 <
 \frac{3\bar\Theta^2}{8},
 \qquad
 \mu_{\eta,\rm max}\simeq
 \frac{\bar\Theta^2\cmass}{16}.
 \label{eq:parametric-band}
\end{equation}
Adiabatic conservation of the homogeneous comoving number gives
\begin{equation}
 \bar\Theta^2\cmass
 =\bar\Theta_s^2m_A(R_s)R_s
 \left(\frac{R_s}{R}\right)^2,
 \qquad
 \bar\Theta_s^2\equiv2(1-\cos\Theta_s).
 \label{eq:parametric-amplitude}
\end{equation}
It is useful to define the comoving mass scale at stopping,
\begin{equation}
 k_s\equiv m_A(R_s)R_s.
 \label{eq:parametric-ks}
\end{equation}
The instantaneous Floquet rate then decreases as $R^{-2}$. At the same time,
the center of the band drifts according to
\begin{equation}
 k^2\simeq
 \frac{\bar\Theta^2\cmass^2}{4}
 =\frac{\bar\Theta_s^2k_s^2}{4}
 \left(\frac{R}{R_s}\right)^{(\n-2)/2}.
 \label{eq:parametric-drift}
\end{equation}
Consequently, the band crosses a mode $k$ at
\begin{equation}
 \frac{R_k}{R_s}\simeq
 \left(\frac{4k^2}{\bar\Theta_s^2k_s^2}\right)^{2/(\n-2)}.
 \label{eq:parametric-crossing}
\end{equation}
This expression describes a crossing during the harmonic evolution only
when $R_k\gtrsim R_s$, or
$k\gtrsim\bar\Theta_s k_s/2$. Thus its domain is parametrically
$k\gtrsim k_s$ for the generic stopping amplitudes
$\bar\Theta_s=\mathcal O(1)$; it is not a prediction for modes whose formal
crossing lies before stopping. Each eligible comoving mode remains resonant
for only $\Delta R/R\sim2/(\n-2)$.

The dimensionless parameter controlling the available amplification is
\begin{equation}
 c_s\equiv\frac{m_A(R_s)}{H_s}=k_s\eta_s
 =\tau_s^{(\n+4)/2}
 \simeq\left(\frac{v}{\sqrt2}\right)^{(\n+4)/(\n+6)}.
 \label{eq:parametric-cs}
\end{equation}
For $\n=8$, $c_s=\tau_s^6=(v/\sqrt2)^{6/7}$. Integration of the maximal
Floquet rate over the band crossing gives
\begin{equation}
 {\cal B}_k\lesssim
 \exp\left[
 \frac{\bar\Theta_s^2c_s}{8(\n-2)}
 \left(
 \frac{\bar\Theta_s^2k_s^2}{4k^2}
 \right)^{2/(\n-2)}
 \right].
 \label{eq:parametric-boost}
\end{equation}
In particular, for modes of order the stopping scale,
\begin{equation}
 \ln{\cal B}_{k_s}\lesssim
 \frac{\bar\Theta_s^2c_s}{8(\n-2)}
 \left(\frac{\bar\Theta_s^2}{4}\right)^{2/(\n-2)}
 =\mathcal O(c_s)
 \qquad(\bar\Theta_s=\mathcal O(1)).
 \label{eq:parametric-boost-ks}
\end{equation}
This makes explicit that parametric amplification can be exponentially
strong around $k\sim k_s$, with $c_s$ setting the parametric size of the
exponent. The growth remains finite because the Floquet rate decreases and
the instability band moves through every fixed comoving mode.

\paragraph{Onset of fragmentation.}
We can now estimate the velocity for which the fluctuations become nonlinear
already during trapping. The last tachyonic window and the first anharmonic
resonance both affect an order-one logarithmic interval around
$k=\mathcal O(k_s)$. Writing $k=\xi k_s$, their combined logarithmic
amplification during the trapping interval is schematically
\begin{equation}
 X_{\rm trap}(\xi)
 \simeq X_{\rm tach}(\xi)
 +\frac{\bar\Theta_s^2c_s}{8(\n-2)}
 \left(\frac{\bar\Theta_s^2}{4\xi^2}\right)^{2/(\n-2)}.
 \label{eq:total-linear-amplification}
\end{equation}
Here $X_{\rm tach}$ denotes the exponent of the last isolated
negative-curvature window, Eq.~\eqref{eq:single-tachyonic-window}; the second
term is the parametric estimate in Eq.~\eqref{eq:parametric-boost}. Equation
\eqref{eq:total-linear-amplification} is an envelope estimate: in the
anharmonic trapping transition the two effects are not cleanly separable, and
their precise overlap depends on the stopping phase. At exactly $k=k_s$ the
tachyonic condition is only marginally attained at $\cos\Theta=-1$; the
finite tachyonic window therefore lies at $k<k_s$. The strongest combined
amplification occurs in the overlap of that window with the first parametric
band, at $k=\mathcal O(k_s)$ but somewhat below $k_s$.

The physical field fluctuation is diluted by $R_1/R_s=1/\tau_s$ before this
amplification takes place. Hence the condition for fragmentation at trapping
is
\begin{equation}
 \sigma_{\theta,\rm trap}\simeq
 \frac{1.67\times10^{-5}v}{\tau_s}
 \sqrt{\Delta\ln k}\,e^{X_{\rm trap}}\gtrsim1,
 \label{eq:trapping-fragmentation-condition}
\end{equation}
where $\Delta\ln k=\mathcal O(1)$ is the width of the amplified band. Notice
that the full logarithm in Eq.~\eqref{eq:theta-rms-seed} must not be used here:
modes far outside $k\sim k_s$ do not receive the trapping amplification.
Equivalently, the required exponent is
\begin{equation}
 X_{\rm trap}\gtrsim
 \ln\left[\frac{\tau_s}
 {1.67\times10^{-5}v\sqrt{\Delta\ln k}}\right].
 \label{eq:required-fragmentation-exponent}
\end{equation}

For $\n=8$,
\begin{equation}
 \tau_s\simeq\left(\frac{v}{\sqrt2}\right)^{1/7},
 \qquad
 c_s=\left(\frac{v}{\sqrt2}\right)^{6/7}
 \simeq39\left(\frac{v}{100}\right)^{6/7}.
 \label{eq:fragmentation-v-scaling}
\end{equation}
To estimate the exponent, write $X_{\rm trap}=\alpha c_s$. Equation
\eqref{eq:parametric-boost} gives the parametric contribution directly,
\begin{equation}
 \alpha_{\rm par}(\xi)=
 \frac{\bar\Theta_s^2}{8(\n-2)}
 \left(\frac{\bar\Theta_s^2}{4\xi^2}\right)^{2/(\n-2)},
 \qquad \xi\equiv\frac{k}{k_s}.
 \label{eq:trapping-alpha-parametric}
\end{equation}
For $\n=8$ and $\bar\Theta_s^2\simeq4$, this ranges from approximately
$0.08$ near the center of the first band to $0.105$ at its low-momentum
edge. The last tachyonic and anharmonic interval typically contributes an
additional $\alpha_{\rm tach}\sim0.02$--$0.04$, although this part is more
sensitive to the stopping phase. We therefore use the representative
estimate $\alpha\simeq0.12$, with an uncertainty of a few hundredths.

Taking $\Delta\ln k\simeq1$, the useful identities
$v/\tau_s=\sqrt2c_s$ and
$v=\sqrt2c_s^{(\n+6)/(\n+4)}$ reduce
Eq.~\eqref{eq:trapping-fragmentation-condition} to
\begin{equation}
 \sqrt2\,(1.67\times10^{-5})c_s e^{\alpha c_s}\simeq1.
 \label{eq:fragmentation-threshold-v100}
\end{equation}
Its formal solution is
\begin{equation}
 v_{\rm frag}\simeq\sqrt2
 \left[\frac1\alpha
 W_0\!\left(\frac{\alpha}
 {\sqrt2\,(1.67\times10^{-5})}\right)
 \right]^{(\n+6)/(\n+4)},
 \label{eq:fragmentation-lambert}
\end{equation}
where $W_0$ is the principal Lambert function. For $\n=8$ and
$\alpha=0.12$, this gives $v_{\rm frag}\simeq153$; varying
$\alpha=0.10$--$0.14$ gives $v_{\rm frag}\simeq184$--$131$. This agrees
well with the independent nonlinear result. Indeed, inserting the fit in
Eqs.~\eqref{eq:D_function} and~\eqref{eq:D_parameters} gives
$D(v)=0.8$ at $v=139.7$. Thus both the analytical estimate and the first $20\%$ dilution of the homogeneous abundance place the onset of significant fragmentation at trapping $v\sim150$. 

\paragraph{Neglecting the source.}
Once the adiabatic initial conditions are fixed, the right-hand side continuously generates an additional perturbation. The largest term in the residual source is the temperature contribution
\begin{equation}
 \frac{\n}{6}(k\eta)^2\Phi_k
 =\frac{3\n}{2}\Phi_k(0)
 \left(\frac{\sin x}{x}-\cos x\right),
 \qquad x=\frac{k\eta}{\sqrt3},
\end{equation}
whose late-time envelope is $3\n|\Phi_k(0)|/2$. To estimate the largest possible response, write the equation for $\hpsi_k$ in terms of $\tau=\eta/\eta_1$ as
\begin{equation}
 \frac{d^2\hpsi_k}{d\tau^2}+\omega_k^2(\tau)\hpsi_k=S_k(\tau).
\end{equation}
In these dimensionless variables we retain the same phase notation,
\begin{equation}
 \Omega_k(\tau)\equiv\int_1^\tau d\tilde\tau\,\omega_k(\tilde\tau).
\end{equation}
The part generated after $\tau=1$ is given by the retarded Green function,
\begin{equation}
 \hpsi_k^G(\tau)=\int_1^\tau d\tau'\,
 G_k^{\rm ret}(\tau,\tau')S_k(\tau'),
 \qquad
 G_k^{\rm ret}(\tau,\tau')=
 \frac{u_{1k}(\tau')u_{2k}(\tau)-u_{2k}(\tau')u_{1k}(\tau)}
 {W_k(\tau')},
\end{equation}
where $u_{1k}$ and $u_{2k}$ are two independent solutions of the homogeneous equation and $W_k$ is their Wronskian. In the adiabatic regime this becomes
\begin{equation}
 G_k^{\rm ret}(\tau,\tau')\simeq
 \frac{\sin\!\left[\Omega_k(\tau)-\Omega_k(\tau')\right]}
 {\sqrt{\omega_k(\tau)\omega_k(\tau')}}.
\end{equation}
We now make the deliberately conservative assumption that the complete source, including its $\sin\Theta$ dependence, remains perfectly coherent with the natural frequency of the mode. Its resonant envelope is then $S_k^{\rm max}=(3\n/2)|\Phi_k(0)|\tau^{\n+3}$, and the Green-function solution gives
\begin{equation}
 |\hpsi_k^G(\tau)|\lesssim
 \frac{1}{2\sqrt{\omega_k(\tau)}}
 \int_1^\tau d\tau'\,
 \frac{S_k^{\rm max}(\tau')}{\sqrt{\omega_k(\tau')}}
 \simeq\frac{3\n}{3\n+14}|\Phi_k(0)|
 \tau^{\frac{\n+6}{2}},
 \label{eq:Gk-maximal}
\end{equation}
where we used $\omega_k\simeq\cmass\eta_1=\tau^{\n/2+1}$. This must be compared with the adiabatic seed,
\begin{equation}
 |\hpsi_k^{\rm ad}|\simeq0.78\,v|\Phi_k(0)|,
\end{equation}
where no subsequent amplification has yet been included. At trapping, $\tau_s^{(\n+6)/2}=v/\sqrt2$, and hence
\begin{equation}
 \frac{|\hpsi_k^{G}|}{|\hpsi_k^{\rm ad}|}
 \lesssim\frac{3\n}{0.78\sqrt2(3\n+14)}
 \simeq0.57
 \qquad(\n=8).
 \label{eq:Gk-vs-adiabatic}
\end{equation}
More generally, before including amplification,
\begin{equation}
 \frac{|\hpsi_k^{G}(\tau)|}{|\hpsi_k^{\rm ad}|}
 \lesssim0.57\left(\frac{\tau}{\tau_s}\right)^7
 \qquad(\n=8).
\end{equation}
The coherent bound would therefore reach the adiabatic seed at $\tau\simeq1.08\tau_s$. Thus the approximation $G_k=0$ is quantitatively justified when fragmentation occurs around trapping, as in the large-$v$ regime studied in our simulations. Any tachyonic or parametric amplification acts on both contributions, but the adiabatic component is present from the beginning and is exposed to the amplification for longer, whereas the sourced contribution is accumulated gradually. Exponential growth consequently strengthens, rather than weakens, the hierarchy above.

A useful feature of this bound is that it does not require us to decide whether a stationary point or the post-crossing endpoint dominates the physical integral. During kination, $\Theta'=v/\tau^2$, and the relevant phase mismatches are
\begin{equation}
 \Delta_\pm(\tau)=\omega_k(\tau)-\frac{v}{\tau^2}
 \pm\frac{k}{\sqrt3}.
\end{equation}
This differs from the usual Sikivie--Xue resonance after trapping~\cite{Sikivie:2021trt}. There, the cancellation between $\omega_k$ and $m_\psi$ improves as the mode becomes nonrelativistic and can sustain a sizeable post-resonant contribution. In kination, instead, the background frequency $v/\tau^2$ decreases, so a mode generally loses phase coherence after crossing.

The rapidly growing source partially counteracts this loss of coherence. Indeed, Eq.~\eqref{eq:Gk-maximal} can be written as
\begin{equation}
 |\hpsi_k^G(\tau)|\lesssim
 \frac{2}{3\n+14}\,
 \frac{S_k^{\rm max}(\tau)}{\omega_k(\tau)}\,\tau.
\end{equation}
Thus even the perfectly coherent integral is controlled by the latest times: the power-law growth of the envelope suppresses the relative importance of any earlier saddle point. With the true phases retained, an isolated crossing gives a finite stationary-phase contribution, while away from the crossing integration by parts gives schematically
\begin{equation}
 |\hpsi_{k,\pm}^{G}(\tau)|
 \sim\frac{S_k(\tau)}
 {\omega_k(\tau)|\Delta_\pm(\tau)|}.
\end{equation}
Depending on the crossing time, either term can be larger: crossings close to trapping favor the saddle, whereas for earlier crossings the rapidly growing envelope can make the latest post-crossing oscillations dominate. Both are bounded by the phase-coherent result above, and oscillatory cancellations can only reduce their amplitude.

\begin{figure}[t]
 \centering
 \includegraphics[width=0.6\columnwidth]{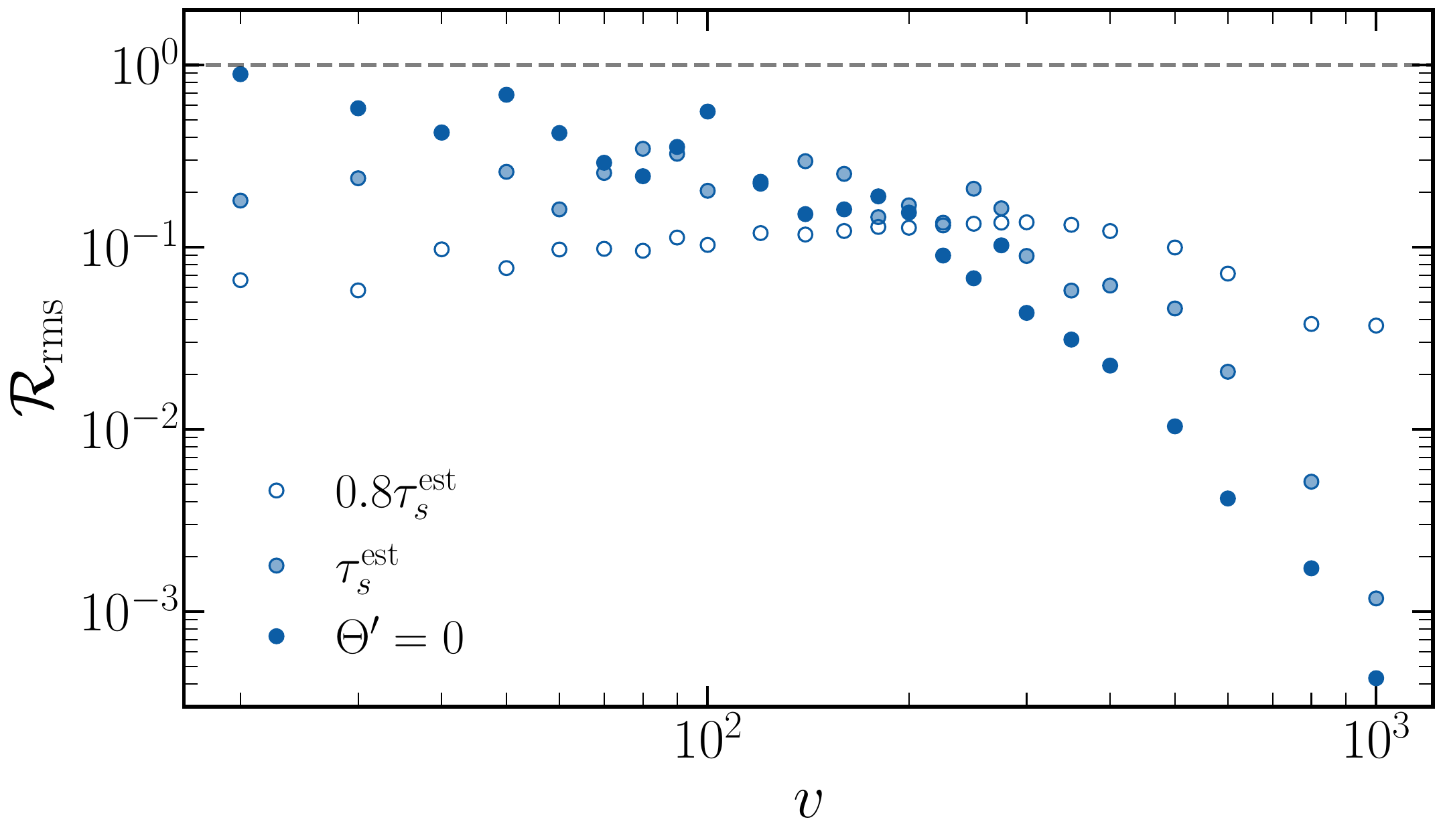}
 \caption{Spectrum-weighted ratio of the source-generated and unsourced linear perturbations, $\mathcal{R}_{\mathrm{rms}}$, cf. Eq.~\eqref{eq:Gk-numerical-ratio}, as a function of the initial velocity. We show the result before trapping (white), at the analytic trapping estimate (light blue), and at the first numerical stopping time (blue). The non-monotonic structure reflects the relative oscillation phases.}
 \label{fig:Gk-linear-scan}
\end{figure}

We test this conclusion by integrating the linear background and mode equations twice, with identical adiabatic initial conditions, once with the complete source in Eq.~\eqref{eq:Gk-full} and once with $G_k=0$. We use $\n=8$, $\Theta_1=1$, and 140 logarithmically spaced modes. Writing $\kappa=k\eta_1$, these cover
\begin{equation}
 10^{-3/2}<\frac{\kappa}{v/(\tau_s^{\rm est})^2}<10^{6/5},
 \qquad
 \tau_s^{\rm est}=\left(\frac{v}{\sqrt{2}}\right)^{1/7}.
 \label{eq:Gk-linear-scan}
\end{equation}
The source-generated solution is isolated as $\hpsi_k^G\equiv\hpsi_k^{\rm full}-\hpsi_k^{G_k=0}$. To avoid artificial divergences when either field crosses zero, we compare the two solutions using the phase-space amplitude
\begin{equation}
 {\cal A}_k^2[\hpsi]
 =|\hpsi_k|^2+
 \frac{|d\hpsi_k/d\tau|^2}{\kappa^2+\tau^{\n+2}},
 \qquad
 {\cal R}_{\rm rms}^2
 =\frac{\displaystyle\int d\ln k\,{\cal A}_k^2[\hpsi^G]}
 {\displaystyle\int d\ln k\,{\cal A}_k^2[\hpsi^{G_k=0}]}.
 \label{eq:Gk-numerical-ratio}
\end{equation}
Figure~\ref{fig:Gk-linear-scan} shows this ratio at a safely linear pre-trapping time, $0.8\tau_s^{\rm est}$, at the estimated trapping time, and at the first numerical stopping time defined by $\Theta'=0$. At low and intermediate velocities the stopping-time result oscillates because the source and the adiabatic mode are sampled at different relative phases. The envelope nevertheless decreases rapidly with velocity. At numerical stopping we find ${\cal R}_{\rm rms}=0.155$, $0.0436$, $0.0224$, $0.0104$, $1.73\times10^{-3}$, and $4.30\times10^{-4}$ for $v=200$, 300, 400, 500, 800, and 1000, respectively. Thus the direct linear evolution confirms that the residual source is negligible in the large-$v$ regime.

We therefore neglect the residual $G_k$ when fragmentation completes around trapping, having already included its early contribution in the initial spectrum. At smaller $v$, however, fragmentation can be delayed or incomplete on this time scale; the bound then no longer justifies setting $G_k=0$, and the Sikivie--Xue sourced fluctuations can provide the dominant seed for the subsequent parametric resonance~\cite{Sikivie:2021trt}. This lower-velocity regime requires an evolution including the gravitational source and is not considered here.

Our analysis therefore identifies $v=\mathcal{O}(10^2)$ as the crossover above which the gravitationally sourced contribution becomes subdominant to the primordial adiabatic fluctuations. This crossover is tied to the fragmentation history: at sufficiently large $v$, tachyonic amplification and fragmentation occur close to trapping, before the gravitational source can accumulate a comparable perturbation.  

\section{Simulation Details }\label{app:simulations}
In this section we present the details of the numerical simulations underlying the main results of this Letter.

\subsubsection{Fragmentation at High Velocities}
First, let us discuss simulation results for the case with large initial velocities.
Parameters used in the simulations are summarized in Table~\ref{tab:sim_parameters}.
We perform simulations for two different values of the axion decay constant, $f_A = 10^9\,\mathrm{GeV}$ and $10^{10}\,\mathrm{GeV}$,
and for each value of $f_A$ we choose 15 different values of the initial velocity shown in the table.
In all cases we set the initial value of the zero mode to $\Theta_1 = 0$, since the results turn out to be almost insensitive to $\Theta_1$ for the range of initial velocities considered here.
Furthermore, in these simulations we put a momentum cutoff at $k = 10^5k_* = 10^5 k_1(\dot{\Theta}_1/2H_1)^{-2/(\n+6)}$ in the initial spectrum of the axion field fluctuations.
Modes with $k>10^5k_*$ were already subhorizon at the onset of kination,
$R=R_{\rm kin}$, assuming the maximal kination duration
$R_s/R_{\rm kin}\simeq10^5$ allowed by the perturbative treatment of
Ref.~\cite{Eroncel:2022vjg}.
The initial spectrum of those UV modes depends on the model of the pre-kination evolution and we simply omit
such extra contributions in this work.

\begin{table}[tbh]
\centering
\normalsize
\renewcommand{\arraystretch}{1.5}
\setlength{\tabcolsep}{10pt}
\begin{tabular}{c c S c}
\hline\hline
$\dot{\Theta}_1/H_1$ & $N$ & {$L/L_1$} & Fit ranges ($\eta_{\mathrm{fit},i}, \eta_{\mathrm{fit},e}$) \\
\hline
$100$, $120$, $140$, $170$, $200$ & 2048 & 2.0 & NA \\
$300$, $400$ & 2048 & 1.0 & NA \\
$500$ & 2048 & 0.7 & $(3\eta_1, 5\eta_1)$, $(4\eta_1, 5\eta_1)$ \\
$1000$ & 1536 & 0.3 & $(4\eta_1, 6\eta_1)$, $(5\eta_1, 6\eta_1)$ \\
$2000$, $5000$ & 1536 & 0.2 & $(4\eta_1, 6\eta_1)$, $(5\eta_1, 6\eta_1)$ \\
$10000$, $20000$ & 1024 & 0.1 & $(5\eta_1, 7\eta_1)$, $(6\eta_1, 7\eta_1)$ \\
$50000$, $100000$ & 1024 & 0.06 & $(5\eta_1, 7\eta_1)$, $(6\eta_1, 7\eta_1)$ \\
\hline
\hline
\end{tabular}
\caption{Parameters used in the simulations to determine the axion relic density from kinetic fragmentation.
The side length of the simulation box is shown in units of $L_1 \equiv 1/\mathcal{H}_1$.
Ranges of the conformal time $\eta_{\mathrm{fit},i} < \eta < \eta_{\mathrm{fit},e}$ over which we perform the fits of the data for the evolution of the axion number
are also shown in the last column.
}\label{tab:sim_parameters}
\end{table}

As shown in Table~\ref{tab:sim_parameters}, we find that the optimal choice for the side length $L$ of the simulation box differs according to the value of the initial velocity.
For simulations with larger initial velocities, smaller values of $L$ are required: For such simulations, the time $\eta_s$ at which the field stops and becomes trapped in the potential is delayed, 
and consequently the characteristic momentum $k/R_{s} \sim m_A(T_{s})$ of the amplified modes becomes larger, requiring a higher resolution in the UV range.
On the other hand, if the initial velocity is decreased, we need to increase the box size $L$ in order to resolve the peak momentum, which is shifted to the infrared (IR).
However, increasing $L$ at fixed $N$ also exacerbates discretisation errors
associated with the rapidly growing axion mass, $m_A\propto\eta^{\n/2}$, so
$N$ must be increased to evolve the system reliably for sufficiently long
times.

\begin{figure}[tbp]
\centering
$\begin{array}{c c}
\subfigure{
\includegraphics[width=0.46\textwidth]{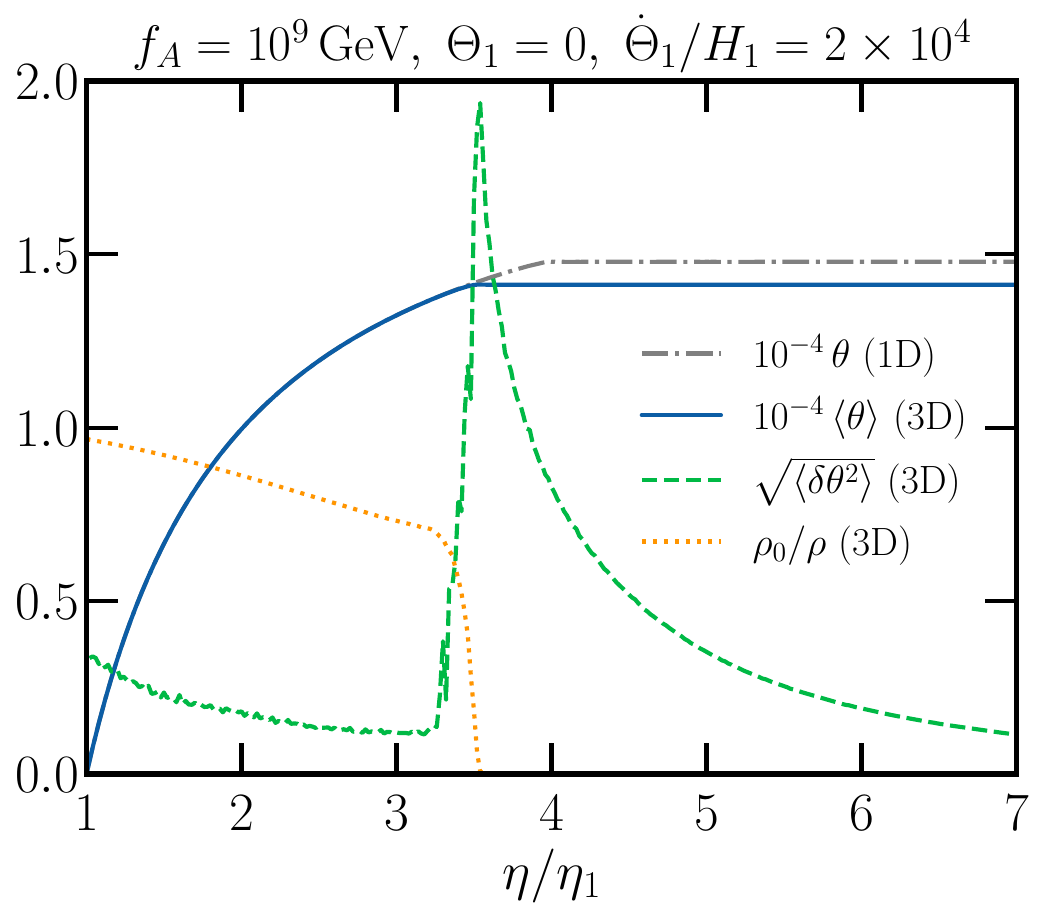}}
&
\subfigure{
\includegraphics[width=0.49\textwidth]{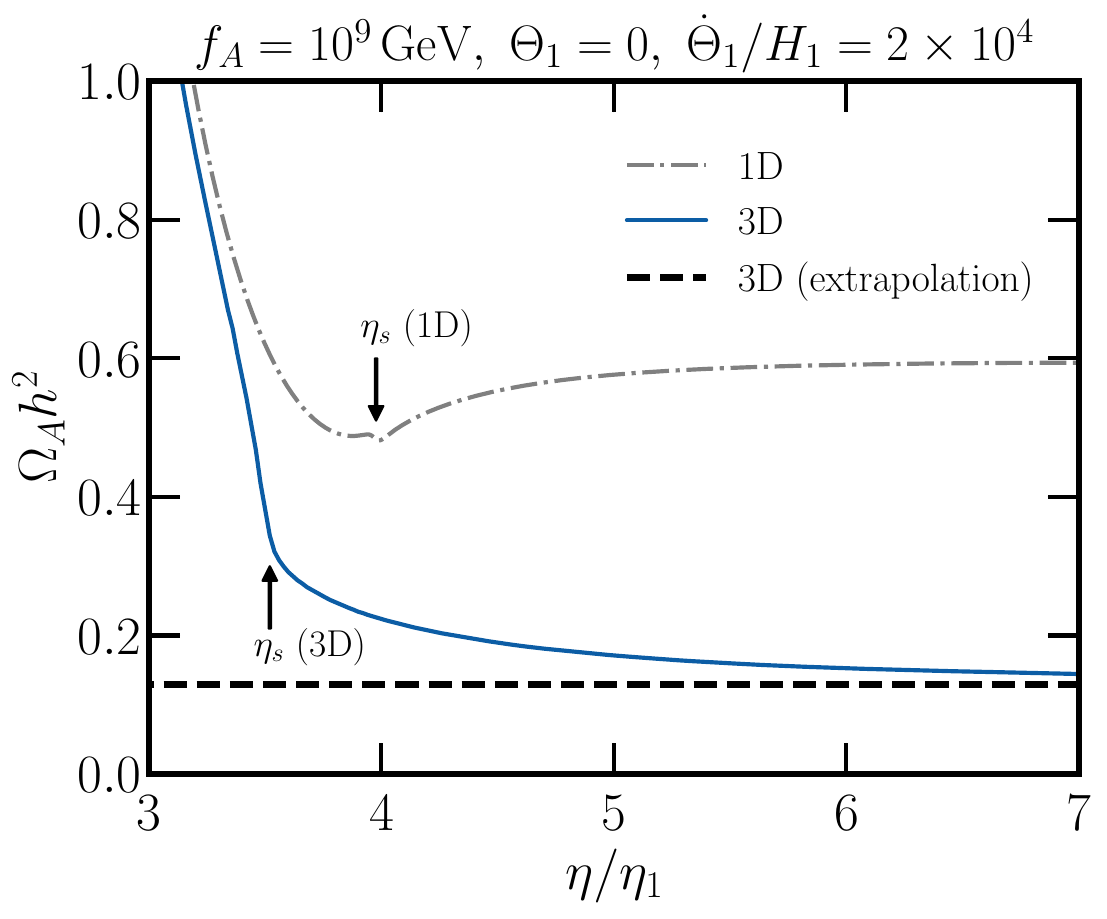}}
\end{array}$
\caption{{\emph{Left:}} Evolution of the axion field value $\theta$ in the 1D simulation (gray dash-dot line) and its spatial average $\langle\theta\rangle$ obtained from the 3D simulation (blue solid line). 
Evolutions of the average square amplitude of the field fluctuations (green dashed line) and the fraction of the energy density of the zero mode (orange dotted line) are also shown.
Note that the values of $\theta$ (1D) and $\langle\theta\rangle$ (3D) are multiplied by $10^{-4}$, as they change by a factor of $\mathcal{O}(10^4)$ due to the high initial velocity.
\emph{Right:} Evolution of the axion number (translated to the present day abundance) in the 1D simulation (gray dash-dot line) and the 3D simulation (blue solid line).
The value of $\Omega_Ah^2$ obtained by extrapolating the 3D result is also shown as black dashed line.}
\label{fig:abundance_evolution}
\end{figure}

Fig.~\ref{fig:abundance_evolution} shows the results of 1D (zero mode only) and 3D simulations for $f_A = 10^9\,\mathrm{GeV}$ with the initial velocity $v=2\times 10^4$.
In the left panel, we see that the average value of $\langle\theta\rangle$ from the 3D simulation stops growing at $\eta_{s} \simeq 3.5\eta_1$, 
which is earlier than $\eta_{s} \simeq 4\eta_1$ for the 1D result.
This early stopping is a consequence of the fragmentation, and the fraction of the energy density of the zero mode $\rho_0/\rho$ actually drops rapidly at that time as shown in the figure.
We also see that the average square amplitude of the field fluctuations $\sqrt{\langle\delta\theta^2\rangle}$ grows rapidly at $\eta_{s}$, implying that the non-linear effect starts to be relevant at that time.
Note that there is a slow decrease in $\rho_0/\rho$, which takes place at times even earlier than $\eta_{s} \simeq 3.5\eta_1$.
This can be attributed to the fact that the energy density of the zero mode $\rho_{0} \propto R^{-6}$ redshifts faster than that of fluctuations $\propto R^{-4}$ before the trapping.

In the right panel of Fig.~\ref{fig:abundance_evolution}, we show the evolution of the axion number obtained from the 3D simulation and compare it to the 1D result.
In the figure, we multiply the comoving axion number $R^3n_A$ by a constant factor such that it represents the present day abundance,
\begin{align}
\Omega_Ah^2 = \frac{m_A}{\rho_{\rm crit}/h^2}\left(\frac{R(\eta)}{R_0}\right)^3n_A(\eta),
\end{align}
where $m_A$ is the zero-temperature mass, $\rho_{\rm crit}$ is the critical density, and $R_0$ represents the scale factor at the present time.
The relativistic code evolves the normalized conformal field
$\hpsi=(R/R_1)\theta=\psi/R_1$. The number density reconstructed from this
field is
\begin{align}
n_A(\eta) &= \frac{1}{L^6}\frac{1}{R^4(\eta)}\sum_{\bf k}\frac{f_A^2}{\sqrt{k^2/R^2(\eta)+m_A^2(T)}} \nonumber\\
&\quad\times\left[\frac{R_1^2}{2}\left|\widetilde{\partial_{\eta}\hpsi}({\bf k},\eta)-\mathcal{H}\widetilde{\hpsi}({\bf k},\eta)\right|^2 
+ \frac{R_1^2}{2}k^2\left|\widetilde{\hpsi}({\bf k},\eta)\right|^2 
+ 2R^4(\eta)m_A^2(T)\left|\widetilde{\sin(R_1\hpsi/2R)}({\bf k},\eta)\right|^2\right],
\label{nA_from_simulations}
\end{align}
where a tilde denotes the Fourier transform.
As clearly seen in Fig.~\ref{fig:abundance_evolution}, there is a suppression in the axion number for the 3D simulation due to the fragmentation of the zero mode into fluctuations.
This number-changing effect is expected to diminish at late times when non-linear interactions become negligible, and the axion number should approach a constant value afterwards.
With this expectation, we fit a function $\Omega_A h^2 = \Omega_A h^2(\eta_{\mathrm{fit},i})/(1+p_1(1-(\eta/\eta_{\mathrm{fit},i})^{p_2}))$ (where $p_1$ and $p_2$ are constant parameters 
and $\eta_{\mathrm{fit},i}$ is the initial time of the fit interval) 
to the data and extrapolate it to $\eta \to \infty$ to extract the asymptotic value.
We perform such a fit for each simulation data with $v \ge 500$. The result of the extrapolation slightly differs according to the choice of the time interval for the fit,
and we assign the uncertainty of the extrapolation based on the difference between the results obtained by performing fits over the two different time intervals shown in Table~\ref{tab:sim_parameters}.
For simulations with $v \le 400$, the extrapolation was not reliably performed because of the discretisation effects at late times,
and we simply estimate the uncertainty in $\Omega_A h^2$ from the difference between the maximum and minimum values of the data within $5 \le \eta/\eta_1 \le 5.5$.

\begin{figure}[tbp]
\centering
$\begin{array}{c c}
\subfigure{
\includegraphics[width=0.48\textwidth]{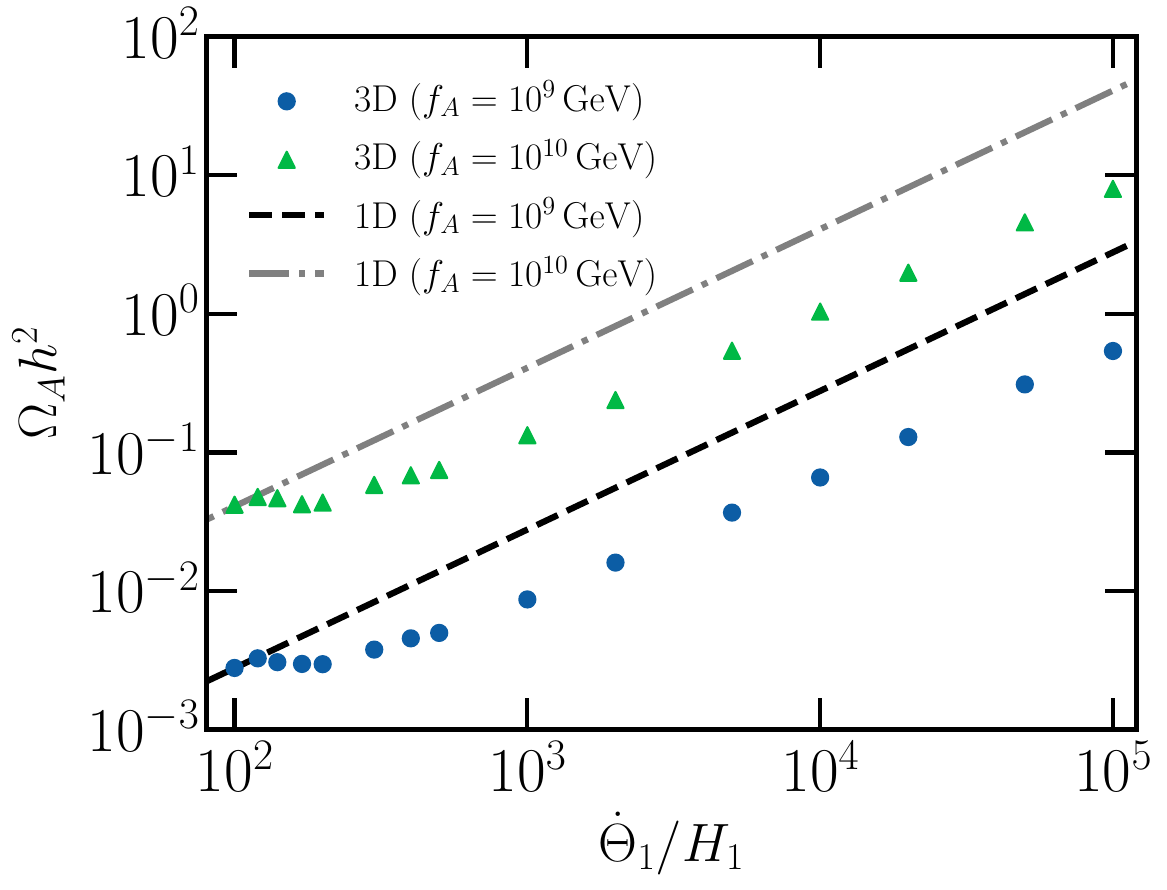}}
&
\subfigure{
\includegraphics[width=0.48\textwidth]{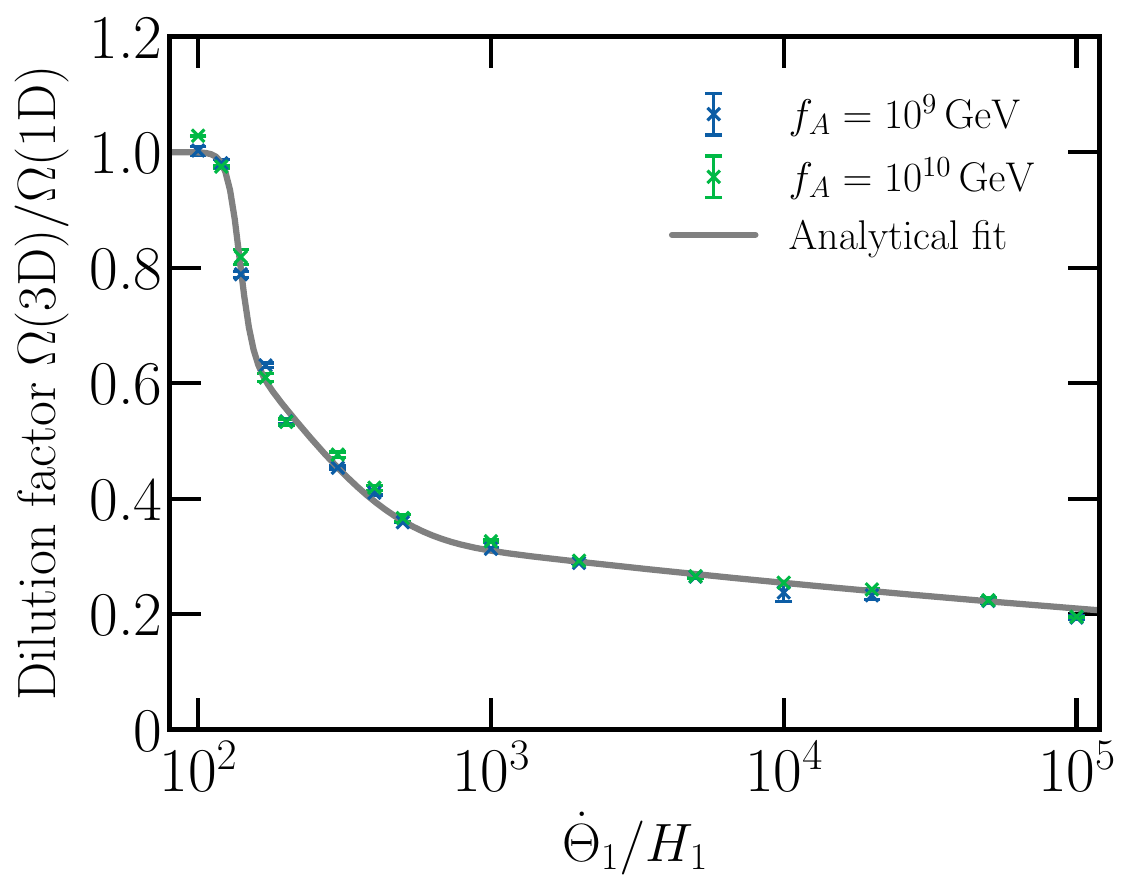}}
\end{array}$
\caption{The axion relic density (\emph{left}) and dilution factor (\emph{right}) as a function of the initial field velocity
obtained from simulations with $f_A = 10^9\,\mathrm{GeV}$ and $10^{10}\,\mathrm{GeV}$.
In the right panel, the fitting function $D(v)$ given by Eq.~\eqref{eq:D_function} with parameters~\eqref{eq:D_parameters} is also shown as gray solid line.
}
\label{fig:Dfactor}
\end{figure}

The left panel of Fig.~\ref{fig:Dfactor} shows the relic density obtained from 3D simulations for a range of initial field velocities.
The figure also shows the estimate from 1D calculations for the sake of comparison.
We see that the fragmentation becomes relevant for $v \gtrsim 100$, in accordance with our estimate from the main Letter.
The comparison between 3D and 1D results allows us to define a dilution factor $D \equiv \Omega_A({\rm 3D})/\Omega_A({\rm 1D})$, which is plotted in the right panel of Fig.~\ref{fig:Dfactor}.
The figure shows that the amount of dilution depends on the initial velocity, taking a value $D \sim 0.2\text{--}0.3$ at large velocities with a slightly decreasing trend.
We also see that the value of $D$ does not significantly depend on $f_A$.
This can be understood from the fact that in units of $\eta \to \eta \mathcal{H}_1$ and $x \to x\mathcal{H}_1$ the equation of motion can be written as
\begin{align}
\theta'' + \frac{2}{\eta}\theta' - \nabla^2\theta + \eta^{2+\n}\sin\theta = 0,
\end{align}
which is independent of $f_A$, if we ignore the change in the effective degrees of freedom of background radiations and a possible deviation from
the pure power law behavior of the temperature-dependent axion mass $m_A(T)$.
Therefore, the value of $f_A$ hardly changes how the initial velocity affects the evolution of the system,
and mainly influences the overall multiplicative factor of $\Omega_A h^2$.

Assuming that the dependence on $f_A$ is negligible, we model the dependence of the dilution factor $D$ on the initial field velocity as
\begin{align}
D(v) = S(v) + \left(1-S(v)\right)\left[A\exp\left(-\frac{v}{v_1}\right) + Bv^{-\alpha}\right], \quad S(v) = \frac{1}{1+(v/v_0)^n}. \label{eq:D_function}
\end{align}
This function approaches $D \to 1$ for $v \ll v_0$ and $D \propto v^{-\alpha}$ for large velocities.
We fit the above function to the simulation results and determine the parameters to
\begin{align}
v_0 = 137.7(4), \quad n = 20.0(9),\quad A = 0.65(1),\quad v_1 = 170.7(2),\quad B = 0.548(2),\quad \alpha = 0.0834(6). \label{eq:D_parameters}
\end{align}
The quoted uncertainties are statistical errors from the fit and do not
include the systematic uncertainties associated with the extrapolation and
lattice discretisation.
Using this fitting function, we can estimate the present-day axion relic abundance including the effect of the fragmentation as
\begin{align}
\Omega_Ah^2 = D(v)\Omega^{\rm 1D}_Ah^2 \simeq 4.08\times 10^{-4}\,\left(\frac{f_A}{10^{10}\,\mathrm{GeV}}\right)^{\frac{\n+6}{\n+4}}D(v)v,
\end{align}
where $\Omega^{\rm 1D}_Ah^2$ is the relic density obtained from 1D calculations.
Note that the condition $\Omega_Ah^2(f_A,v) = 0.12$ is not necessarily single-valued because of the rapid decrease of the function $D(v)$ for $v\gtrsim 100$.
We find that the observed CDM relic density could be explained by multiple values of $v$ for $2.15\times 10^{10}\,\mathrm{GeV} \lesssim f_A \lesssim 2.48\times 10^{10}\,\mathrm{GeV}$,
and the corresponding region is indicated by dotted lines in Fig.~\ref{fig:finetuning} of the main Letter.

\subsubsection{Summit Landing}
Let us consider the case where the field stops at the top of the potential for tuned values of $\Theta_1$ and $v$ with lower initial velocities in more detail.
By scanning a grid of $\{\Theta_1,v\}$ with 1D calculations, we find such a summit-landing takes place at various different values of $\Theta_1$ and $v$ (see Fig.~\ref{fig:grid} of the main Letter).
Rather than investigating all possible values of them, here we perform dedicated simulations for one specific choice:
We fix $f_A = 2.7\times 10^{10}\,\mathrm{GeV}$ and $\Theta_1=0$, and scan the values of the initial velocity around $v\sim 66.3$.
For the scan, we perform an array of 3D simulations with $N = 2048$ and $L/L_1 = 2$, changing $v$ from 65.65 to 66.9
to find a value of $v$ closest to the point at which the amplification of fluctuations becomes most significant.
After finding the tuned value of $v$, we also perform a simulation with a larger grid size on that exact point to study the feature of the power spectrum, which will be discussed in the following subsections.

\begin{figure*}[tb]
    \centering
    \hfill
    \includegraphics[width=0.48\textwidth]{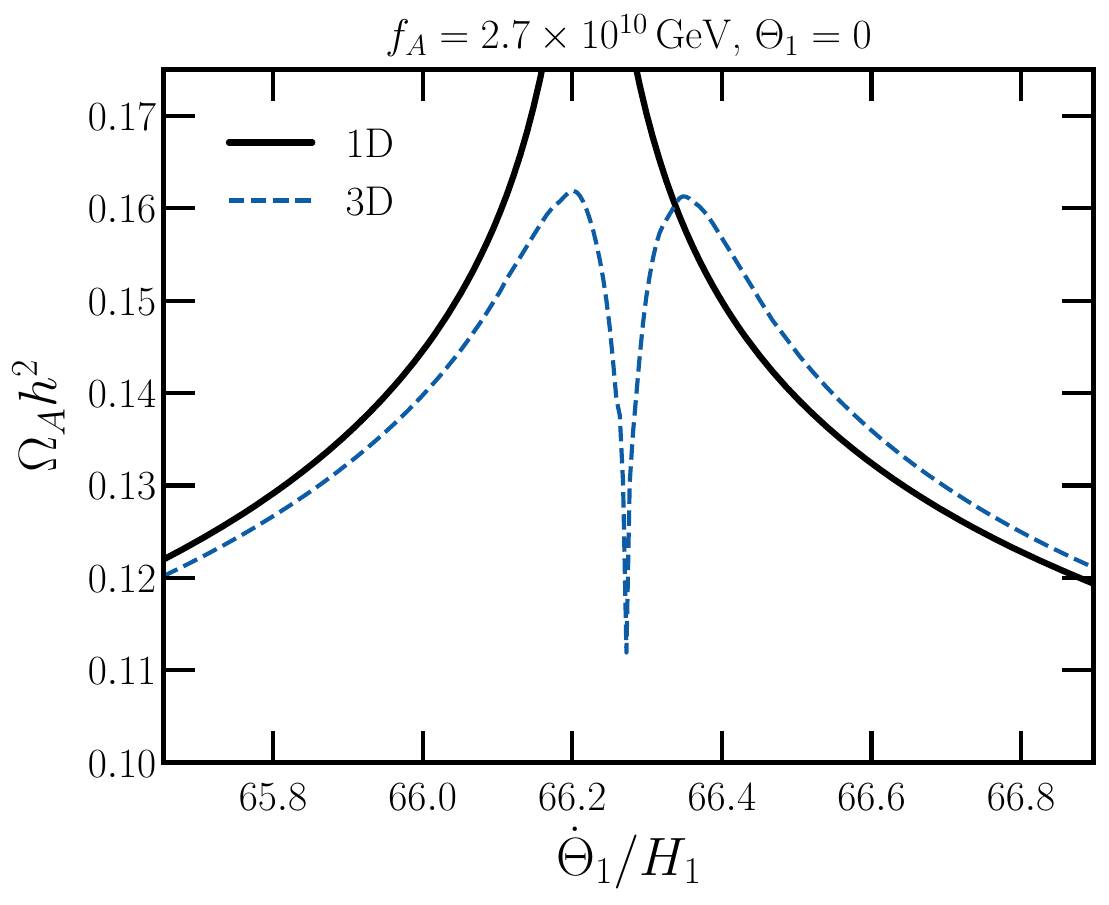}
    \hfill
    \includegraphics[width=0.48\textwidth]{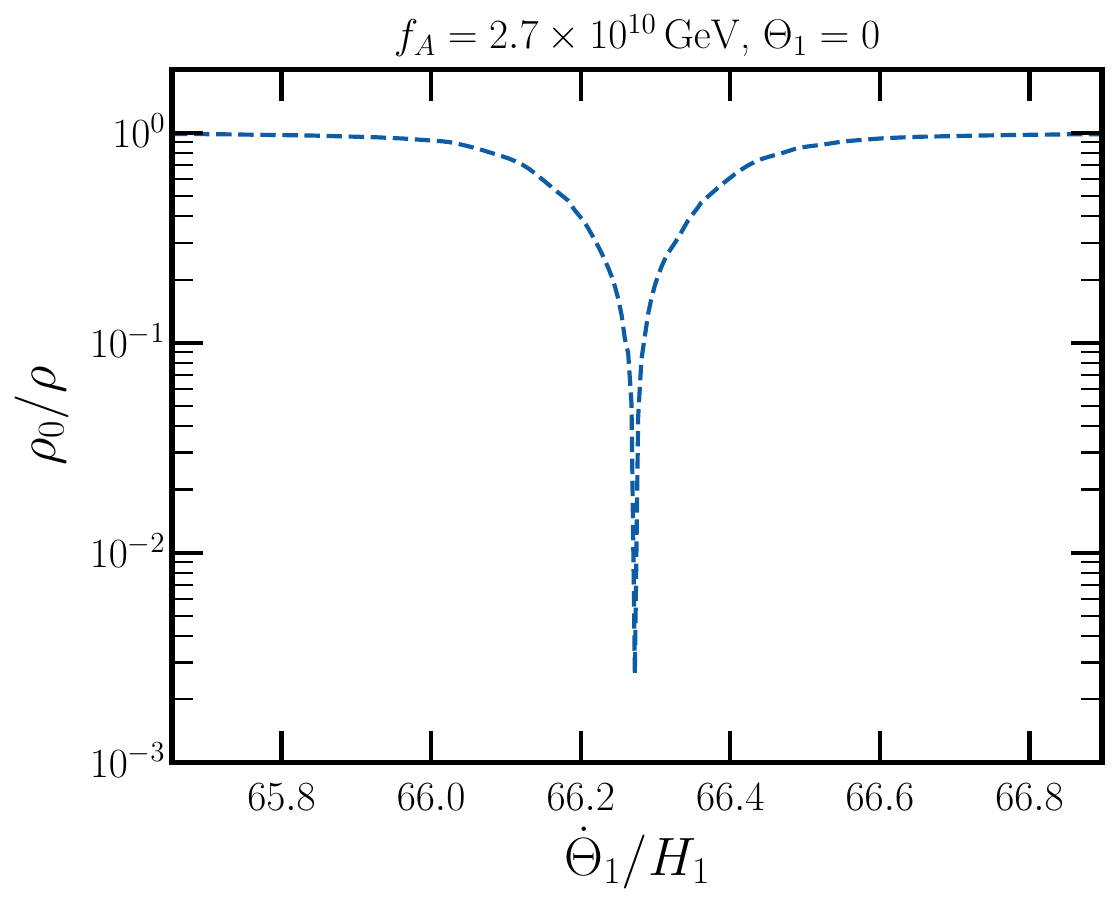}
    \hfill
    \caption{Axion relic density (\emph{left}) and fraction of the energy density of the zero mode (\emph{right}) as a function of the initial field velocity around 
a critical point $\dot{\Theta}/H_1 \sim 66.3$ for the summit-landing.
Blue dashed lines represent the results of 3D simulations evaluated at $\eta = 5\eta_1$. 
In the left panel, the results of 1D calculations are also shown as black solid line.}
\label{fig:summitlanding_abundance}
\end{figure*}

The left panel of Fig.~\ref{fig:summitlanding_abundance} shows the comparison of the axion relic density as a function of $v$
obtained from 3D simulations with that from the 1D scan.
For $v \lesssim 66.2$ and $v \gtrsim 66.3$, both 1D and 3D results show that there is an enhancement of $\Omega_Ah^2$, 
but the curve obtained from the 3D calculation is shifted overall toward higher values of $v$ compared with that obtained from the 1D calculation.
This is because in the 3D system, the zero mode carries slightly less kinetic energy than the 1D case due to the production of fluctuations, and this tiny difference in the initial energy affects the behavior of the zero mode at the summit-landing. For $v \lesssim 66.2$, the field climbs up the potential and turns around at $\Theta \lesssim \pi$. In the 3D case, this turnaround occurs earlier than the 1D case due to the smaller kinetic energy, and it leads to less oscillation energy afterwards. On the other hand, for $v \gtrsim 66.3$, the field overshoots $\Theta = \pi$ after the summit-landing. Hence, in the 3D case, the field stays for a longer time at $\Theta = \pi$ and starts to oscillate later than the 1D case. This makes the axion more abundant in 3D simulations at $v \gtrsim 66.3$.

We also see that the difference between 1D and 3D results becomes most pronounced at the critical point $v\sim 66.3$,
where the abundance is significantly suppressed in the 3D simulations.
This suppression is attributed to the fact that the field fluctuations are significantly amplified due to the tachyonic instability when the field stops precisely at $\Theta = \pi$,
which leads to the abundant production of relativistic axions from the homogeneous mode.
We actually see that the energy density of the homogeneous mode is considerably suppressed at the end of the simulation, 
if we tune the value of $v$ to the critical point (see the right panel of Fig.~\ref{fig:summitlanding_abundance}).
The drastic decrease in the fraction of the zero mode near the critical value implies that fluctuations in this region cannot be analysed within linear theory and require a non-perturbative treatment.

Unfortunately, it turns out that performing fully controlled simulations at the critical point requires enormous computational resources, making it highly challenging in practice.
Since the tachyonic instability is relevant to the lower momentum modes, we need to take a larger value of $L$ in order to follow the evolution of relevant IR modes.
However, we find that higher momentum modes are produced due to the non-linear effect at late times (see Fig.~\ref{fig:ps_evol} for the power spectrum discussed in the following subsections), 
which requires extremely high resolution for the UV region of the spectrum.
This is practically unfeasible because of the rapid increase in the axion mass $m_A \propto \eta^{\n/2}$, 
which causes severe discretization errors at late times of the simulation if we take a larger value of $L$.
Typically, we could not find a clear convergence of $\Omega_Ah^2$ at $\eta \gtrsim 4\eta_1$
with respect to feasible choices of simulation parameters ($L$ and $N$) for the range of $v$ that leads to $\rho_0/\rho \lesssim 0.1$, 
and accordingly the values of $\Omega_Ah^2$ at $v \sim 66.2\text{--}66.3$ shown in the left panel of Fig.~\ref{fig:summitlanding_abundance} potentially involve huge uncertainties.

\subsubsection{Large Scale Simulations}

After obtaining the estimate of relic density as a function of $f_A$ and $v$, we perform large scale simulations to measure the evolution of the power spectrum of axion density fluctuations for selected values of $f_A$. Parameters used for the simulations are summarised in Table~\ref{sim_parameters_large}. We choose eight different values of $f_A$ ranging from $3.16\times 10^8\,\mathrm{GeV}$ to $1.78\times 10^{10}\,\mathrm{GeV}$, and for each case the initial velocity is set to a value such that the relic density satisfies $\Omega_Ah^2 \simeq 0.12$ when extrapolated to $\eta \to \infty$. In addition to these high velocity simulations, we also perform a simulation with $f_A= 2.7 \times 10^{10}\,\mathrm{GeV}$ and $v = 66.2724$, which lies near the center of the dip shown in Fig.~\ref{fig:summitlanding_abundance} and represents the summit-landing case. In all simulations, we set the initial value of the zero mode to $\Theta_1 = 0$. Larger values of the grid size $N$ allow us to simulate the system for longer times, up to the final time $\eta_{\rm end}$ listed in the table. The final time is chosen as a conformal time at which the ratio of the physical lattice spacing $a = RL/N$ to the Compton wavelength of the axion $m_A^{-1}$, which serves as a proxy for the magnitude of discretisation errors, becomes of $\mathcal{O}(1)$ (see the last column of Table~\ref{sim_parameters_large}).

\begin{table}[tbh]
\centering
\normalsize
\renewcommand{\arraystretch}{1.5}
\setlength{\tabcolsep}{10pt}

\begin{tabular}{
S[table-format=2.2e2]
@{\,}
l
c
c
S[table-format=1.3]
c
c
}
\hline\hline

\multicolumn{2}{c}{$f_A\,[\mathrm{GeV}]\ (m_A\,[\mathrm{meV}])$}
& $\dot{\Theta}_1/H_1$
& $N$
& {$L/L_1$}
& $\eta_{\rm end}/\eta_1$
& $m_Aa(\eta_{\rm end})$
\\
\hline
3.16e8  & (18.0)  & $8.35\times10^4$ & 2048 & 0.085 & 8.0 & 1.11 \\
5.62e8  & (10.1)  & $3.74\times10^4$ & 2048 & 0.085 & 8.0 & 1.05 \\
1.00e9  & (5.70)  & $1.85\times10^4$ & 3072 & 0.17  & 7.6 & 1.05 \\
1.78e9  & (3.21)  & $9.10\times10^3$ & 3072 & 0.17  & 7.6 & 1.00 \\
3.16e9  & (1.80)  & $4.16\times10^3$ & 3072 & 0.28  & 7.0 & 1.07 \\
5.62e9  & (1.01)  & $1.98\times10^3$ & 3072 & 0.28  & 7.0 & 1.00 \\
1.00e10 & \ (0.570) & $9.06\times10^2$ & 4096 & 0.50  & 7.0 & 1.27 \\
1.78e10 & \ (0.321) & $3.32\times10^2$ & 8192 & 2.0   & 6.0 & 1.19 \\
2.70e10 & \ (0.211) & 66.2724          & 8192 & 3.0   & 5.5 & 1.16 \\
\hline
\hline
\end{tabular}
\caption{Parameters used for large scale simulations. The ratio of the physical lattice spacing $a$ to the Compton wavelength of the axion at the final time of the simulation is also shown in the last column.}
\label{sim_parameters_large}
\end{table}

\begin{figure*}[tbh]
    \centering
    \includegraphics[width=0.8\textwidth, raise=1.7mm]{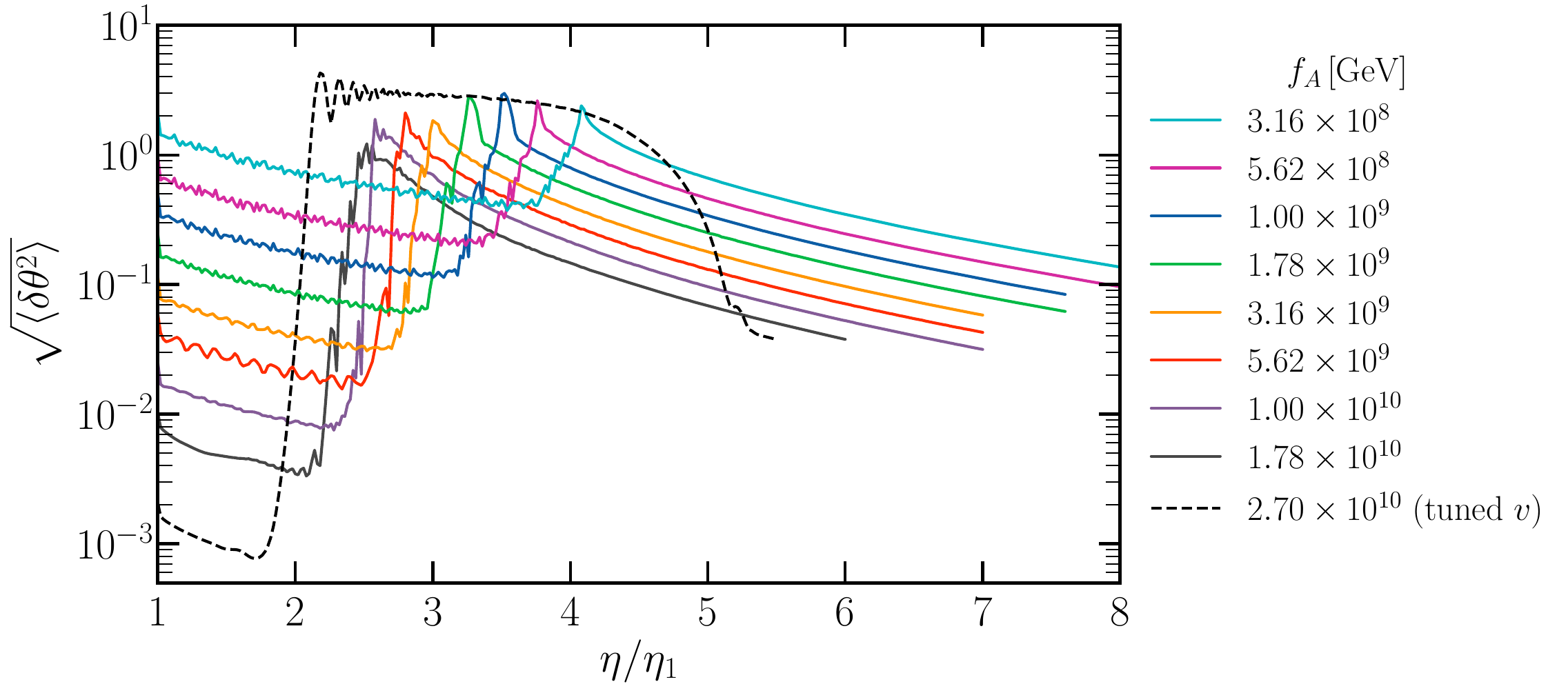}
    \caption{Evolution of the average square amplitude of the axion field fluctuations for different values of $f_A$ and initial velocities.}
\label{fig:summitlanding_deltatheta}
\end{figure*}

Figure~\ref{fig:summitlanding_deltatheta} shows the evolution of the average square amplitude of the field fluctuations obtained from simulations specified in Table~\ref{sim_parameters_large}. Except for the simulation with $f_A = 2.7 \times 10^{10}\,\mathrm{GeV}$, all simulations exhibit qualitatively similar behavior: The fluctuation is amplified around $\eta_{s}$, after which it gradually decays. For larger initial velocities (smaller values of $f_A$), the time of the amplification is delayed as $\eta_{s} \propto v^{2/(\n + 6)}$. On the other hand, the simulation for the summit-landing case (black dashed line) shows a qualitatively different behavior. In this case, the fluctuations are significantly amplified at $\eta \sim 2\eta_1$ due to the tachyonic instability, and remain highly non-linear ($\sqrt{\langle\delta\theta^2\rangle} \gtrsim \mathcal{O}(1)$) for a relatively long time interval. It should be noted, however, that the late-time behavior may be affected by numerical artifacts, as discussed in connection with the relic density in the previous subsection, and should therefore be interpreted with caution.

\subsubsection{Power Spectra}

The power spectrum $P_{\delta}(k)$ is defined as
\begin{align}
\langle\tilde{\delta}({\bf k})\tilde{\delta}^*({\bf k'})\rangle = (2\pi)^3\delta^{(3)}({\bf k} - {\bf k'})P_{\delta}(k),
\end{align}
where $\tilde{\delta}({\bf k})$ is the Fourier transform of the density contrast, $\delta({\bf x}) = \rho({\bf x})/\bar{\rho} - 1$ with $\bar{\rho}$ the spatial average of the energy density $\rho({\bf x})$. Accordingly, the dimensionless power spectrum is given by
\begin{align}
\Delta^2(k) = \frac{k^3}{2\pi^2}P_{\delta}(k) = \frac{k^3}{2\pi^2L^3\bar{\rho}^2}\frac{1}{N_k}
\sum_{|{\bf k}|=k}\left|\tilde{\rho}({\bf k})\right|^2,
\label{eq:Delta2_from_simulations}
\end{align}
where $N_k$ is the number of Fourier modes in the bin centered at $k$.
For the simulations specified in Table~\ref{sim_parameters_large}, $\Delta^2(k)$ was measured and recorded every $\Delta\eta = 0.02\eta_1$ time units.

\begin{figure*}[tbp]
\centering
\hfill
\includegraphics[width=.48\textwidth]{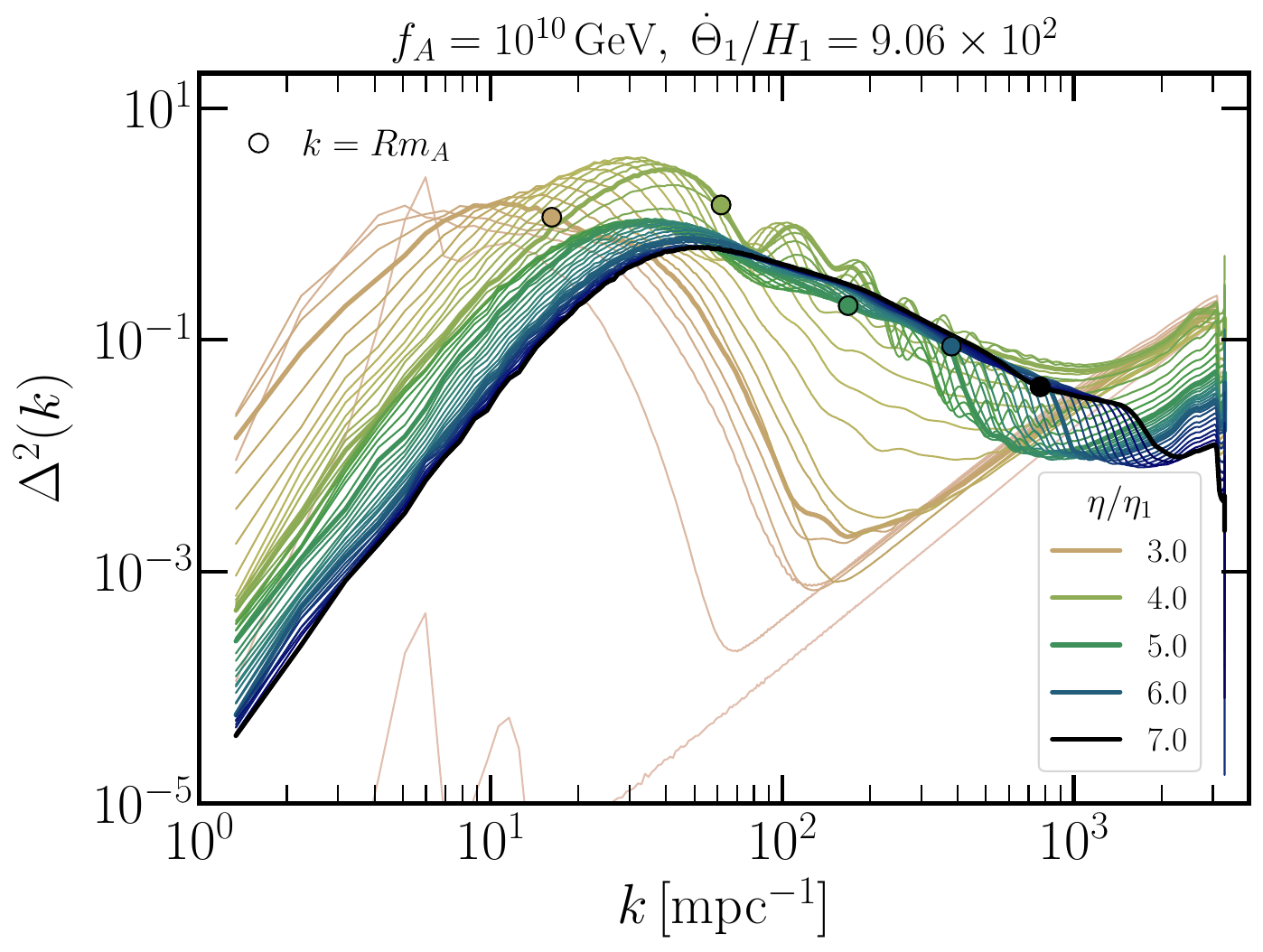}
\hfill
\includegraphics[width=.48\textwidth]{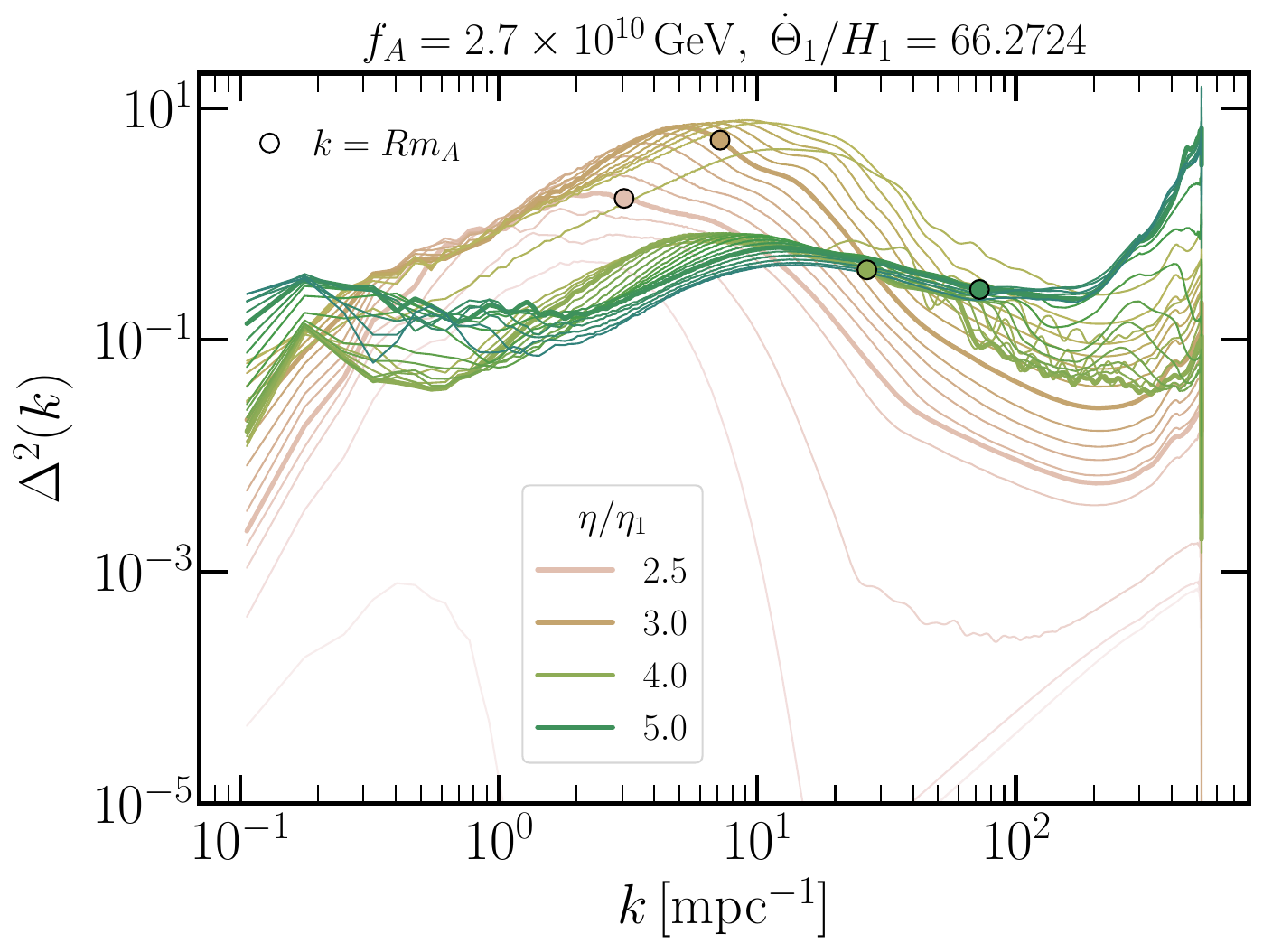}
\hfill
\caption{Evolution of the power spectrum for selected values of $f_A$ and $\dot{\Theta}_1/H_1$, corresponding to the kinetic fragmentation case (\emph{left}) and the summit-landing case (\emph{right}). Time evolution is plotted at intervals of $\Delta \eta = 0.1\eta_1$ using different curves. The comoving wavenumber corresponding to the axion mass scale $k = Rm_A$ is also indicated by circles.}
\label{fig:ps_evol}
\end{figure*}

Figure~\ref{fig:ps_evol} shows the evolution of the power spectra obtained from simulations. For the kinetic fragmentation case (left panel), the peak develops at the scale $k \sim R(\eta_{s})m_A(\eta_{s})$ when the field gets trapped in the QCD potential, and slowly moves to UV due to the self-interactions discussed in the main Letter. Note that the axion mass scale $Rm_A \propto \eta^{\n/2+1}$ shifts to UV much faster than this peak as indicated by circles in the figure. 
A similar behavior of the peak amplification and evolution can also be seen in the summit-landing case (right panel), but in that case the spectra exhibit additional features at both the UV and IR regions at late times. These features are found to depend on simulation parameters, such as the box size and lattice spacing, which suggests that they may arise from discretisation errors, although no definitive conclusion can be drawn at present.

\begin{figure*}[tb]
    \centering
    \includegraphics[width=0.6\textwidth]{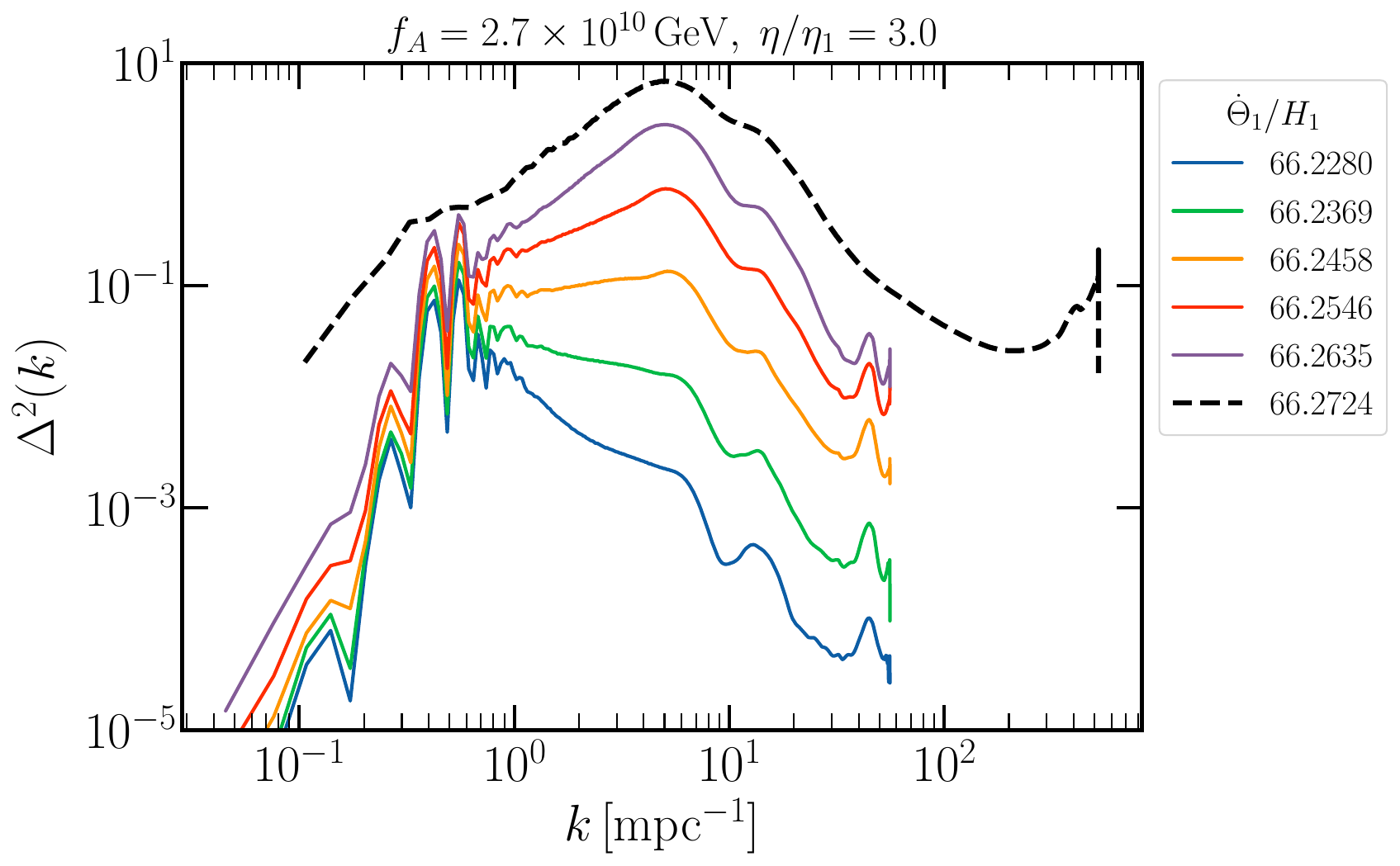}
    \caption{Power spectra evaluated at the conformal time $\eta = 3\eta_1$
for different values of the initial field velocity $\dot{\Theta}_1/H_1$ around a critical point for the summit-landing.
Solid lines are obtained from simulations with $N=2048$, 
while dashed line is from the simulation with $N=8192$ specified in the last row of Table~\ref{sim_parameters_large}.}
\label{fig:summitlanding_ps}
\end{figure*}

The peculiarity of the summit-landing scenario is also shown in Fig.~\ref{fig:summitlanding_ps}, where the power spectra with different initial velocities around the critical value are plotted.
In all cases shown in the figure, the tachyonic effect taking place at $\eta/\eta_1 \sim 2$ amplifies modes around $k \lesssim Rm_A(\eta/\eta_1\sim 2) \sim 1\,\mathrm{mpc}^{-1}$, 
leaving characteristic features in the spectrum at the corresponding wavenumbers. 
Furthermore, as the initial velocity approaches its critical value ($v \to 66.2724$), an additional peak appears in the UV region of the spectrum. 
The position of this peak almost coincides with the axion mass scale at that time ($k \lesssim Rm_A(\eta/\eta_1 = 3) \sim 7\,\mathrm{mpc}^{-1}$ in Fig.~\ref{fig:summitlanding_ps}). 
We interpret this behavior as arising from the extremely strong nonlinearity at the critical point, which continuously produces higher-momentum modes (mildly relativistic axions).
These results suggest that the power spectrum in this scenario is highly sensitive to the initial value and velocity of the zero mode, 
although a detailed analysis of this dependence is beyond the scope of the present work.

In addition to the power spectrum of axion density fluctuations, we also compute the comoving number spectrum, defined by
\begin{align}
N(k) = \frac{(R/R_1)^3}{f_A^2\dot{\Theta}_1}\frac{\partial n_A}{\partial\log k},
\end{align}
where $n_A$ can be computed from the lattice data as in Eq~\eqref{nA_from_simulations}.
Since the typical value of the axion number is of the same order of the PQ charge induced by the field velocity, $n_A \sim f_A^2\dot{\Theta}$, the pre-factor $1/f_A^2\dot{\Theta}_1$ makes $N(k)$ a dimensionless quantity of $\mathcal{O}(1)$ at the maximum.
The number spectrum also exhibits a peak at a scale comparable to that of the peak in the power spectrum.
We identify the peak position from a local maximum in the spectral data and follow its evolution over time. 
The resulting time evolution is presented in Fig.~\ref{fig:kpeak_fit} of the main Letter.
We typically find that the peak position of the power spectrum fluctuates because of the wavy features in the spectrum,
whereas the peak in the number spectrum exhibits a smoother evolution.

\subsubsection{Masking the Axitons}

The nonlinear process associated with the axion fragmentation typically leads to the formation of dense pseudo-soliton configurations, called \emph{axitons}~\cite{Kolb:1993hw}.
The existence of axitons can contaminate the power spectra, which should be treated with care.

In the calculation of the power spectra, we mask the contribution of axitons in a similar way as we masked the string core in the calculation of the spectrum of axions radiated from strings in Ref.~\cite{Saikawa:2024bta}.
We define a mask field $M({\bf x})$, which vanishes around the axiton core and equals 1 elsewhere, and estimate the energy density (in the position space) and its spatial average appearing in Eq.~\eqref{eq:Delta2_from_simulations} as
\begin{align}
\rho({\bf x}) \to M({\bf x})\rho({\bf x}), \quad \bar{\rho} \to \frac{1}{N^3 - N_{\rm mask}}\sum_{\bf x}M({\bf x})\rho({\bf x}),
\end{align}
where $N_{\rm mask}$ is the number of masked points.
We identify the axiton core as the location where the energy density exceeds a prescribed multiple of its spatial average, 
$\rho({\bf x}_{\rm core}) > c_{\rm thres}\bar{\rho}$, where $\bar{\rho}$ here is evaluated without the mask (the average over the whole grid).
We then define $M({\bf x}) = 0$ for all points lying within a distance $r_{\rm mask}$ (in units of $L/N$) from the identified core position.
Here, the energy density threshold $c_{\rm thres}$ and the radius $r_{\rm mask}$ are treated as input parameters and are specified manually.
After testing several choices, we adopted $c_{\rm thres}=100$ and $r_{\rm mask}=6$ as our fiducial values for the threshold and radius, respectively.

\begin{figure}[tbp]
\centering
\includegraphics[width=.5\textwidth]{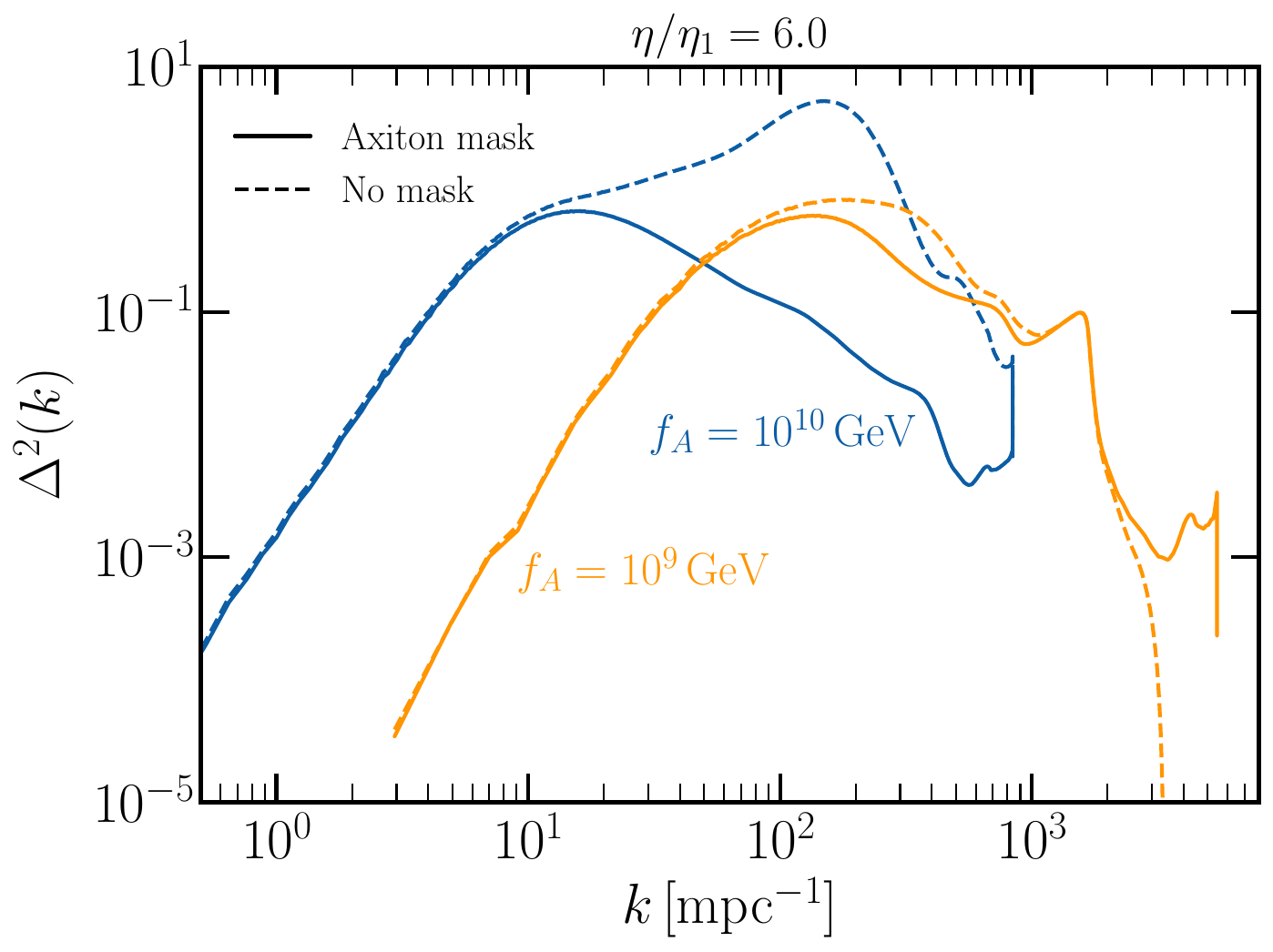}
\caption{Power spectra obtained from the data with the axiton mask (solid line) and those without the mask (dashed line) for two different values of $f_A$.}
\label{fig:axiton_mask}
\end{figure}

Figure~\ref{fig:axiton_mask} shows the comparison between the power spectra obtained by applying the axiton mask and those obtained without the mask.
We see that an additional peak emerges at a scale $k \sim Rm_A$ corresponding to the axiton core size when no mask is applied.
This peak continues to move to the UV as $Rm_A \propto \eta^{\n/2+1}$ since the axitons typically do not decay within the duration of the simulation.
As shown in the figure, the difference between the power spectra with and without the mask is more pronounced for larger values of $f_A$ where the peak of interest is located at a scale well separated from this axiton peak.
These results demonstrate that failing to properly mask the axiton core can lead to incorrect estimates of both the peak position and amplitude of the power spectrum.

\section{Non-relativistic Evolution}\label{app:nrevol}
The evolution of the conformal axion field, 
\begin{equation}
\ctheta''-\nabla^2 \ctheta -\frac{R''}{R}\ctheta + \cmass^2
R\sin\left(\ctheta/R\right) = 0,
\end{equation}
becomes numerically prohibitive due to the increasingly large axion mass $\cmass=m_A R$, which drives $\psi$-oscillations. 
We can factor out these oscillations in the definition of the non-relativistic (NR) field $\Upsilon$, 
\begin{equation}
\label{CPAXION}
\ctheta = \frac{1}{\sqrt{2\omega_0}}
\left(\Upsilon e^{-i\Omega_0}+\mathrm{h.c.}\right)
\end{equation}
with 
\begin{equation}
 \Omega_0(\eta)=\int^\eta d\eta'\,\omega_0(\eta'),
 \qquad
 \omega_0^2=\cmass^2-\frac{R''}{R}\simeq\cmass^2.
\end{equation}
This leads to 
\begin{equation}
\label{GPEU}
i \Upsilon' = -\frac{\nabla^2 \Upsilon}{2 \cmass} + 
\frac{\cmass }{2}
\left[\frac{J_1(2x) }{x}-1\right]\Upsilon \quad ,
\quad x= \frac{|\Upsilon|}{\sqrt{2 \cmass}R}.
\end{equation}
For $x\ll1$, $J_1(2x)/x=1-x^2/2+\mathcal O(x^4)$, and
Eq.~\eqref{GPEU} reduces to the attractive Gross--Pitaevskii equation
\begin{equation}
 i\Upsilon'=-\frac{\nabla^2\Upsilon}{2\cmass}
 -\frac{1}{8R^2}|\Upsilon|^2\Upsilon
 +\mathcal O\!\left(\frac{|\Upsilon|^4}{\cmass R^4}\Upsilon\right).
 \label{eq:GPEU-small-x}
\end{equation}
The NR description assumes that the envelope varies slowly compared with the
carrier oscillation and that the occupied modes have small physical momentum,
\begin{equation}
 \frac{|\Upsilon'|}{\cmass|\Upsilon|}\ll1,
 \qquad
 \frac{k}{\cmass}\ll1,
 \qquad
 \frac{\mathcal H}{\cmass}\ll1.
 \label{eq:NR-validity}
\end{equation}
Under these conditions we neglect second derivatives of the envelope and
rapidly oscillating number-changing terms. We also set
$\omega_0\simeq\cmass$, thereby neglecting the small $R''/R$ correction in
the NR evolution.

We choose the phase convention $\Omega_0(\eta_{\rm NR}^{\rm init})=0$ and
initialize $\Upsilon$ from the relativistic fields as
\begin{equation}
 \Upsilon=\sqrt{\frac{\omega_0}{2}}
 \left[\ctheta(1+i\zeta)+i\frac{\ctheta'}{\omega_0}\right],
 \qquad
 \zeta\equiv\frac{\omega_0'}{2\omega_0^2}.
 \label{eq:relativistic-NR-matching}
\end{equation}
In the adiabatic limit, $\omega_0\simeq\cmass$ and
\begin{equation}
 \zeta\simeq\frac{\cmass'}{2\cmass^2}
 =\frac{(\cmass^2)'}{4\cmass^3}
 =\frac{\n+2}{4\cmass}\frac{R'}{R}.
 \label{eq:NR-adiabaticity}
\end{equation}
The NR equation is invariant under a global phase rotation of $\Upsilon$ and
therefore conserves its norm. With our dimensionless axion-field convention,
the physical comoving axion number is
\begin{equation}
 n_AR^3=f_A^2\langle|\Upsilon|^2\rangle.
 \label{eq:NR-number-norm}
\end{equation}
We fix this norm by requiring the present axion density to equal the observed
dark-matter density,
\begin{equation}
 f_A^2\langle|\Upsilon|^2\rangle
 =\frac{(\Omega_Ah^2)\rho_c}{m_{A,0}},
 \qquad \Omega_Ah^2=0.12.
 \label{eq:NR-relic-normalization}
\end{equation}

These runs are an independent test of the late peak drift rather than a
configuration-by-configuration continuation of the relativistic simulations.
We generate Gaussian $\{\psi,\psi'\}$ fields with random phases and target
spectrum
\begin{equation}
 \frac{d\langle\theta^2\rangle}{d\ln k}
 =F_0\frac{(k/k_0)^a}{1+(k/k_0)^b},
 \qquad a=3,
 \qquad b=4,
 \label{eq:NR-initial-spectrum}
\end{equation}
using the relativistic dispersion relation to set the two field quadratures.
The number-spectrum peak is placed at the value inferred by extrapolating the
relativistic simulations; for the above shape,
$k_p=k_0[a/(b-a)]^{1/b}=3^{1/4}k_0$. The overall amplitude $F_0$ is then fixed
by Eq.~\eqref{eq:NR-relic-normalization}.

We perform small simulations with $N^3=256^3$. For smaller
$f_A$, the peak lies further in the UV, requiring smaller box sizes and making
the evolution considerably more expensive.
The time-stepping follows a Potential$\to$Laplacian$\to$Potential split scheme, with the Laplacian evolution implemented through a four-stage split operator acting on $\Upsilon_i$ and $\Upsilon_r$. The Laplacian is calculated with four neighbours to minimise the norm drift. In the NR theory axion number is conserved by construction and can only drift because of numerical errors. We monitor the norm in Eq.~\eqref{eq:NR-number-norm} and find deviations below $1\%$ in every run.

The attractive interaction causes sufficiently overdense regions to undergo
wave collapse. Once their cores approach the lattice cutoff, unresolved
high-frequency power can contaminate the spectrum. We regulate
this collapse with the fattening (FAT) prescription, replacing $\cmass$ only
in the resummed Bessel potential of
Eq.~\eqref{GPEU} by
\begin{equation}
 m_{\psi,V}=\frac{\mu_{\rm fat}^2}{\delta x^2\cmass},
 \qquad
 x_V=\frac{|\Upsilon|}{\sqrt{2m_{\psi,V}}R},
 \label{eq:NR-fattening}
\end{equation}
where $\delta x$ is the lattice spacing and we use
values $\mu_{\rm fat}= 2\sim3$. The regulated potential term is therefore
\begin{equation}
 \frac{m_{\psi,V}}{2}
 \left[\frac{J_1(2x_V)}{x_V}-1\right]\Upsilon.
\end{equation}
Its small-$x_V$ limit remains
$-|\Upsilon|^2\Upsilon/(8R^2)$ and hence leaves the dilute GP evolution
unchanged. Indeed, the corresponding dilute Hamiltonian contains
\begin{equation}
 \mathcal{H}_{\rm GP}=\int d^3x\left[
 \frac{|\nabla\Upsilon|^2}{2\cmass}
 -\frac{|\Upsilon|^4}{16R^2}\right].
 \label{eq:NR-GP-Hamiltonian}
\end{equation}
For an overdensity of characteristic comoving momentum $k$, wave collapse
begins when the attractive energy overcomes gradient pressure,
\begin{equation}
 \frac{|{\cal H}_{\rm int}|}{{\cal H}_{\rm grad}}
 \sim\frac{\cmass|\Upsilon|^2}{8R^2k^2}\gtrsim1,
 \qquad\text{i.e.}\qquad
 |\Upsilon|^2\gtrsim\frac{8R^2k^2}{\cmass},
 \label{eq:NR-collapse-threshold}
\end{equation}
up to a profile-dependent factor of order unity. Since
$m_{\psi,V}[J_1(2x_V)/x_V-1]/2=-|\Upsilon|^2/(8R^2)
+\mathcal O(m_{\psi,V}x_V^4)$,
the FAT prescription leaves both terms in
Eq.~\eqref{eq:NR-GP-Hamiltonian}, and therefore the collapse threshold,
unchanged. It modifies the dynamics only once the collapsing core reaches
large $x_V$. The FAT prescription does not impose a unique DAS size. Instead,
it regulates the characteristic core density and introduces a minimum
comoving radius,
\begin{equation}
 r_{\rm DAS}\gtrsim r_{\rm min}\sim\frac{\delta x}{\mu_{\rm fat}},
 \qquad
 k_{\rm DAS}\lesssim\frac{\mu_{\rm fat}}{\delta x},
 \label{eq:NR-DAS-minimum-radius}
\end{equation}
The dense branch has an approximately constant characteristic density~\cite{Visinelli:2017ooc}; additional mass is therefore accommodated mainly
by increasing the radius. The regulated population can consequently span a
range of masses and radii larger than $r_{\rm min}$, producing an extended UV
feature rather than a peak fixed at a single lattice momentum. These compact
objects are the so-called dense axion stars, long-lived states of the NR
equation~\cite{Braaten:2015eeu,Visinelli:2017ooc}. Such states are intrinsically
inconsistent endpoints of the NR description: their cores necessarily enter
a regime in which field gradients and binding energies become relativistic,
outside the validity of the NR equation.

The relativistic completion of wave collapse is instead the axiton/oscillon
dynamics observed in our relativistic simulations. These objects are
long-lived and may undergo repeated episodes of collapse and explosive
emission. Such bursts can convert parts of the NR population into relativistic
axions and thereby reduce both the comoving axion number and the final
cold-dark-matter yield. Since our relativistic simulations show convergence
of the comoving number after this number-changing stage, we assume that the
axiton-induced number loss itself is converging. It is then consistent to
continue with the NR prescription, which is number conserving by
construction: both the kinetic and nonlinear evolution operators conserve
$\int d^3x\,|\Upsilon|^2$. We subsequently use the NR simulations only to
follow the number-conserving spectral drift. Comparing regulated and
unmodified NR runs, we find that the drift is relatively insensitive to the
treatment of the compact UV component, although the size of the effect
depends on the simulation parameters. In particular, the finite width of the
DAS mass distribution can shift the extracted peak, but this does not alter
our qualitative conclusions.

We follow the peak of the comoving number spectrum,
$N(k)\propto dn_A/d\ln k\propto k^3|\Upsilon(k)|^2$. Before measuring the
spectrum, we mask the largest overdensities,
\begin{equation}
 \delta_\Upsilon\equiv
 \frac{|\Upsilon|^2}{\langle|\Upsilon|^2\rangle}-1
 \gtrsim5,
 \label{eq:NR-DAS-mask}
\end{equation}
following the masking strategy described in the preceding section. This mask
removes the largest overdensities in order to suppress the effect of the DAS
population on the measured spectrum as much as possible, while retaining the
bulk of the field.

\begin{figure*}[t]
    \centering
    \includegraphics[width=0.65\linewidth]{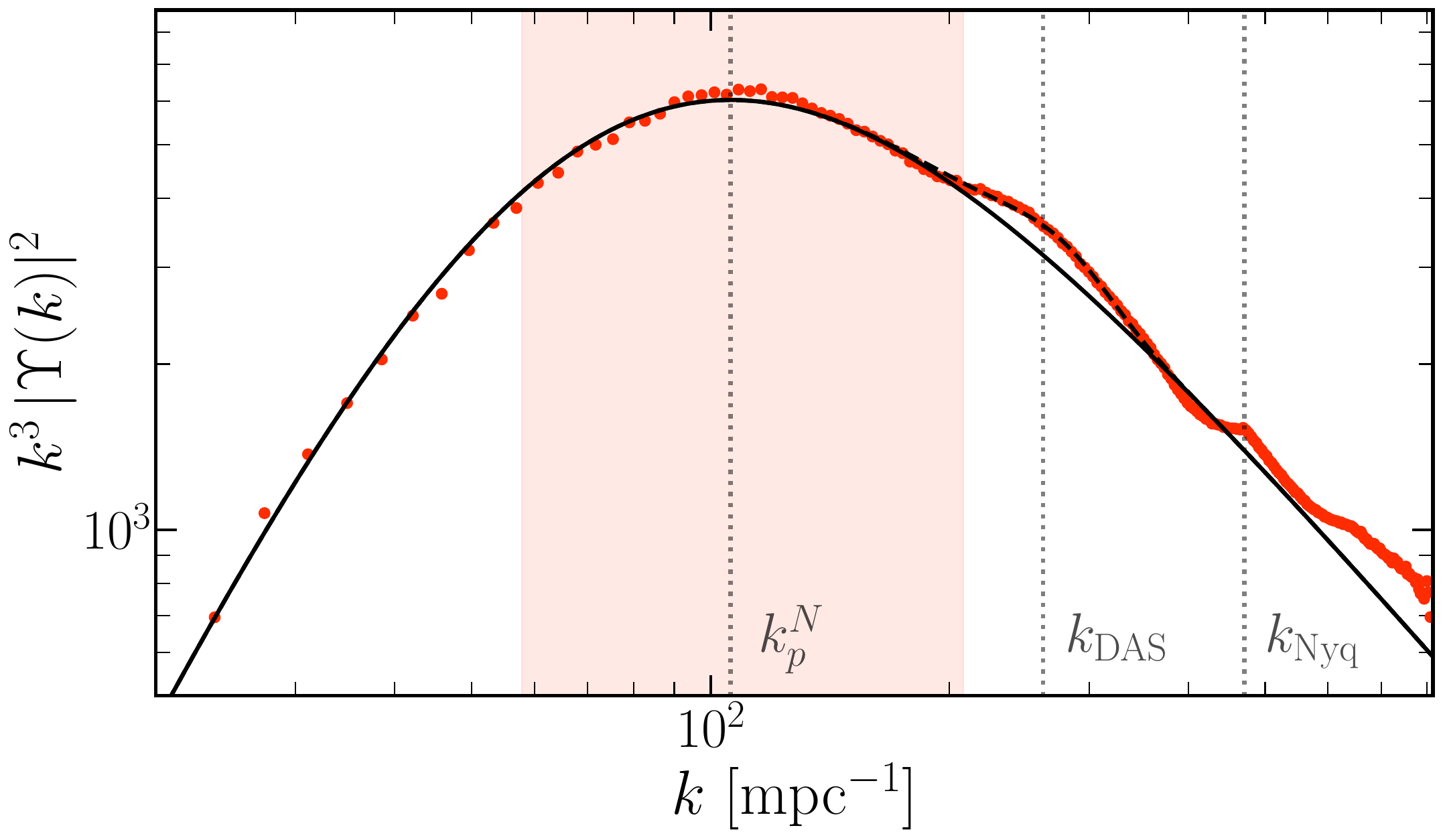}
    \caption{Extraction of the peak position $k_p^N$ from the masked, $k^3$-weighted  number spectrum $k^3|\Upsilon(k)|^2$ using the DAS-subtraction method for $f_A=5.62\times 10^9$ GeV. We fit a two-component model accounting for both the physical peak at $k_p^N$ and the artificial peak at $k_{\mathrm{DAS}}$ and evaluate it with the DAS contribution switched off. Both scales, $k_p^N$ and $k_{\mathrm{DAS}}$, together with the Nyquist scale $k_{\mathrm{Nyq}}$ of the grid, are marked by dotted vertical lines and the fit with the DAS contribution turned on (off) is shown as a dashed (solid) line. The shaded band indicates the width of the peak at $68\%$ of its maximum.
    }
    \label{fig:DAS_substraction}
\end{figure*}

At each snapshot we fit the masked spectrum in logarithmic momentum (see Fig.~\ref{fig:DAS_substraction}). The
spectral peak is described by a broken power law whose infrared behavior is
close to $k^3$ and which turns over at $k_p^N$.
Once the DAS population develops, the fit includes a second component: a
Gaussian peak centered at the characteristic DAS momentum $k_{\rm DAS}$,
close to the grid scale, which describes the residual dense-axion-star
contribution. Although the DAS contribution is also expected to develop a
$k^3$ infrared tail, our estimates show that this tail is much smaller than
the main spectral peak over the fitted range, and we therefore neglect it.
We choose
the optimal box size $L_{\rm opt}$ such that
\begin{equation}
 k_{\rm DAS}\simeq4k_p^N,
 \label{eq:NR-peak-separation}
\end{equation}
which keeps the two fitted peaks well separated. We determine $k_p^N$
from the maximum of the broken-power-law component, obtained by setting the
Gaussian DAS peak to zero. The FWHM is determined from the two half-maximum
roots of the same component, allowing its infrared and
ultraviolet tails to have different slopes.

\begin{table}[tbh]
\centering
\normalsize
\renewcommand{\arraystretch}{1.5}
\setlength{\tabcolsep}{8pt}
\begin{tabular}{S[table-format=2.2e2] @{\,} l S c S S c}
\hline\hline
\multicolumn{2}{c}{$f_A\,[\mathrm{GeV}]\ (m_A\,[\mathrm{meV}])$}
& {$\eta_{\mathrm{NR}}^\mathrm{init}/\eta_1$}
& {$k_{\mathrm{max}}/m_\psi^{\mathrm{init}} \left[\times10^{-2}\right]$}
& {$k_{p}^{N,\mathrm{init}}$ [mpc$^{-1}$]}
& {$k_{p}^{N,\mathrm{final}}$ [mpc$^{-1}$]}
& $L/L_1 \left[\times10^{-2}\right]$\\
\hline
$^*$3.16e8  & (18.0)  &  25.4 & 8.3 & 3161.6 & 5002.6 & 0.17\\
$^*$5.62e8  & (10.1)  &  22.4 & 8.0 & 1459.2 & 2320.8 & 0.33\\
1.00e9  & (5.70)  & 17.7 & 7.0  & 513.1 & 670.9 & 1.1, 1.7, 2.2\\
1.78e9  & (3.21)  & 15.6 & 6.8 & 231.9 & 300.1 & 2.1, 3.2, 4.2 \\
3.16e9  & (1.80)  & 13.7 & 6.7 & 107.7 & 161.6 & 4.1, 6.1, 8.1 \\
5.62e9  & (1.01)  & 12.1 & 6.5 & 49.6 & 79.5 & 7.8, 11.7, 15.6  \\
1.00e10 & \ (0.570) & 10.6 & 6.3 & 23.0 & 42.1 & 15, 22, 30 \\
1.78e10 & \ (0.321) & 9.3 & 6.1 & 10.5 & 21.1 & 29, 43, 58 \\
\hline
\hline
\end{tabular}
\caption{Parameters used for the NR simulations. For each unstarred value of
$f_A$, we ran simulations with $N^3=256^3$ grid sites at the three reported
box sizes $L/L_1$ and five different fattening scales,
$\mu_{\mathrm{fat}}\in\{2.0, 2.2, 2.5, 2.7, 3.0\}$, giving 15 runs per
$f_A$. Here $k_{\mathrm{max}}=\sqrt{3}k_{\mathrm{Nyq}}
=\sqrt{3}\pi N/L$, and $1\,\mathrm{mpc}\equiv10^{-3}\,\mathrm{pc}$.
$^*$ Due to the very small box sizes required for these values of $f_A$, we
run only individual test simulations as crosschecks, rather than the full
series for run-to-run statistics. We therefore exclude them from
Figs.~\ref{fig:kpeak_fit} and~\ref{fig:kpeak_panel}.}
\label{tab:pax_parameters}
\end{table}

\begin{figure}[h]
 \centering
 \includegraphics[width=\linewidth]{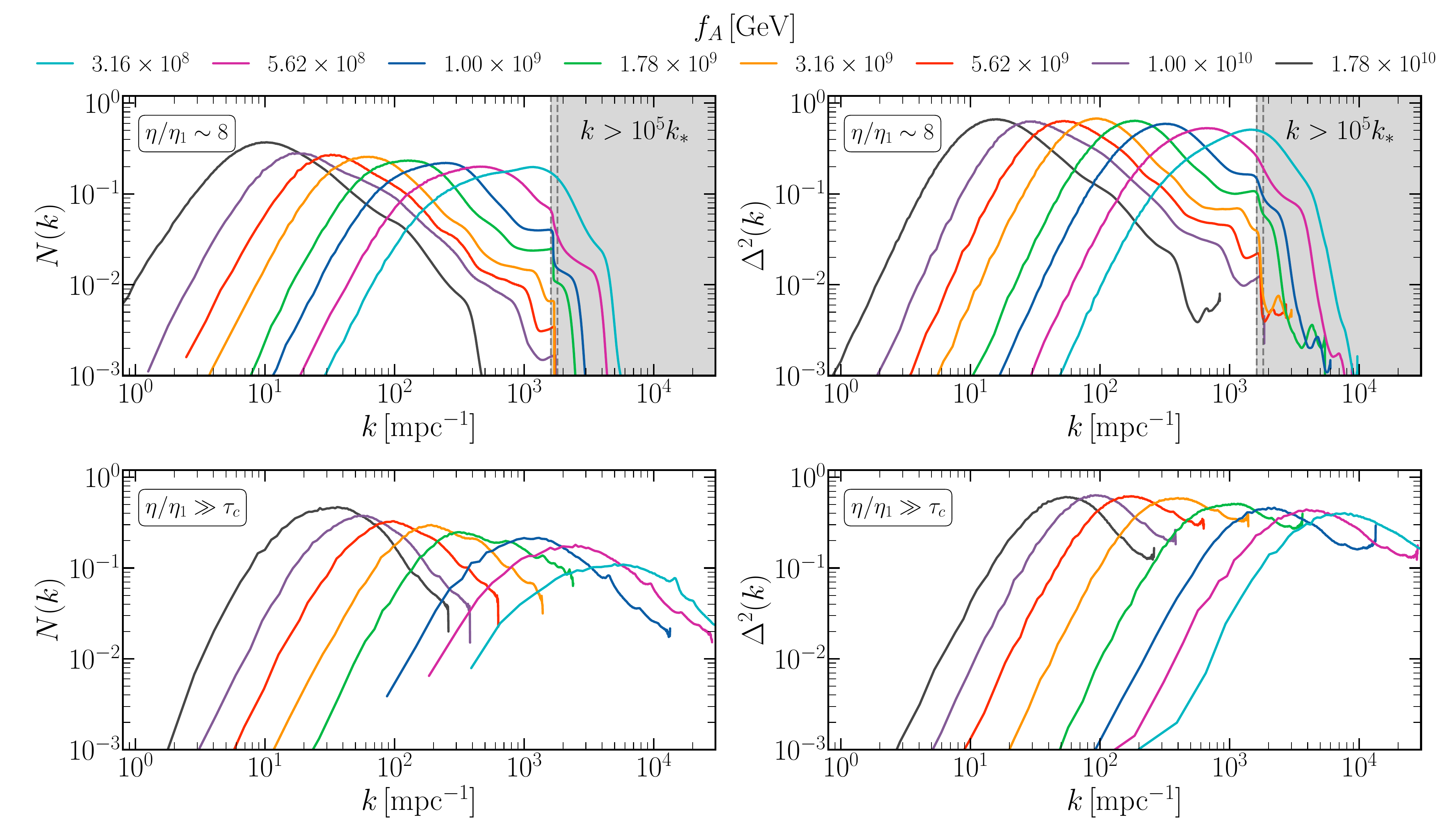}
 \caption{Comoving axion-number spectrum $N(k)$ (\emph{left}) and dimensionless
 density power spectrum $\Delta^2(k)$ (\emph{right}) for the indicated values of
 $f_A$. The upper panels show the spectra near the beginning of the late
 evolution, at $\eta/\eta_1\simeq8$, while the lower panels show their
 asymptotic form after the QCD crossover, $\eta/\eta_1\gg\tau_c$. The gray
 region in the upper panels lies above the primordial cutoff
 $k=10^5k_*$ and is not used to interpret the physical spectrum.}
 \label{fig:spectra_panel}
\end{figure}

\begin{figure*}[tbh]
 \centering
 \includegraphics[width=\linewidth]{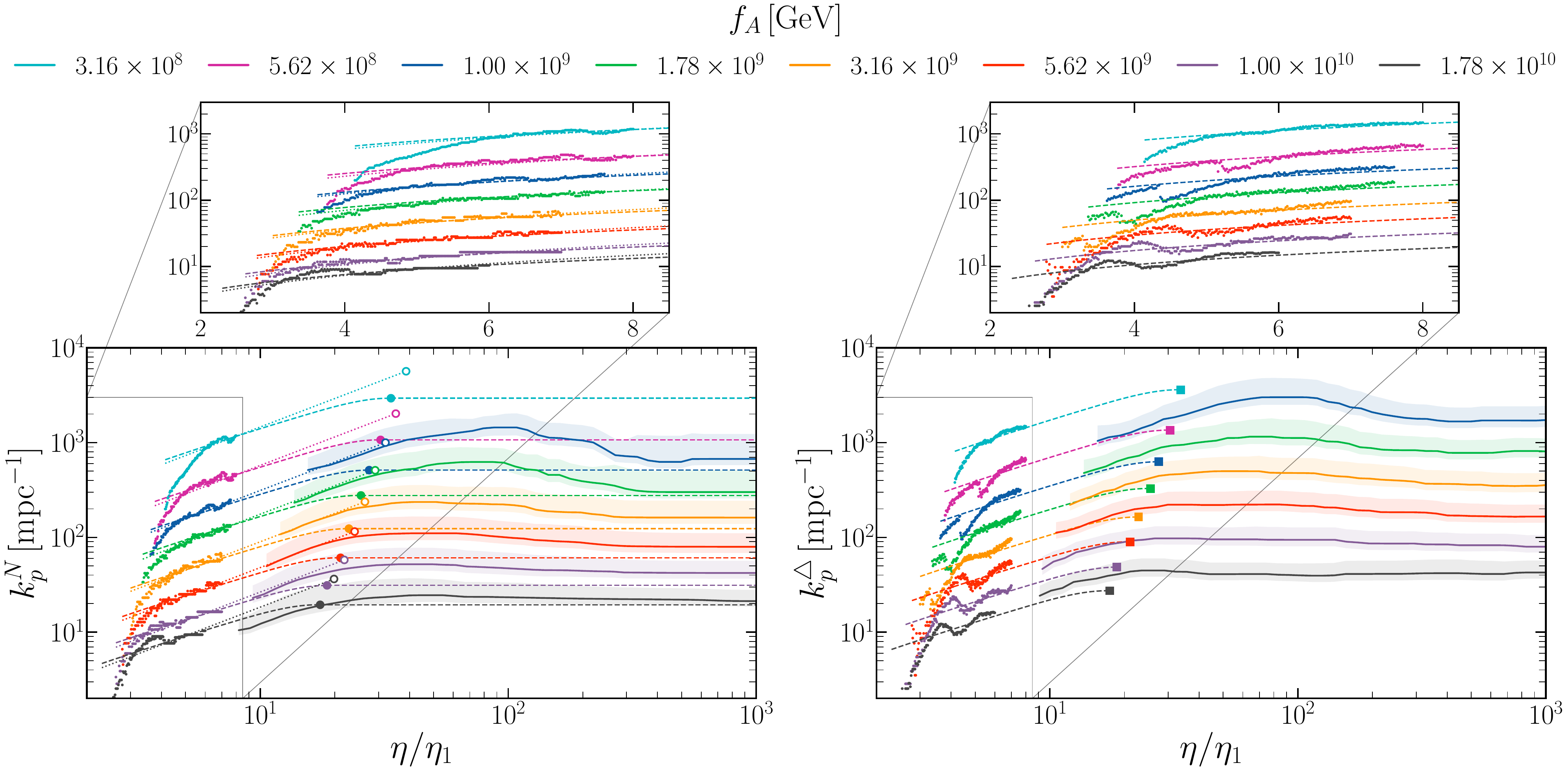}
 \caption{Evolution of the peak of the comoving number spectrum,
 $k_p^N$ (\emph{left}), and of the dimensionless density power spectrum,
 $k_p^\Delta$ (\emph{right}), for the indicated values of $f_A$. Points at early
 times are obtained from the relativistic simulations, while the solid curves
 show the NR continuations; the upper insets magnify the relativistic interval
 $2\lesssim\eta/\eta_1\lesssim8$. Dashed curves show the prediction
 $k_p\propto\sqrt{m_A}/R$ from marginal self-interaction scattering, and
 dotted curves its linear high-temperature approximation. The circle and
 square markers indicate the extrapolated QCD-crossover time for the number
 and density spectra, respectively. Shaded bands show $25\%$ of the fitted
 full width at half maximum (FWHM).}
 \label{fig:kpeak_panel}
\end{figure*}

For each unstarred value of $f_A$ we perform 15 runs, varying the box size around
$L_{\rm opt}$ over the values listed in Table~\ref{tab:pax_parameters} and
varying the fattening parameter. We quote the run-to-run median as the
representative $k_p^N$ trajectory and use the width of a typical fitted peak
to quantify the intrinsic spectral uncertainty. The corresponding number and
density spectra are shown in Fig.~\ref{fig:spectra_panel}, and the resulting
peak histories are shown as solid curves in the left panel of
Fig.~\ref{fig:kpeak_panel}. The peak grows approximately linearly,
$k_p^N\propto\eta$, until the QCD crossover and subsequently settles to its
late-time value. The following section derives this behavior from the
self-interaction rate.

We apply the same procedure to the dimensionless density power spectrum
$\Delta^2(k)$, defined in Eq.~\eqref{eq:Delta2_from_simulations}. In this case
the infrared slope of the broken power law is left free, and $k_p^\Delta$ is
obtained analytically from the corresponding fitted component. The results
are shown in the right panel of Fig.~\ref{fig:kpeak_panel}. Compact coherent
objects contribute disproportionately to density fluctuations and therefore
make the residual DAS feature more prominent in $\Delta^2$ than in the number
spectrum. 

Moreover, since $k_p^\Delta\simeq2k_p^N$ is expected
~\cite{Gorghetto:2024vnp}, the density peak lies closer to this compact UV
feature. Its extraction consequently becomes more difficult at late times.
Varying the box size and regulator provides the corresponding run-to-run
stability check.

\section{Understanding the Peak Drift}\label{app:peak}
We derive the motion of the spectral peak from the non-relativistic
self-interaction rate quoted in the main Letter. The kinetic calculation of
Ref.~\cite{Gorghetto:2024vnp} gives
\begin{equation}
 \Gamma_\lambda\simeq
 \frac{\lambda^2 n_A^2R^2}{64m_A^3k_p^2},
 \qquad
 \lambda=\frac{m_A^2}{f_A^2},
 \label{eq:peak-scattering-rate}
\end{equation}
where $k_p=Rp_p$ is the comoving peak momentum. Equivalently,
\begin{equation}
 \frac{\Gamma_\lambda}{H}
 =\frac{n_A^2}{64f_A^4m_AH}
 \left(\frac{k_p}{m_AR}\right)^{-2}.
 \label{eq:peak-rate-mass-ratio}
\end{equation}

The crucial normalization is fixed by requiring axions to constitute the
observed dark matter. Once number-changing processes have frozen, comoving
number conservation and the present abundance imply
\begin{equation}
 n_A(R)R^3=n_{A,0},
 \qquad
 n_{A,0}=\frac{\rho_{A,0}}{m_{A,0}}
 =\frac{(\Omega_Ah^2)\rho_c}{m_{A,0}},
 \qquad \Omega_Ah^2=0.12,
 \label{eq:peak-number-normalization}
\end{equation}
where $m_{A,0}$ is the zero-temperature mass and
$\rho_c\equiv\rho_{\rm crit}/h^2$; here and below we normalize the present
scale factor to $R_0=1$. Substitution into
Eq.~\eqref{eq:peak-scattering-rate} gives
\begin{align}
 \frac{\Gamma_\lambda}{H}
 &=\frac{(\Omega_Ah^2)^2\rho_c^2}{64f_A^4m_{A,0}^4}
 \frac{m_{A,0}^2}{H R^2}
 \frac{m_A}{k_p^2R^2}
 \notag\\
 &\simeq4.94\times10^3
 \left(\frac{\Omega_Ah^2}{0.12}\right)^2
 \left(\frac{10^{10}\,{\rm GeV}}{f_A}\right)^{2.3512}
 \frac{k_1^2}{k_p^2}
 \left(\frac{R}{R_1}\right)^{\n/2-2}.
 \label{eq:peak-rate-abundance}
\end{align}
In the first line we have only regrouped the exact result, using
$\chi_0=(f_Am_{A,0})^2$. In the second we used
$HR^2=H_1R_1^2$, $m_A=H_1(R/R_1)^{\n/2}$,
$k_1=H_1R_1$, and
$m_{A,0}=5.7\,\mu\mathrm{eV}\,(10^{12}\,\mathrm{GeV}/f_A)$, together with the
values of $H_1$, $R_1$, and $\rho_c$ specified in
Eqs.~\eqref{eq:H1_QCD},~\eqref{eq:R1_QCD}, and~\eqref{eq:rho_c_definition}.

Equation~\eqref{eq:peak-rate-abundance} makes the efficient-scattering region
explicit. The condition $\Gamma_\lambda/H\gtrsim1$ is
\begin{equation}
 k_p\lesssim70.3\,k_1
 \left(\frac{\Omega_Ah^2}{0.12}\right)
 \left(\frac{10^{10}\,\mathrm{GeV}}{f_A}\right)^{1.1756}
 \left(\frac{R}{R_1}\right)^{\n/4-1}.
 \label{eq:peak-efficient-scattering}
\end{equation}
Modes below this boundary are redistributed within a Hubble time, whereas
for larger $k_p$ the quartic scattering is inefficient.

If $k_p$ were fixed, this ratio would grow while the QCD axion mass turns on.
In the kinetic picture, scattering preferentially redistributes axions
towards larger momenta, where a much larger phase-space volume is available,
thereby increasing $k_p$. This UV drift can be viewed as part of the kinetic
relaxation towards thermalisation, accompanied by an increase of the
coarse-grained entropy. The shift also reduces the scattering rate itself.
Efficient evolution therefore approaches the
marginal trajectory $\Gamma_\lambda/H\sim1$, which gives
\begin{equation}
 k_p\simeq
 \frac{(\Omega_Ah^2)\rho_c}{8f_A^2m_{A,0}}
 \left(\frac{m_A}{H R^4}\right)^{1/2}.
 \label{eqn:peak_pos}
\end{equation}
During radiation domination,
$H=H_1(R_1/R)^2$, and hence
\begin{equation}
 k_p\simeq
 \frac{(\Omega_Ah^2)\rho_c}
 {8f_A^2m_{A,0}\sqrt{H_1}R_1}
 \frac{\sqrt{m_A}}{R}
 \propto\frac{\sqrt{m_A}}{R}.
 \label{eq:peak-drift-attractor}
\end{equation}
Using the QCD-cosmology relations summarized above, its numerical form is
\begin{equation}
 k_p\simeq70.5\,k_1
 \left(\frac{10^{10}\,{\rm GeV}}{f_A}\right)^{1.1756}
 \left(\frac{m_A}{H_1}\right)^{1/2}\frac{R_1}{R},
 \label{eq:peak-drift-numerical}
\end{equation}
or in physical units
\begin{equation}
k_p\simeq 2.93\, \mathrm{mpc}^{-1}
   \left(\frac{10^{10}\, \rm GeV}{f_A}\right)^{1.3426}
    \left(\frac{m_A}{H_1}\right)^{1/2}\frac{R_1}{R}, \label{eq:peak-drift-numerical_phys}
\end{equation}
which is the result quoted in the main Letter.

\begin{figure*}[h]
    \centering
    \includegraphics[width=\linewidth]{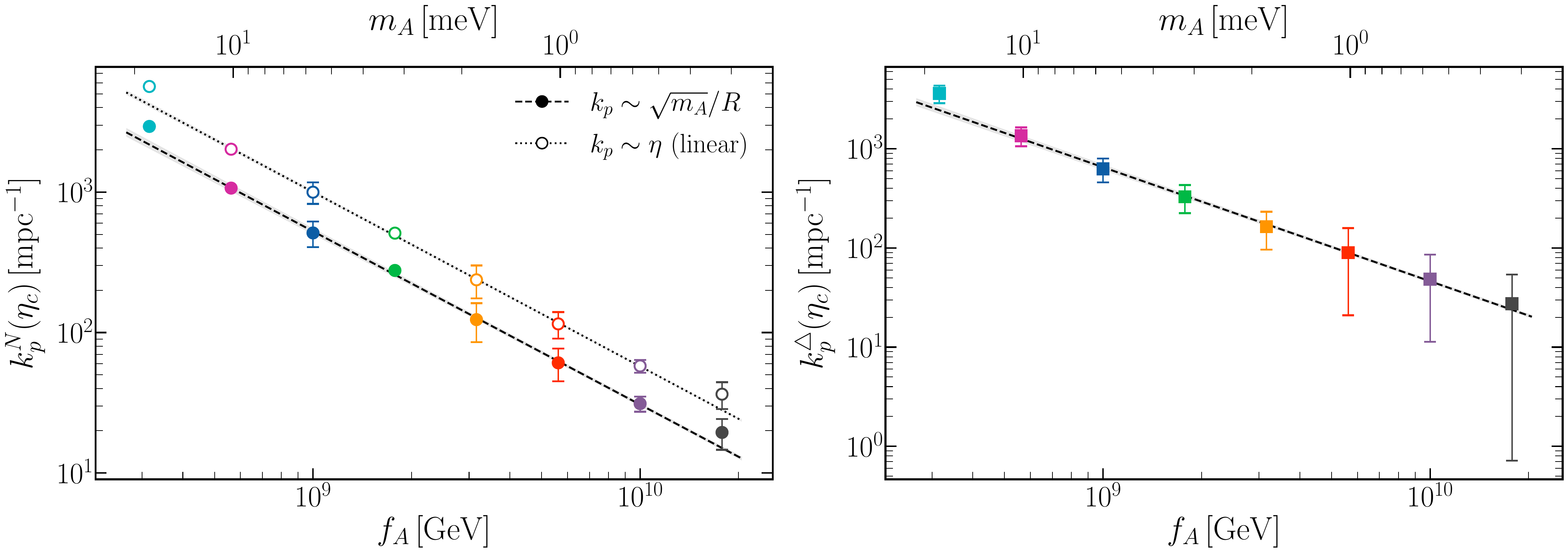}
    \caption{Peak momenta at the QCD crossover, $\eta_c$, as functions of
    $f_A$ (\emph{lower axes}) and $m_{A,0}$ (\emph{upper axes}). The left panel shows the
    peak $k_p^N$ of the comoving number spectrum. Filled circles extrapolate
    its evolution using the marginal-scattering prediction $k_p\propto\sqrt{m_A}/R$, while open circles show the strictly linear
    high-temperature extrapolation $k_p\propto\eta$; the corresponding
    scalings are shown by the dashed and dotted curves. The right panel shows
    the peak $k_p^\Delta$ of the density power spectrum, with the dashed curve
    indicating its fitted mass dependence. Vertical bars quantify the systematic error coming from the fit of the late time relativistic simulation data to a pure linear or $\sqrt{m_A}/R$ trend. }
    \label{fig:kpeak_vs_fA}
\end{figure*}

At high temperature $m_A\propto R^{\n/2}$, so the marginal trajectory obeys
\begin{equation}
 k_p\propto R^{\n/4-1}.
 \label{eq:peak-drift-power-law}
\end{equation}
For $\n\simeq8$ this gives $k_p\propto R\propto\eta$. The drift moderates
when the effective mass exponent falls below four and freezes after the QCD
cross-over, when $m_A$ saturates and $\Gamma_\lambda/H$ decreases. Evaluated
near that endpoint, Eq.~\eqref{eq:peak-drift-numerical} also gives the
approximately $m_{A,0}^{1.3}$ dependence measured in the simulations, as
shown in Fig.~\ref{fig:kpeak_vs_fA}.

\begin{figure*}[h]
 \centering
 \includegraphics[width=0.65\linewidth]{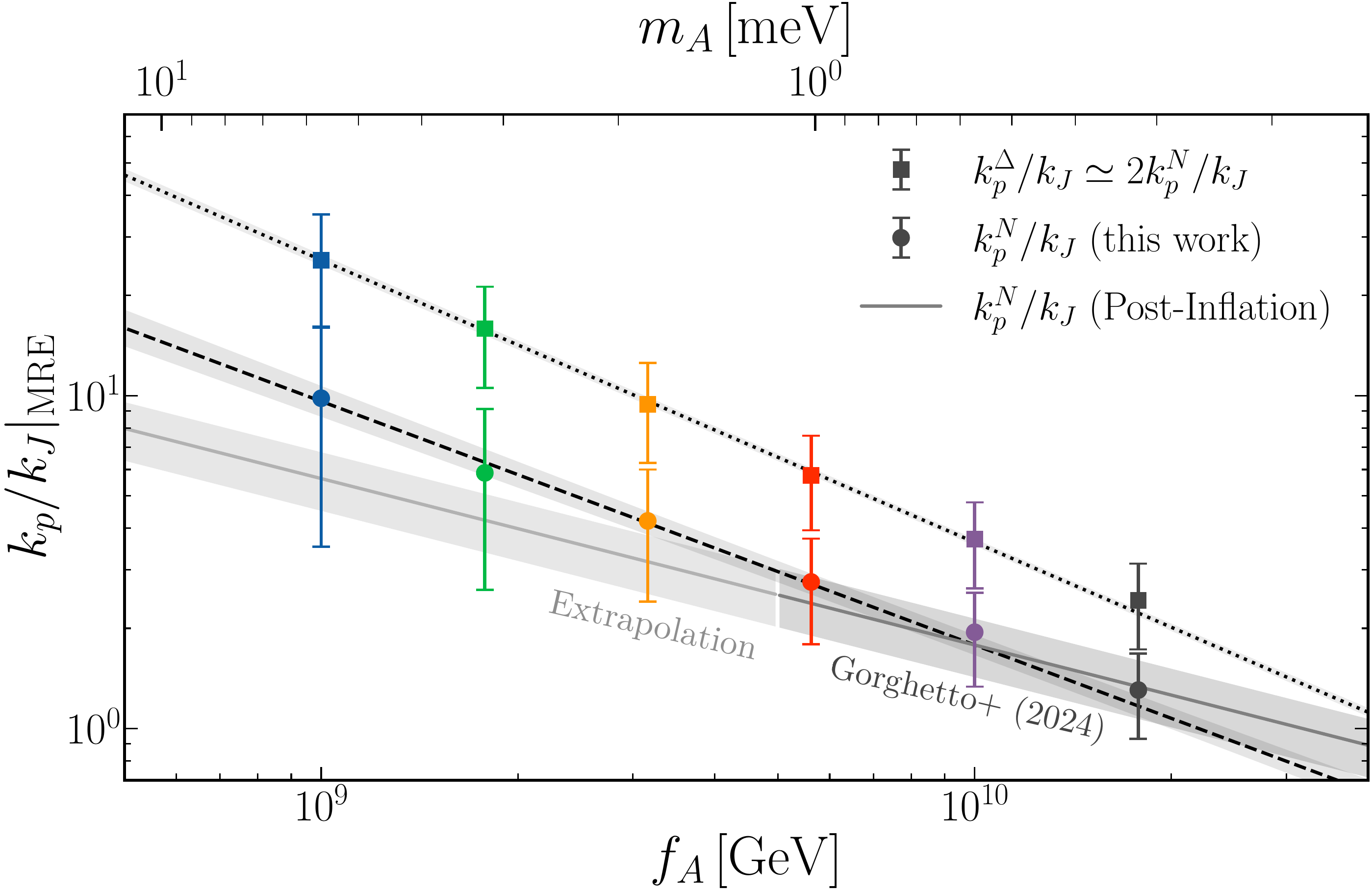}
 \caption{Peak momentum at matter--radiation equality in units of the axion
 Jeans scale $k_J$, as a function of $f_A$ (\emph{lower axis}) and $m_{A,0}$ (\emph{upper axis}).
 Circles show $k_p^N/k_J$ from the number spectrum and squares show
 $k_p^\Delta/k_J\simeq2k_p^N/k_J$ from the density spectrum; vertical bars
 indicate 25\% of the fitted peak widths (FWHM). The gray line and band show the
 post-inflationary prediction of Ref.~\cite{Gorghetto:2024vnp}, while the
 dashed and dotted curves extrapolate the corresponding comparison to the
 number- and density-spectrum peaks.}
 \label{fig:comparison}
\end{figure*}

\section{Summary of scales}

\begin{figure*}[ht]
    \centering
    \includegraphics[width=0.9\linewidth]{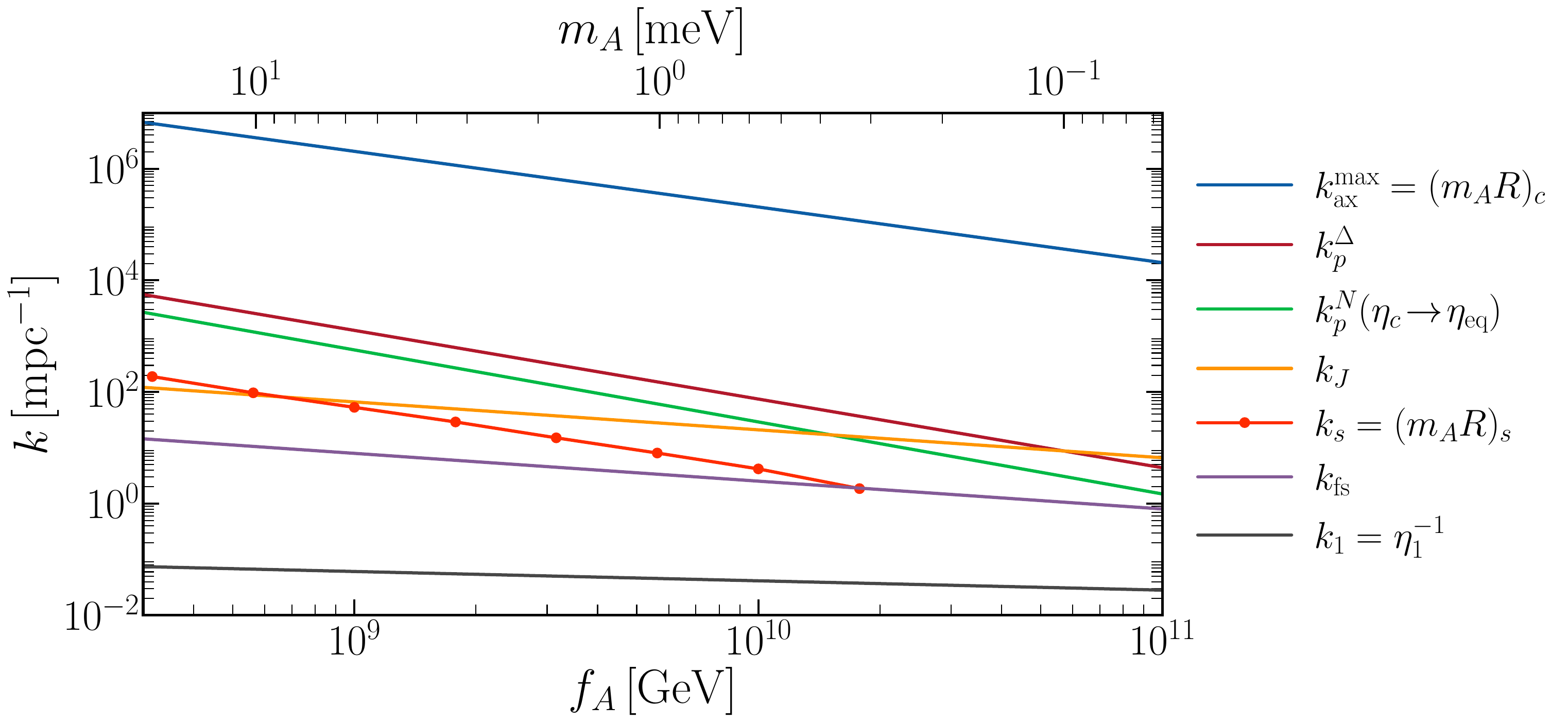}
    \caption{Summary of the scales relevant for this work. See the text for details. 
    }
    \label{fig:scale_summary}
\end{figure*}

Fig.~\ref{fig:scale_summary} shows a summary of the different length scales relevant in this work by their wavenumber. 
The base scale is the wavenumber entering the horizon when the QCD potential is $k_1=1/\eta_1$, defined as $m_A(\eta_1)=H(\eta_1)$. 

Fragmentation at large $v$ happens around trapping/stopping, $\eta_s$, due to amplification mechanisms operating by the QCD potential at momenta below the comoving axion mass at that time, $k_s=(m_AR)_s$. Due to the step time dependence of $m_AR$, this is much larger than $k_1$. 

Self/interactions are active roughly until the temperature drops to the QCD cross-over. The main wavenumber of axions, $k_p^N$, has rosen by a factor of a few at this time and will keep at this value until MRE. The peak of axion density fluctuations, $k_p^\Delta$ is a factor of $\sim2$ above. The characteristic size of axitons at that time is however much smaller, $k_{\rm ax}^{\rm max}\sim (m_A R)_c$. 

The decoupling of self-interactions due to the expansion of the Universe, makes objects which were coherently produced/held by self-interactions diffuse and eventually free-stream to a comoving distance
\be
\label{eq:freestream}
\lambda_k \sim \int_{\eta_c}^{\eta} \frac{k}{m_A R}d\eta = \frac{k\eta_c}{m_{A,0}R_c}\log \frac{R}{R_c}. 
\ee
Objects of size $\lambda_k=1/k$ diffuse more than their size for
\be
k> k_{\rm fs} \sim 2.5 {\rm mpc}^{-1} \(\frac{10^{10}\rm GeV}{f_A}\)^{-0.498}. 
\ee
This is a relatively small scale. Axitons, or our useful dense-axion-star regulators of wave-collapse, are mostly coherent objects, in the sense that they are made of few axion waves (they are a subdominant contribution to $n_A$ with very careful phase tunings. These phases are de-tuned by the increasingly free evolution, and thus these objects diffuse away. Free diffusion does not alter $N(k)$, but it erases coherent features in $\Delta^2$. 

We note here that smaller objects diffuse the longest distances, cf.  Eq.~\ref{eq:freestream}. Both axitons and the DAS of our simulations diffuse to scales longer than $1/k_1$ and are not shown. 

Note that $k_p^N$ is above the free-streaming scale, and thus the typical axion lump will diffuse away more than its size. However, these lumps seem to be incoherently placed, and the void left by one is filled by others, so these density fluctuations are not affected.  

Free-streaming merges into the onset of matter domination, where the gravitational potential of axion dark matter starts to drag axion waves to form bound objects. The wave nature of the axion field as dark matter fluid implies a $k$-dependent effective sound-speed, which prevents gravitational collapse above the so-called quantum Jeans scale $k_J$ (orange). For virtually all our simulated parameter space $f_A<6\times 10^{10}$ GeV, the peak of the density fluctuations is above the Jeans wavenumber, and this ``quantum pressure'' affects the early gravitational collapse as described in \cite{Gorghetto:2024vnp}.

\end{document}